\documentclass[useAMS,usenatbib]{mn2e}
\usepackage{natbib}
\usepackage{graphicx}
\usepackage{setspace}
\usepackage{mathtools}
\usepackage{pdflscape, longtable}
\usepackage{breqn}
\usepackage[toc,page]{appendix}
\renewenvironment{thebibliography}[1]{%
\begin{oldthebibliography}{#1}%
\setlength{\itemsep}{-.3ex}%
}%
{%
\end{oldthebibliography}%
}

\newcommand{\SpitzerGB}{SGBS}
\newcommand{\SpitzerYC}{SYC}

\def \arcsec {$^{\prime\prime}$}
\def \arcmin {$^\prime$}

\usepackage{multirow}

\usepackage[labelfont=it,textfont={bf,it}]{caption}
 
\title[The JCMT Gould Belt Survey: Evidence for radiative heating in Serpens MWC 297 and its influence on local star formation]{The JCMT Gould Belt Survey: Evidence for radiative heating in Serpens MWC 297 and its influence on local star formation}

\author[D. J. Rumble and J. Hatchell]{D. Rumble$^{1}$, J. Hatchell$^{1}$, R.A. Gutermuth$^{2}$, H. Kirk$^{3}$, J. Buckle$^{4, 5}$, S.F. Beaulieu$^{6}$, \newauthor D.S. Berry$^{7}$, H. Broekhoven-Fiene$^{8}$, M.J. Currie$^{7}$, M. Fich$^{6}$, T. Jenness$^{7, 9}$, \newauthor D. Johnstone$^{7, 3, 8}$, J.C. Mottram$^{10}$, D. Nutter$^{11}$, K. Pattle$^{12}$, J.E. Pineda$^{13, 14, 15}$, \newauthor C. Quinn$^{11}$, C. Salji$^{4, 5}$, S. Tisi$^{6}$, S. Walker-Smith$^{4, 5}$, J. Di Francesco$^{3, 8}$, \newauthor M.R. Hogerheijde$^{10}$, D. Ward-Thompson$^{12}$, L.E. Allen$^{16}$, L.A. Cieza$^{17}$, \newauthor M.M. Dunham$^{18}$, P.M. Harvey$^{19}$, K.R. Stapelfeldt$^{20}$, P. Bastien$^{21}$, H. Butner$^{22}$, \newauthor M. Chen$^{8}$, A. Chrysostomou$^{23}$, S. Coude$^{21}$, C.J. Davis$^{24}$, E. Drabek-Maunder$^{25}$, \newauthor A. Duarte-Cabral$^{1}$, J. Fiege$^{26}$, P. Friberg$^{7}$, R. Friesen$^{27}$, G.A. Fuller$^{14}$, \newauthor S. Graves$^{4, 5}$, J. Greaves$^{28}$, J. Gregson$^{29, 30}$, W. Holland$^{31, 32}$, G. Joncas$^{33}$, \newauthor J.M. Kirk$^{12}$, L.B.G. Knee$^{3}$, S. Mairs$^{8}$, K. Marsh$^{11}$, B.C. Matthews$^{3, 8}$, \newauthor G. Moriarty-Schieven$^{3}$, J. Rawlings$^{34}$, J. Richer$^{4, 5}$, D. Robertson$^{35}$, \newauthor E. Rosolowsky$^{36}$, S. Sadavoy$^{37}$, H. Thomas$^{7}$, N. Tothill$^{38}$, S. Viti$^{34}$, G.J. White$^{29, 30}$, \newauthor C.D. Wilson$^{35}$, J. Wouterloot$^{7}$, J. Yates$^{34}$, M. Zhu$^{39}$ \\ 
$^{1}$Physics and Astronomy, University of Exeter, Stocker Road, Exeter EX4 4QL, UK\\
\\
$^{2}$Department of Astronomy, University of Massachusetts, Amherst, MA, USA\\
$^{3}$National Research Council of Canada, 5071 West Saanich Rd, Victoria, BC, V9E 2E7, Canada\\
$^{4}$Astrophysics Group, Cavendish Laboratory, J J Thomson Avenue, Cambridge, CB3 0HE\\
$^{5}$Kavli Institute for Cosmology, Institute of Astronomy, University of Cambridge, Madingley Road, Cambridge, CB3 0HA, UK\\
$^{6}$Department of Physics and Astronomy, University of Waterloo, Waterloo, Ontario, N2L 3G1, Canada  \\
$^{7}$Joint Astronomy Centre, 660 N. A`oh\={o}k\={u} Place, University Park, Hilo, Hawaii 96720, USA\\
$^{8}$Department of Physics and Astronomy, University of Victoria, Victoria, BC, V8P 1A1, Canada\\
$^{9}$Department of Astronomy, Cornell University, Ithaca, NY 14853, USA\\
$^{10}$Leiden Observatory, Leiden University, PO Box 9513, 2300 RA Leiden, the Netherlands\\
$^{11}$School of Physics and Astronomy, Cardiff University, the Parade, Cardiff, CF24 3AA, UK\\
$^{12}$Jeremiah Horrocks Institute, University of Central Lancashire, Preston, Lancashire, PR1 2HE, UK\\
$^{13}$European Southern Observatory (ESO), Garching, Germany\\
$^{14}$Jodrell Bank Centre for Astrophysics, Alan Turing Building, School of Physics and Astronomy, University of Manchester, \\Oxford Road, Manchester, M13 9PL, UK\\
$^{15}$Current address: Institute for Astronomy, ETH Zurich, Wolfgang-Pauli-Strasse 27, CH-8093 Zurich, Switzerland\\
$^{16}$National Optical Astronomy Observatories, Tucson, AZ, USA\\
$^{17}$Facultad de Ingenier{\'i}a, Universidad Diego Portales, Av. Ej{\'e}rcito 441, Santiago, Chile\\
$^{18}$Harvard-Smithsonian Center for Astrophysics, 60 Garden Street, MS 78, Cambridge, MA 02138, USA\\
$^{19}$Department of Astronomy, the University of Texas at Austin, 2515 Speedway, Stop C1400, Austin, TX 78712-1205, USA\\
$^{20}$NASA Goddard Space Flight Center, Exoplanets and Stellar Astrophysics Laboratory, Code 667, Greenbelt, MD 20771, USA\\
$^{21}$Universit\'e de Montr\'eal, Centre de Recherche en Astrophysique du Qu\'ebec et d\'epartement de physique, C.P. 6128, succ. centre-ville, \\Montr\'eal, QC, H3C 3J7, Canada\\
$^{22}$James Madison University, Harrisonburg, Virginia 22807, USA\\
$^{23}$School of Physics, Astronomy \& Mathematics, University of Hertfordshire, College Lane, Hatfield, HERTS AL10 9AB, UK\\
$^{24}$Astrophysics Research Institute, Liverpool John Moores University, Egerton Warf, Birkenhead, CH41 1LD, UK\\
$^{25}$Imperial College London, Blackett Laboratory, Prince Consort Rd, London SW7 2BB, UK\\
$^{26}$Dept of Physics \& Astronomy, University of Manitoba, Winnipeg, Manitoba, R3T 2N2 Canada\\
$^{27}$Dunlap Institute for Astronomy \& Astrophysics, University of Toronto, 50 St. George St., Toronto ON M5S 3H4 Canada\\
$^{28}$Physics \& Astronomy, University of St Andrews, North Haugh, St Andrews, Fife KY16 9SS, UK\\
$^{29}$Dept. of Physical Sciences, the Open University, Milton Keynes MK7 6AA, UK\\
$^{30}$The Rutherford Appleton Laboratory, Chilton, Didcot, OX11 0NL, UK.\\
$^{31}$UK Astronomy Technology Centre, Royal Observatory, Blackford Hill, Edinburgh EH9 3HJ, UK\\
$^{32}$Institute for Astronomy, Royal Observatory, University of Edinburgh, Blackford Hill, Edinburgh EH9 3HJ, UK\\
$^{33}$Centre de recherche en astrophysique du Qu\'ebec et D\'epartement de physique, de g\'enie physique et d'optique, Universit\'e Laval, \\1045 avenue de la m\'edecine, Qu\'ebec, G1V 0A6, Canada\\
$^{34}$Department of Physics and Astronomy, UCL, Gower St, London, WC1E 6BT, UK\\
$^{35}$Department of Physics and Astronomy, McMaster University, Hamilton, ON, L8S 4M1, Canada\\
$^{36}$Department of Physics, University of Alberta, Edmonton, AB T6G 2E1, Canada\\
$^{37}$Max Planck Institute for Astronomy, K\"{o}nigstuhl 17, D-69117 Heidelberg, Germany\\
$^{38}$University of Western Sydney, Locked Bag 1797, Penrith NSW 2751, Australia\\
$^{39}$National Astronomical Observatory of China, 20A Datun Road, Chaoyang District, Beijing 100012, China\\}
\begin{document}

\date{Accepted 2014 Sept. 1. Received 2014 July 25.}

\pagerange{\pageref{firstpage}--\pageref{lastpage}} \pubyear{2014}

\maketitle 

\label{firstpage}

\begin{abstract}

%Insert abstract here - 1 paragraph - 9 sentences - words 260

%Aims/Observations:

We present SCUBA-2 450$\micron$ and 850$\micron$ observations of the Serpens MWC 297 region, part of the JCMT Gould Belt Survey of nearby star-forming regions. %The area covered includes the B1.5Ve Herbig AeBe star MWC 297 and a collection of low mass dense clouds. 
Simulations suggest that radiative feedback influences the star-formation process and we investigate observational evidence for this by constructing temperature maps. %Methods: 
Maps are derived from the ratio of SCUBA-2 fluxes and a two component model of the JCMT beam for a fixed dust opacity spectral index of $\beta$ = 1.8. 
Within 40\arcsec\ of the B1.5Ve Herbig star MWC 297, the submillimetre fluxes are contaminated by free-free emission with a spectral index of 1.03$\pm$0.02, consistent with an ultra-compact HII region and polar winds/jets. Contamination accounts for 73$\pm$5\,per cent and 82$\pm$4\,per cent of peak flux at 450$\micron$ and 850$\micron$ respectively. The residual thermal disk of the star is almost undetectable at these wavelengths. %, a finding that has consequences for submillimetre observations of regions where O/B stars are forming. 
Young Stellar Objects are confirmed where SCUBA-2 850$\micron$ clumps identified by the \textsc{fellwalker} algorithm coincide with \emph{Spitzer} Gould Belt Survey detections. %SCUBA-2 YSOs are classified through calculation of $\alpha_{\mathrm{IR}}$, $T_{\mathrm{bol}}$ and $L_{\mathrm{smm}}$/$L_{\mathrm{bol}}$. %construction of spectral energy distributions and . %Results: 
We identify 23 objects and use $T_{\mathrm{bol}}$ to classify nine YSOs with masses 0.09 to 5.1$M_{\odot}$. We find two Class 0, one Class 0/I, three Class I and three Class II sources. %, calculated using real temperature data where available. 
The mean temperature is 15$\pm$2K for the nine YSOs and 32$\pm$4K for the 14 starless clumps. %In the later case our result is consistent with the assumptions and models in the literature. 
We observe a starless clump with an abnormally high mean temperature of 46$\pm$2K and conclude that it is radiatively heated by the star MWC 297. %, located 0.025\,pc to the SW. %Conclusions: 
Jeans stability provides evidence that radiative heating by the star MWC 297 may be suppressing clump collapse.   

\end{abstract}


%%%%%%%%%%%%%%%%%%%%%%%%%%%%%%%%%%%%%%%%%%%%%%%%%%%

\section{Introduction}

%9 paragraph - 4 sentences per para - 90 per para - 4 ref per para

%\subsection{PARA1 - Motivation within context of the role of radiative feedback within star formation.}
%\subsection{PARA2 - alt. methods of calculating temperature}
%\subsection{PARA3 - Previous methods of calculating using dust}
%\subsection{PARA4 - SCUBA-2 GBS ((Figure~\ref{fig:minimaps1})-Thompson)}
%\subsection{PARA5 - Serpens Aquila (distance to, etc)}
%\sbsection{PARA6/7 - What is MWC 297 (x2)}
%\subsection{PARA8 - Summarise paper}

The temperature of gas and dust in dense, star-forming clouds is vital in determining whether or not clumps undergo collapse 
and potentially form stars \citep{Jeans:1902dz}. Dense clouds can be heated by a number of mechanisms: heating from 
the interstellar radiation field (ISRF) \citep{Mathis:1983dq, Shirley:2000uq, Shirley:2002vn}, evolved OB stars with HII regions 
\citep{Koenig:2008jo, Deharveng:2012fk} or strong stellar winds \citep{Canto:1984dq, Ziener:1999kl, Malbet:2007zr}; and 
internally through gravitational collapse of the Young Stellar Object (YSO) and accretion onto its surface \citep{calvet98}. 
%Outflows are also observed in the form of jets \citep{Skinner:1993bh, Bontemps:1996fu, Gomez:2004cr} and stellar winds 
%\citep{Drew:1997qf} as a feedback mechanism as well as radiative heating from the photosphere of the star \citep{Hatchell:2013ij}. 
Radiative feedback is thought to play an important role in the formation of the most massive stars through the suppression of 
core fragmentation \citep{Bate:2009uq, Offner:2009pt, Hennebelle:2011ly}. 
%This paper looks at observational evidence of radiative feedback directly influencing the formation process.

\begin{figure*}
\begin{centering}
\includegraphics[scale=0.7]{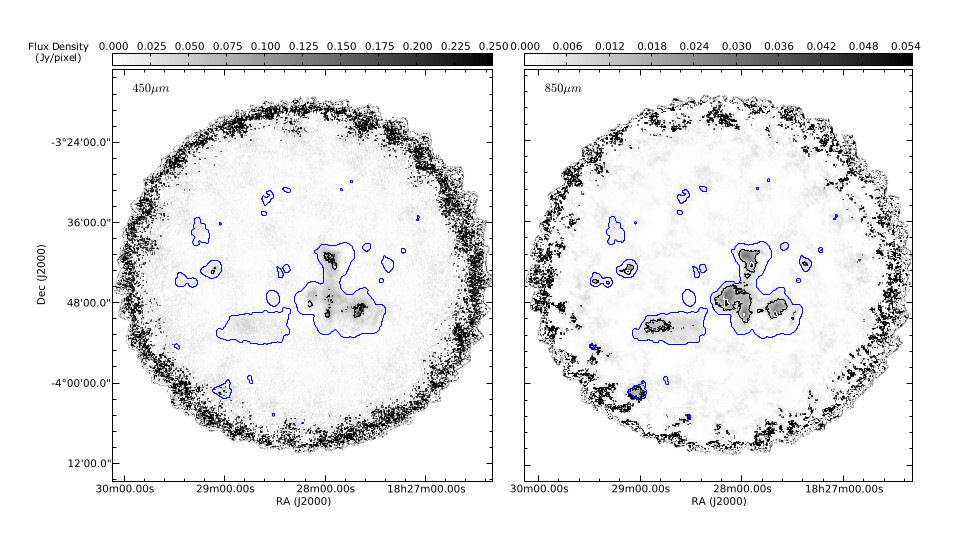}
\caption{SCUBA-2 450\,$\micron$ (\emph{left}) and 850\,$\micron$ (\emph{right}) data. Contours show 5$\sigma$ and 15$\sigma$ levels in both cases: levels are at 0.082, 0.25 Jy/ 4\arcsec\ pixel and 0.011, 0.033 Jy/ 6\arcsec\ pixel  at 450\,\micron\ and 850\,\micron\ respectively. The blue outer contour shows the data reduction mask for the region, based on \emph{Herschel} 500\,$\micron$ observations. Noise levels increase towards the edges of the map on account of the mapping method outlined in Section 2.1.} \label{fig:maps}
\end{centering}
\end{figure*} 

The temperature of star-forming regions has been observed and calculated using a variety of different methods and data. 
Some methods utilise line emission from the clouds: for example, \cite{Ladd:1994ly} and \cite{Curtis:2010zr} examine the 
CO excitation temperature and \cite{Huttemeister:1993ve} looked at a multilevel study of ammonia lines. Often temperature 
assumptions are made in line with models of Jeans instability and Bonnor-Ebert Spheres \citep{Ebert:1955vn, Bonnor:1956vn, 
Johnstone:2000fk}. An alternative method is to fit a single temperature greybody model to an observed Spectral Energy Distribution (SED) 
of dust continuum emission for the YSO \citep{Hildebrand:1983fy}; however, this method is sensitive to the completeness 
of the spectrum, the emission models and local fluctuations in dust properties \citep{Konyves:2010oq, Bontemps:2010fk}.

Where multiple submillimetre observations exist, low temperatures (less than 20\,K), which favour cloud collapse, can be 
inferred by the relative intensity of longer wavelengths over shorter wavelengths. For example, \emph{Herschel}  provides FIR and 
submillimetre data through PACS bands 70\,$\micron$, 100\,$\micron$ and 160\,$\micron$ and SPIRE bands 250\,$\micron$, 
350\,$\micron$ and 500\,$\micron$ \citep{Pilbratt:2010fk}. \cite{Menshchikov:2010kl, Andre:2010kx} use \emph{Herschel} 
data to construct a low resolution temperature map for the Aquila and Polaris region through fitting a greybody to dust 
continuum fluxes (an opacity-modified blackbody spectrum). \emph{Herschel} data offers five bands of FIR and submillimetre 
observations and low noise levels; however, it lacks the resolution of the JCMT which can study structure on a scale of 
7.9\arcsec (450\,$\micron$) and 13.0\arcsec (850\,$\micron$) \citep{Dempsey:2013uq} as opposed to 25.0\arcsec\ and larger 
for 350\,$\micron$ or greater submillimetre wavelengths. \cite{Sadavoy:2013qf} combine \emph{Herschel} and SCUBA-2 
data to constrain both $\beta$ and temperature.

%5.8\arcsec (70\,$\micron$), 7.1\arcsec (100\,$\micron$), 11.2\arcsec (160\,$\micron$), 18.2\arcsec (250\,$\micron$), 25.0\arcsec (350\,$\micron$) and 36.4\arcsec (500\,$\micron$) with \emph{Herschel} \citep{Aniano:2011fk}. 

This work develops a method which takes the ratio of fluxes at submillimetre wavelengths when insufficient data points exist 
to construct a complete SED. The ratio method allows the constraint of temperature or $\beta$, but not both simultaneously. 
Throughout this paper we used a fixed $\beta$. The value and justification for this are discussed in Section 3. Similar methods 
have been applied by \cite{Wood:1994qf}, \cite{Arce:1999bh} and \cite{Font:2001cr} and used by \cite{Kraemer:2003uf} 
at 12.5\,$\micron$ and 20.6\,$\micron$ and by \cite{Schnee:2005zr} at 60\,$\micron$ and 100\,$\micron$. \cite{Mitchell:2001ve} 
first used 450\,$\micron$ and 850\,$\micron$ fluxes from SCUBA, though full analysis was limited by the quality and quantity of 
450\,$\micron$ data. A more rigorous analysis of SCUBA data was completed by \cite{Reid:2005ly} who are able to constrain 
errors on the temperature maps from sky opacity and the error beam components. Most recently similar methods have 
been used by \cite{Hatchell:2013ij} to analyse heating in NGC1333. This work looks to utilise these methods to further 
investigate radiative feedback in star-forming regions. 

This study uses data from the JCMT Gould Belt Survey (GBS) of nearby star-forming regions \citep{WardThompson:2007ve}. 
The survey maps all major low and intermediate-mass star-forming regions within 0.5 kpc. The JCMT GBS provides some of the 
deepest maps of star forming regions where $A_v > 3$  with a target sensitivity of 3 mJy beam$^{-1}$ at 850\,$\micron$ and 12 mJy 
beam$^{-1}$ at 450\,$\micron$. The improved resolution of the JCMT also allows for more detailed study of large scale 
structures such as filaments, protostellar envelopes, extended cloud structure and morphology down to the Jeans length.
%and with the aim of plotting the location Class 0 and I protostars and better determine their evolutionary timescale and processes. 

This paper focuses on the Serpens MWC 297 region, a region of low mass star formation associated with the B star MWC 297 and 
part of the larger Serpens-Aquila star forming complex. The exact distance to the star MWC 297 is a matter of debate. Preliminary 
estimates of the distance to the star were put at $450\hbox{ pc}$ by \cite{Canto:1984dq} and $530\pm70\hbox{ pc}$ 
by \cite{Bergner:1988bh}. \cite{Drew:1997qf} used a revised spectral class of B1.5Ve to calculate a closer distance of 
$250 \pm 50\hbox{ pc}$ which is in line with the value of $225\pm55\hbox{ pc}$ derived by \cite{Straizys:2003nx} for 
the minimum distance to the extinction wall of the whole Serpens-Aquila rift of which the star MWC 297 is thought to be a part. 
The distance to the Serpens-Aquila rift was originally put at a distance of $250 \pm 50\hbox{ pc}$ due to association with 
Serpens Main, a well constrained star forming region the north of MWC 297; however, recent work by \cite{Dzib:2010dq,Dzib:2011cr} 
has placed Serpens Main at $429\pm2 \hbox{ pc}$ using parallax. \cite{Maury:2011ys} argues that previous methods measured 
the foreground part of the rift and that Serpens Main is part of a separate star forming region positioned further back. 
On this basis, we adopt a distance of $d = 250\pm50\hbox{ pc}$ to the Aquila rift and the Serpens MWC 297 region \citep{Sandell:2011dz}. 

The star MWC 297 is an isolated, intermediate mass Zero Age Main Sequence (ZAMS) star at RA(J2000) = $18^{h}$ 
$27^{m}$ $40^{s}.6$, Dec. (J2000) = $-03^{\circ}$ $50'$ 11\arcsec. \cite{Drew:1997qf} noted that MWC 297 has strong reddening 
due to foreground extinction ($A_V$ = 8) and particularly strong Balmer line emission. The star has been much studied as an 
example of a classic Herbig AeBe star, defined by \cite{Herbig:1960eu}, \cite{Hillenbrand:1992kl} and \cite{Mannings:1994kx} 
as an intermediate mass (1.5 to 10\,$M_{\odot}$) equivalent of classical T-Tauri star, typically a Class III pre-main sequence 
star of spectral type A or B. 

Herbig AeBe stars are strongly associated with circumstellar gas and dust with a wide range of temperatures. \cite{Berrilli:1992cr} 
and \cite{di-Francesco:1994dq,  di-Francesco:1998fk} find evidence of an extended disk/circumstellar envelope around the star MWC 297. 
Radio observations constrain disk size to $< 100$\,AU and also find evidence for free-free emission at the poles that suggest the 
presence of polar winds or jets \citep{Skinner:1993bh, Malbet:2007zr, Manoj:2007ly}. MWC 297 is in a loose binary system 
with an A2 star, hereafter referred to as \emph{OSCA}, which has been identified as a source of X-ray emission \citep{Vink:2005uq, 
Damiani:2006ve}. There is evidence for an optical nebulae, SH2-62, which is coincident with MWC 297 \citep{Sharpless:1959hc}. 

This paper is structured as follows. In Section 2 we describe the observations of the Serpens MWC 297 region by SCUBA-2 and \emph{Spitzer}. 
In Section 3 we apply our method for producing temperature maps from the flux ratio and asses possible sources of contamination of the 
submillimetre data. In Section 4 we identify clumps in the region and calculate masses. In Section 5 we examine 
external catalogues of YSO candidates for the region and produce our own SCUBA-2 catalogue of star-forming cores. 
In Section 6 we discuss our findings in the context of radiative feedback and global star formation within the region and ask if there is any evidence 
that radiative feedback from previous generations of stars is influencing present day and future star formation. %In Section 7 we conclude our findings.

%%%%%%%%%%%%%%%%%%%%%%%%%%%%%%%%%%%%%%%%%%%%%%%%%%%

\section{Observations and Data Reduction}

\subsection{SCUBA-2}

Serpens MWC 297 was observed with SCUBA-2 \citep{Holland:2013fk} on the 5th and 
8th of July 2012 as part of the JCMT Gould Belt Survey (GBS, \citeauthor{WardThompson:2007ve} 
\citeyear{WardThompson:2007ve}) MJLSG33 SCUBA-2 Serpens Campaign 
\citep{Holland:2013fk}. One scan was taken on the 5th at 12:55 UT in good 
Band 2 with 225 GHz opacity $\tau_{\mathrm{225}} = 0.04-0.06$. Five further scans 
taken on the 8th between 07:23 and 11:31 UT in poor Band 2, $\tau_{\mathrm{225}} 
= 0.07-0.11$.

Continuum observations at 850\,$\micron$ and 450\,$\micron$ were made using 
fully sampled 30\arcmin\ diameter circular regions (PONG1800 mapping mode, 
\citeauthor{Chapin:2013vn} \citeyear{Chapin:2013vn}) centered on RA(J2000) = 
$18^{h}$ $28^{m}$ $13^{s}.8$, Dec. (J2000) = $-03^{\circ}$ $44'$ 1.7\arcsec.

The data were reduced using an iterative map-making technique (\texttt{makemap} 
in {\sc smurf}, \citeauthor{Chapin:2013vn} \citeyear{Chapin:2013vn}, 
\citeauthor{Jenness:2013fk} \citeyear{Jenness:2013fk}), and gridded to 6\arcsec\ 
pixels at 850\,$\micron$, 4\arcsec\ pixels at 450\,$\micron$.  
The iterations were halted when the map pixels, on average, changed by 
$<$0.1\,per cent of the estimated map rms. The initial reductions of each individual 
scan were coadded to form a mosaic from which a signal-to-noise mask was 
produced for each region.  This was combined with \emph{Herschel} 500\,$\micron$ 
emission at greater than 2\,Jy/beam to include all potential emission regions. 
The final mosaic was produced from a second reduction using this mask to define 
areas of emission. Detection of emission structure and calibration accuracy
are robust within the masked regions, and are uncertain outside of the masked 
region. The reduced map and mask are shown in Figure \ref{fig:maps}.
% The mask used in the reduction can be seen in the quality array in the reduced datafile (Figure \ref{fig:maps}).

A spatial filter of 600\arcsec\ is used in the reduction, which means that flux 
recovery is robust for sources with a Gaussian Full Width Half Maximum (FWHM) 
less than 2.5\arcmin. Sources between 2.5\arcmin\ and 7.5\arcmin\ will be 
detected, but both the flux and the size are underestimated because Fourier 
components with scales greater than 5\arcmin\ are removed by the filtering 
process. Detection of sources larger than 7.5\arcmin\ is dependent on the 
mask used for reduction.

The data presented in Figure~\ref{fig:maps} are initially calibrated in units of pW 
and are converted to Jy per pixel using Flux Conversion Factors (FCFs) derived 
by \cite{Dempsey:2013uq} from the average values of JCMT calibrators. 
By correcting for the pixel area, it is possible to convert maps of units Jy/pixel 
to Jy/beam using 
\begin{equation}
S_{\textup{beam}} = S_{\textup{pixel}}\frac{\textup{FCF}_{\textup{peak}}}{\textup{FCF}_{\textup{arcsec}}}\frac{1}{\textup{Pixel area}}.
\label{eqn:pixelFCF} 
\end{equation}
%\begin{equation}
%S_{\textup{pixel}} = S_{\textup{beam}}\frac{\textup{Pixel area}}{\textup{Beam area}}
%\label{eqn:pixelgen} 
%\end{equation}
FCF$_{\mathrm{arcsec}}$ = 2.34$\pm$0.08 and 4.71$\pm$0.5 Jy/pW/arcsec$^{2}$, 
at 850\,$\micron$ and 450\,$\micron$ respectively, and FCF$_{\mathrm{peak}}$ = 537$\pm$26 
and 491$\pm$67 Jy/pW at 850\,$\micron$ and 450\,$\micron$ respectively. The PONG 
scan pattern leads to lower noise in the map centre and overlap regions, while data 
reduction and emission artefacts can lead to small variations in the noise over the whole 
map. Typical noise levels were 0.0165 and 0.0022 Jy per pixel at 450\,$\micron$ and 
850\,$\micron$ respectively.

The JCMT beam can be modelled as two Gaussian components \citep{Drabek:2012uq, 
Dempsey:2013uq}. The primary (or main) beam contains the bulk of the signal and is 
well described by a Gaussian, $G_{\mathrm{MB}}$, but in addition to this there is also 
a secondary beam which is much wider and lower in amplitude, $G_{\mathrm{SB}}$. 
Together they make up the 2-component beam of the telescope, 

\begin{equation}
G_{\mathrm{total}}=a G_{\mathrm{MB}} + b G_{\mathrm{SB}}, \label{eqn:effbeam} 
\end{equation}
where $a$ and $b$ are relative amplitude, listed in Table~\ref{table:beams} 
alongside the FWHM, $\theta$, of the primary (MB) and secondary (SB) beams.   

\begin{table}
\caption{JCMT beam properties}
\label{table:beams}
\begin{center}
\centering
\begin{tabular}{l | p{1cm} | p{1cm}  }
		&	450\,$\micron$ &	850\,$\micron$ \\
	\hline
	$\theta_{\mathrm{MB}}$ &	7.9\arcsec\	&	13.0\arcsec\		\\
	$\theta_{\mathrm{SB}}$	&	25.0\arcsec\	&	48.0\arcsec\		\\	
	$a$	&	0.94	&	0.98		\\
	$b$	&	0.06	&	0.02		\\
	Pixel size&	4\arcsec\	&	6\arcsec\		\\
	\hline
\end{tabular}
\end{center}

JCMT beam Full Width Half Maximum ($\theta$) and relative amplitudes from \cite{Dempsey:2013uq} Table 1. Pixel sizes are those chosen by the JCMT SGBS data reduction team.

\end{table}

\subsection{\emph{Spitzer} catalogues}

%\setcounter{table*}{3} % replace with the previous table number in the paper
%\footnotesize
%\begin{center}
%\begin{longtable}{@{}lccccccccc@{}}
% \centering
%\begin{minipage}{230mm}

\begin{table*}
\caption{A sample of \emph{Spitzer} YSO candidates (YSOc) from the \SpitzerGB. The full version appears as supplementary material online.}
\label{table:SGBS_YSO_small} 
\centering
\begin{tabular}{@{}lccccccccc@{}}
\hline
%  \begin{tabular}{@{}lcccccccccccr@{}}
ID & SSTgbs &\multicolumn{4}{c}{ \hrulefill\quad \emph{Spitzer} IRAC \quad \hrulefill }&\multicolumn{2}{c}{\hrulefill\quad \emph{Spitzer} MIPS\quad \hrulefill}\\

	 &&{$S_{\mathrm{3.6}}$ }&{$S_{\mathrm{4.5}}$ }&{$S_{\mathrm{5.8}}$ }&{$S_{\mathrm{8.0}}$ }&{$S_{\mathrm{24}}$ }&{$S_{\mathrm{70}}$ } &$\alpha_{\mathrm{IR}}$\footnotemark \\
&&{ mJy }&{mJy }&{mJy }&{mJy }&{mJy }&{mJy } &\\

\hline

%\footnotetext{Spectral index calculated from a fit between K$_S$ and MIPS $24\umu$m}

YSOc2 &J18271323-0340146 &$193.0\pm10.4$ &$220.0\pm11.3$ &$258.00\pm13.40$ &$354.00\pm17.20$ &$1170.0\pm110.0$ &$1610\pm 172$ &$-0.17$ \\
YSOc11 &J18272664-0344459 &$76.2\pm 3.7$ &$87.4\pm 4.5$ &$92.30\pm4.33$ &$116.00\pm6.00$ &$198.0\pm18.4$ &$ 312\pm  41$ &$-0.49$ \\
YSOc15 &J18273641-0349133 &$ 5.4\pm 0.3$ &$ 3.5\pm 0.2$ &$10.60\pm1.77$ &$38.60\pm8.78$ &$107.0\pm28.4$ &\ldots &$0.03$ \\
YSOc16 &J18273671-0350047 &$11.0\pm 0.6$ &$16.7\pm 0.9$ &$24.10\pm2.55$ &$29.00\pm2.68$ &$340.0\pm84.2$ &\ldots &$0.75$ \\
YSOc17 &J18273710-0349386 &$868.0\pm43.7$ &$985.0\pm52.2$ &$1100.00\pm57.60$ &$1230.00\pm67.20$ &$1780.0\pm262.0$ &\ldots &$-0.43$ \\
YSOc21 &J18273921-0348241 &$ 1.0\pm 0.3$ &$ 1.3\pm 0.1$ &$10.90\pm1.48$ &$38.70\pm7.55$ &$73.5\pm15.2$ &\ldots &$1.51$ \\
YSOc38 &J18275223-0344173 &$ 0.1\pm 0.0$ &$ 0.0\pm 0.0$ &$0.14\pm0.05$ &$0.38\pm0.11$ &$ 3.8\pm 1.2$ &$ 454\pm 195$ &$1.17$ \\
YSOc32 &J18275019-0349140 &$44.2\pm 2.2$ &$63.7\pm 3.1$ &$84.40\pm4.01$ &$96.70\pm4.79$ &$427.0\pm39.6$ &\ldots &$0.17$ \\
YSOc41 &J18275472-0342386 &$153.0\pm 7.7$ &$254.0\pm12.7$ &$370.00\pm17.80$ &$571.00\pm27.20$ &$2100.0\pm202.0$ &$3480\pm 446$ &$0.56$ \\
YSOc47 &J18280541-0346598 &$11.9\pm 0.6$ &$37.0\pm 1.9$ &$59.80\pm2.80$ &$70.10\pm3.36$ &$344.0\pm32.1$ &$3560\pm 386$ &$0.96$ \\
YSOc73 &J18290545-0342456 &$ 4.7\pm 0.2$ &$14.2\pm 0.7$ &$21.80\pm1.04$ &$25.00\pm1.20$ &$49.4\pm 4.6$ &$ 662\pm  71$ &$0.30$ \\

\end{tabular}
%\end{tabular}
%\end{minipage}
\end{table*}

% Damian's paper table 3
% Gutermuth et al. (2008) Serpens South (cluster grinder x c2d)
% Gutermuth et al. (2009) clusters (cluster grinder)

%\emph{Spitzer}GB Description based on (but much cut down from) ophN_submitted.tex and Harvey et al. 2008 with reference to gutermuth et al. 2008
The MWC 297 region was observed twice by \emph{Spitzer} in the 
mid-infrared, first as part of the \emph{Spitzer} Young Clusters Survey 
(\SpitzerYC; \citealt{gutermuth09}) and secondly as part of the 
\emph{Spitzer} legacy program ``Gould's Belt: star formation in the solar 
neighbourhood'' (\SpitzerGB, PID: 30574).   

In both surveys, mapping observations were taken at 3.6, 4.5, 5.8 and 8.0\,\micron\ 
with the Infrared Array Camera (IRAC; \citealt{fazio04}) and at 24\,\micron\
with the Multiband Imaging Photometer for \emph{Spitzer} (MIPS;
\citealt{rieke04}).  The \SpitzerGB\ also provided MIPS 70 and 
160\,\micron\ coverage, although the latter saturates towards
MWC 297.  The IRAC observations have an angular resolution of 
2\arcsec\ whereas MIPS is diffraction limited with 6\arcsec, 18\arcsec\ 
and 40\arcsec\ resolution at 24, 70 and 160\,\micron\ respectively. 

The \SpitzerYC\ targeted 36 young, nearby, star-forming clusters.  
Specifically, a 15\arcmin\ $\times$ 15\arcmin\ area centred on the star 
MWC 297 was observed as part of this survey.  Observations, data 
reduction and source classification were carried out using ClusterGrinder 
as described in \citet{gutermuth09}.
   
The \SpitzerGB\ program aimed to complete the mapping of local 
star formation started by the \emph{Spitzer} ``From Molecular Cores to 
Planet-forming Disks'' (c2d) project \citep{c2d,evans09} by targeting 
the regions IC5146, CrA, Scorpius (renamed Ophiuchus~North), 
Lupus II/V/VI, Auriga, Cepheus Flare, Aquila (including MWC 297), 
Musca, and Chameleon to the same sensitivity and using the same 
reduction pipeline \citep{gutermuth08,harvey08,kirk09,peterson11,spezzi11,hatchell12}. 
The Serpens MWC 297 region was mapped as part of the Aquila rift 
molecular cloud that also includes the Serpens~South cluster 
\citep{gutermuth08} and Aquila~W40 regions.  
%in the `Aquila' extension to the c2d-mapped Serpens~Main / NH$_3$ / VV~Ser region, an extension 
The observational setup, data reduction and source classification used 
the c2d pipeline as described in detail in \citet{harvey07}, \cite{harvey08}, \cite{gutermuth08} 
and the c2d~delivery document \citep{c2ddel}.

As a result of these two \emph{Spitzer} survey programmes, two independent 
lists of Young Stellar Object candidates (YSOc) exist for the MWC 297 
region. We refer to \citet{gutermuth09} for the \SpitzerYC\ observations 
and \SpitzerGB\ for the \emph{Spitzer} Gould's Belt survey. The \SpitzerGB\ 
catalogue (Table \ref{table:SGBS_YSO_small}) covers the entire region 
mapped by SCUBA-2 whereas the \SpitzerYC\ extent is 15\arcmin\ $\times$ 
15 \arcmin\ around MWC 297. YSOc from these methods are revisited in Section 5.1.

\begin{figure}
\begin{centering}
\includegraphics[scale=0.35]{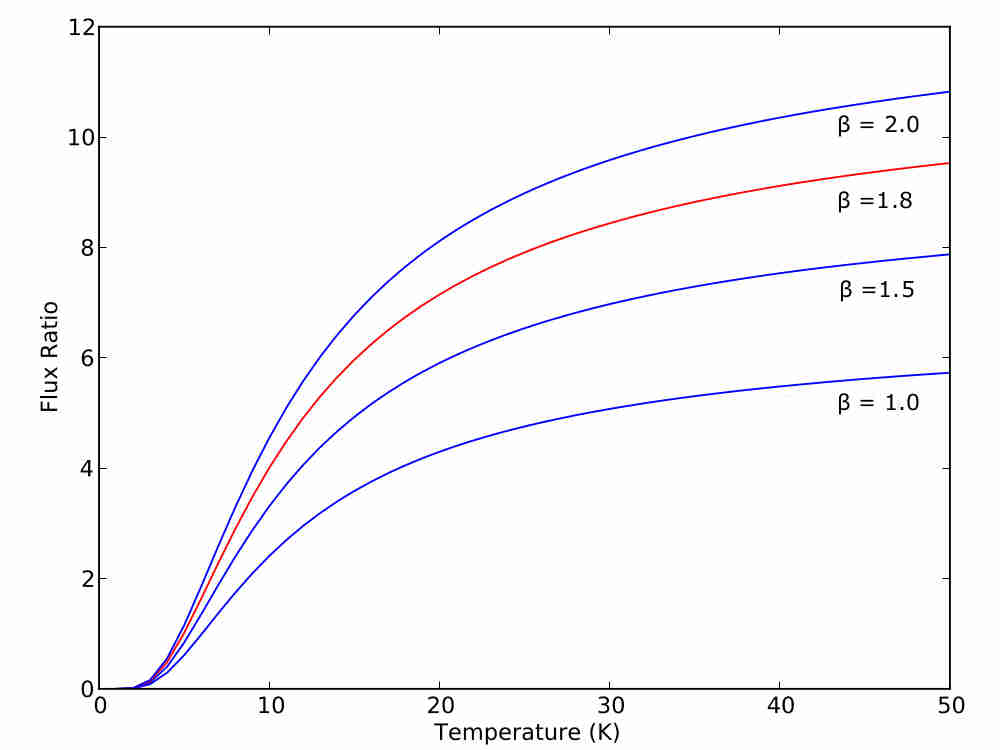}
\caption{Flux ratio as a function of temperature as described by Equation \ref{eqn:temp}. The temperature range is that commonly observed in protostellar cores.} \label{fig:SR_temp}
\end{centering}
\end{figure} 

%\begin{figure}
%\begin{center}
%\includegraphics[scale=0.5]{temp_flowchart_simple.jpeg}
%\caption{A flow chart providing a simplified description of the ratio mapping process}
%\label{fig:flow}
%\end{center}
%\end{figure}

\section{Temperature mapping}

Using the ratio of 450\,$\micron$ and 850\,$\micron$ fluxes from SCUBA-2, 
we develop a method that utilises the two frequency observations of the same 
region where the ratio depends partly on the dust temperature ($T_{\mathrm{d}}$) 
via the Planck function and also on the dust opacity spectral index, $\beta$ 
(a dimensionless term dependent on the grain model as proposed by 
\citeauthor{Hildebrand:1983fy}\,\citeyear{Hildebrand:1983fy}), as described by 
\begin{equation}
\frac{S_{\mathrm{450}}}{S_{\mathrm{850}}} = \left ( \frac{850}{450} \right )^{3+\beta }\left ( \frac{\exp(hc/\lambda _{\mathrm{850}}k_{\mathrm{B}}T_{\mathrm{d}})-1}{\exp(hc/\lambda _{\mathrm{450}}k_{\mathrm{B}}T_{\mathrm{d}})-1} \right ), \label{eqn:temp}
\end{equation}
otherwise referred to as `the temperature equation' \citep{Reid:2005ly}.

Temperature is known to influence the process by which dust grains coagulate 
and form icy mantles and therefore the value of $\beta$. Observations by 
\cite{Ubach:2012fk} have shown decreases in $\beta$ in protoplanetary disks 
but for the most part there is little evidence that $\beta$ changes significantly in 
pre/protostellar cores \citep{Schnee:2014uq}. \cite{Sadavoy:2013qf} fitted 
\emph{Herschel} 160\,$\micron$ to 500\,$\micron$ data with SCUBA-2 data in 
the Perseus B1 region  and concluded that $\beta$ is approximately 2.0 in 
extended, filamentary regions whereas it takes a lower value of approximately 
1.6 towards dense protostellar cores. 

Figure~\ref{fig:SR_temp} describes how small changes in $\beta$ lead to a 
large range of flux ratios, especially at higher temperatures. For ratios of 3, 7 
and 9, a $\beta$ of 1.6 would return temperatures of 8.9, 25.4 and 85\,K whereas 
a $\beta$ of 2.0 would return temperatures of 7.6, 15.7 and 25\,K. Higher ratios 
indicate heating above that available from the Interstellar Radiation Field (ISRF) 
for any reasonable value of $\beta$. 

Removing the requirement for the uncertainty in $\beta$ requires data at additional 
wavelengths, for example 250\,$\micron$ and 350\,$\micron$ as observed by 
\emph{Herschel}. Reconciling the angular scales of \emph{Herschel} observations 
with those of SCUBA-2 is a non-trivial process and goes beyond the scope of this paper. 

Smaller values of $\beta$ are found to be consistent with grain growth which only 
occurs sufficiently close to compact structures \citep{Ossenkopf:1994vn}. 
\cite{Stutz:2010hq} used the dominance of extended structure to that of compact 
structure to argue for a uniform, higher value of $\beta$. Likewise \cite{Hatchell:2013ij} 
assumed a constant $\beta$, arguing that variation in temperature dominates to that 
of $\beta$ in NGC1333. On this basis we adopt a uniform $\beta$ of 1.8, a value 
consistent with the popular OH5 dust model proposed by \cite{Ossenkopf:1994vn} 
and studies of dense cores with \emph{Planck}, \emph{Herschel} and SCUBA-2 
\citep{Stutz:2010hq, Juvela:2011ys, Sadavoy:2013qf}. We note that in this regime an 
apparent fall in temperature towards the centre of a core might be symptomatic of 
low $\beta$ values and therefore we cannot be as certain about the temperatures 
at these points.

There is no analytical solution for temperature and so pixel values are inferred from 
a lookup table. The method by which temperature maps are made can be split into 
two distinct parts: creating maps of flux ratio from input 450\,$\micron$ and 850\,$\micron$ 
data and building temperature maps based on the ratio maps. Both methods were 
discussed by \cite{Hatchell:2013ij}, for here on referred to as the H13 method. 
We focus on the development of this method and the additional features that have 
been incorporated.

\subsection{Ratio maps}

\begin{figure*}
\begin{center}
\includegraphics[scale=0.9]{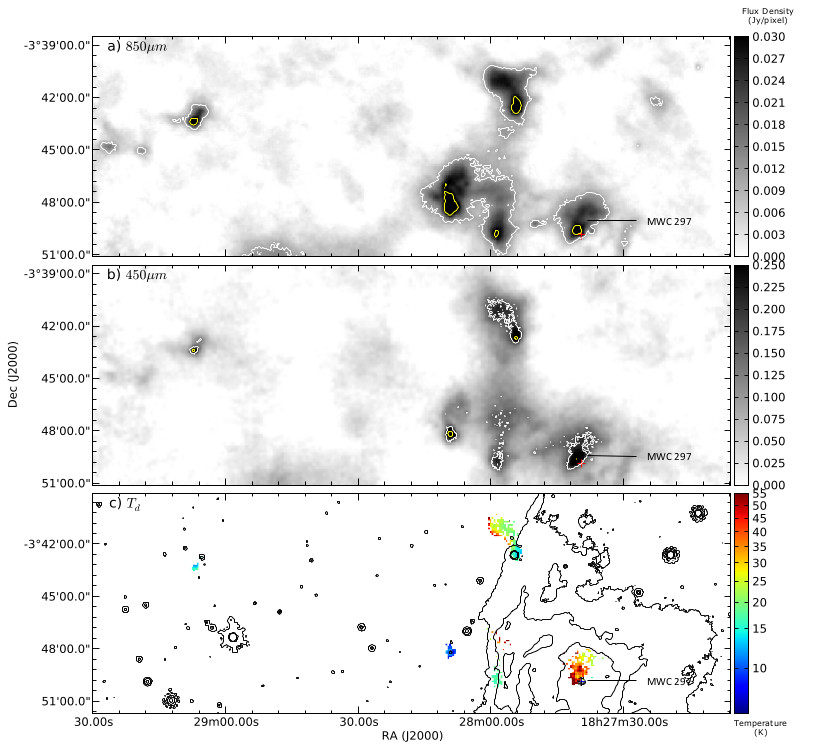}
\caption{Top to bottom: a) SCUBA-2 convolved 850\,$\micron$ flux map of Serpens MWC 297. Contours from the original 850\,$\micron$ data are at 0.011, 0.033\,Jy/pixel (corresponding to 5 and 15 $\sigma$). b) SCUBA-2 convolved and aligned 450\,$\micron$ flux map of Serpens MWC 297 in Jy/pixel. Contours from the original 450\,$\micron$ data are at 0.082, 0.25\,Jy/pixel (corresponding to 5 and 15 $\sigma$). The crosses in a) and b) mark the location of the ZAMS B1.5Ve star MWC 297 and its binary partner \emph{OSCA} (A2v). c) Dust temperature map of Serpens MWC 297 for $\beta$ = 1.8. Contours of \emph{Spitzer} 24\,$\micron$ emission at 32, 40 and 70\,MJy per Sr are overlaid.}
\label{fig:mwc}
\end{center}
\end{figure*}

%Our method is summarised in Figure~\ref{fig:flow}. 

Free parameters of our method are limited to $\beta$ (which we set at 1.8). Input 
450\,$\micron$ and 850\,$\micron$ flux density data (scaled in Jy/pixel) have fixed 
noise levels. Other fixed parameters which are used in the beam convolution 
include: the pixel area per  map, FWHM of the primary ($\theta_{\mathrm{MB}}$) 
and secondary ($\theta_{\mathrm{SB}}$) beams and beam amplitudes all of which 
are measured by \cite{Dempsey:2013uq} and given in Table \ref{table:beams}.

Input maps are first convolved with the JCMT beam (Equation 1) at the alternate 
wavelength to match resolution. Pixel size is taken into account in this process.
The 450\,$\micron$ fluxes are then regridded onto the 850\,$\micron$ pixel grid. 
Data are then masked leaving only 5$\sigma$ detections or higher. 450\,$\micron$ 
fluxes are then divided by 850\,$\micron$ fluxes to create a map of flux ratio.

Whereas the H13 method made a noise cut based on the variance array calculated 
during data reduction, our model introduces a cut based on a single noise estimate, 
following the method introduced by \cite{Salji:2013kx}. The data are masked to remove 
pixels which carry astronomical signal. The remaining pixels are placed in a histogram 
of intensity and a Gaussian is fitted to the distribution, from which a standard deviation, 
$\sigma$,  can be extracted as the noise level. This calculation is a robust form of 
measuring statistical noise that includes residual sky fluctuations.

%TJW calculated flux in units per beam beam which necessititated a calibration factor and an associated error of 5 to 10per cent depending on wavelength \citep{Dempsey:2013uq} however since then GBS Internal Release 1 standard has revised the units of input data array to units per pixel which eliminates this necessary calbriation and source of statisical error. As a result the improved error derived from the noise level can be propagated through the various process with the model to produce new error arrays. 

We introduce a secondary beam component into the H13 method, which previously 
assumed that the secondary component was negligible. This adds complexity 
to the convolution process as it requires convolution of the data with a normalised 
Gaussian of the form of the JCMT beam's primary and secondary components for 
the alternative wavelength. The primary component at 850\,\micron\ is then scaled with 

\begin{equation}
\frac{a_{\mathrm{450}}\theta _{\mathrm{MB_{\mathrm{450}}}}^{2}}{a _{\mathrm{450}}\theta _{\mathrm{MB_{\mathrm{450}}}}^{2}+b _{\mathrm{450}}\theta _{\mathrm{SB_{\mathrm{450}}}}^{2}}, \label{eqn:scaleMB}
\end{equation}

and likewise 

\begin{equation}
\frac{b_{\mathrm{450}}\theta _{\mathrm{SB_{\mathrm{450}}}}^{2}}{a_{\mathrm{450}}\theta _{\mathrm{MB_{\mathrm{450}}}}^{2}+b_{\mathrm{450}}\theta _{\mathrm{SB_{\mathrm{450}}}}^{2}}, \label{eqn:scaleSB}
\end{equation} 
for the secondary component. The 450\,$\micron$ map is convolved with the 850\,$\micron$ 
beam is a similar way. Corresponding parts are then summed together for 450\,$\micron$ and 
850\,$\micron$ data separately to construct the convolved maps with an effective beam size of 
19.9\arcsec\ as shown in Figure~\ref{fig:mwc}.

The inclusion of the secondary beam was found to decrease temperatures by between 5 per cent and 
9 per cent with the coldest regions experiencing the largest drop in temperature and warmest the least. 
 
%The inclusion of the secondary beam was found to decrease peak temperatures 
%in the final temperature maps on average by $4.5\pm0.6\,per cent$ whilst increasing 
%low temperatures on average by $11\pm3\,per cent$. Overall the secondary beam 
%increased the mean temperature across large scale areas of the map by $1.1\pm0.4\,per cent$.

Applying a 5$\sigma$ cut based on the original 450\,$\micron$ data to mask uncertain regions of 
large scale structure after the beam convolution can lead to spuriously high values around the 
edges of our maps where fluxes from pixels below the threshold are contributing to those above, 
producing false positives. These `edge effects' are mitigated by clipping but we advise that 
where the highest temperature pixels meet the map edges these data be regarded with a degree of scepticism.

\begin{figure*}
\begin{center}
\includegraphics[scale=0.7]{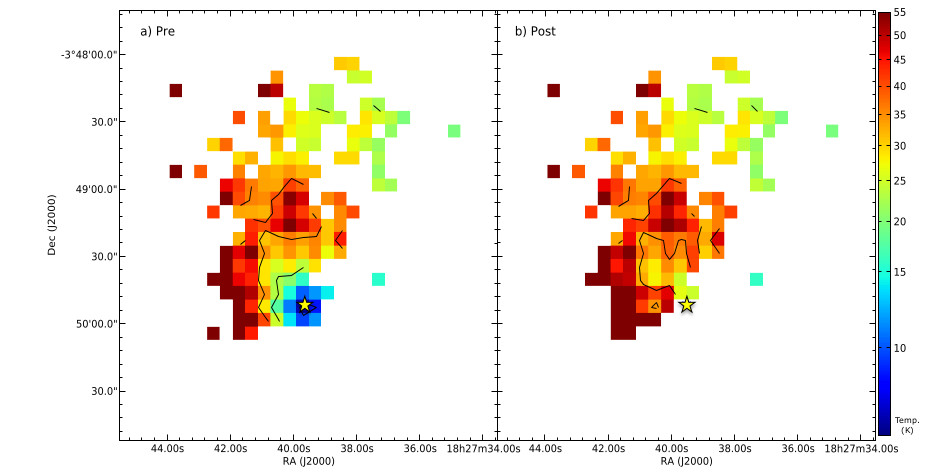}
\caption{Temperature maps of MWC 297 from the ratio of 450\,$\micron$ and 850\,$\micron$ emission pre (\emph{left}) and post (\emph{right}) free-free 
contamination subtraction. Contours are at 11, 25 and 38\,K. The location of MWC 297 is marked with a star.}
\label{fig:Tcomp}
\end{center}
\end{figure*}

\begin{figure*}
\begin{center}
\includegraphics[scale=0.7]{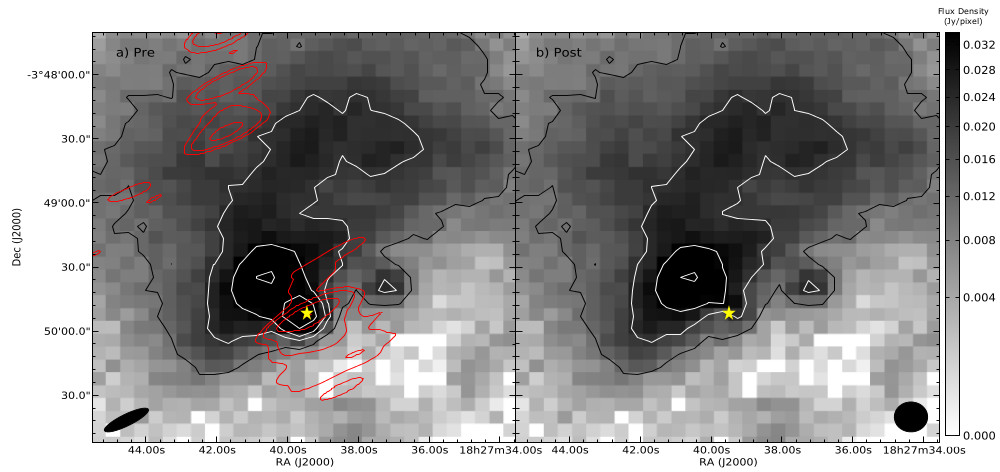}
\caption{IR1 SCUBA-2 850\,$\micron$ data before \emph{left} and after \emph{right} removal of free-free contamination from an UCH\textrm{II} region and polar jets/winds (represented by the point source contours in the \emph{left} plot). SCUBA-2 contours are at 0.011, 0.022, 0.033 and 0.055 Jy/pixel (corresponding to 5, 10, 15 and 25 $\sigma$ detection limits). 6\,cm VLA contours (red) from Sandell (private comm.) at 0.002, 0.005, 0.02, 0.072, 0.083 Jy/beam are overlaid on the left hand panel. The location of MWC 297 is marked with a star. Beam sizes are shown at the bottom of the image (VLA CnD config. \emph{left} and JCMT \emph{right}.) }
\label{fig:contamination}
\end{center}
\end{figure*}

\subsection{Dust temperature maps}

Ratio maps are converted to temperature maps using Equation \ref{eqn:temp} 
implemented as a look-up table as there is no analytical solution. The H13 method 
subsequently cuts pixels with an arbitrary uncertainty in temperature of greater 
than 5.5\,K. We replace this with a cut of pixels of an uncertainty in temperature 
(calculated from the noise level propagated through the method described in 
Section 3.1) of greater than 5 per cent. 

%Dust temperature cannot be calculated analytically from equation \ref{eqn:temp}. 
%Instead a look up table of temperatures is created and corresponding flux 
%ratios are calculated. Maps of temperature are then derived from these tables. 
%the H13 method introduces further cuts post conversion to temperature maps. 
%Anomalously high temperature values for ratio data where noise is excessive 
%at one wavelength are an artefact of the temperature process and are removed. 
%We introduce a cut on the temperature map to remove data with an error 
%greater than a user specified value whereas the H13 method had an arbitrary cut 
%%for pixels with a variance in temperature of greater than 30 K. The optimal 
%percentage cut for low extinction regions was as low as 3\,per cent whereas for 
%more noisy regions such as South and MWC 297 5\,per cent was used. 

The 450\,$\micron$ and 850\,$\micron$ SCUBA-2 data for the MWC 297 region are 
presented in Figure~\ref{fig:mwc} alongside a map of temperature of submillimetre 
dust in that region. These maps show a large diversity in temperature across five 
isolated regions of significant flux (shown in Figure~\ref{fig:mwc}c). Mean cloud 
temperatures range from 10.1$\pm$0.9\,K and 15$\pm$2\,K for regions which are 
relatively cold and isotropic, to 25$\pm$17\,K for warmer regions with a large diversity 
of temperatures. Figure~\ref{fig:Tcomp} shows one cloud that has a temperature of 41$\pm$19\,K 
which is hot to the extent that this would suggest an active heat source. The range in 
temperatures suggests that the regions within the Serpens MWC 297 vary in physical 
conditions. 

%We compare these initial findings to previous attempts to map temperature across star 
%forming regions as introduced in Section 1. 

\cite{Menshchikov:2010kl} infer temperature variation from contrasting strengths of 
350\,$\micron$ flux bands to the shorter 70\,$\micron$ and 160\,$\micron$ 
bands of \emph{Herschel}. They quote a temperature range for dense, starless filaments 
of 7.5 to 15\,K across the whole Aquila rift. However, we do not observe a typical filamentary 
structure in Serpens MWC 297 region (Figure \ref{fig:maps}).

\cite{Konyves:2010oq} and \cite{Bontemps:2010fk} used single-temperature modified black-body 
fitting of SEDs of \emph{Herschel} 500\,$\micron$ data points in Aquila and Polaris. 
Their study includes Serpens MWC 297 and they find temperatures for the 
region ranging between 24 and 26\,K. Though \emph{Herschel} 500\,$\micron$ data is 
at a lower resolution than our effective beam, the general temperatures of the region seem 
consistent with our findings. 

\cite{Hatchell:2013ij} use only the primary beam to study NGC1333, finding typical dust temperatures 
of ranging from 12 to 16\,K. They also argue for a heated region pushing temperatures up as 
high as 35 to 40\,K near the location of the B star SVS3. When the moderating effects of the 
secondary beam are taken into account, these results are largely consistent with our findings 
(Serpens MWC 297 also contains a B star). 

Figure~\ref{fig:mwc}c shows \emph{Spitzer} MIPS 24\,$\micron$ flux for the Serpens MWC 297 
region. These data show hot compact sources associated with individual stellar cores. It also 
shows the morphology of an extended IR nebulosity, associated with SH2-62, that is centred 
on MWC 297. As well as the dust within the immediate vicinity of the star MWC 297 showing clear 
signs of heating, we observe 24\,$\micron$ emission that is coincident with heating in the SCUBA-2 
temperature maps. As 24\,$\micron$ emission provides independent 
evidence of heating, where we observe high temperature pixels that are not coincident with 
24\,$\micron$ emission (for example in the northernmost cloud) we conclude we are likely 
witnessing data reduction artefacts as opposed to warm gas and dust. 

In addition to providing evidence for direct heating by MWC 297, the 24\,$\micron$ data also 
provide strong evidence that the B star is physically connected to the observed clouds. 
The Aquila rift is thought to be a distance of 250$\pm$50 pc \citep{Maury:2011ys} and through 
association we conclude that the distance to MWC 297 matches this figure.

\subsection{Contamination}

Reliable temperatures depend on accurate input fluxes. Systematic contamination of 450\,$\micron$ and 850\,$\micron$ flux 
by molecular lines, in particular CO, is a known problem within SCUBA-2 data \citep{Drabek:2012uq}. We investigate the contribution of CO and 
free-free emission to these bands and attempt to mitigate their effects where necessary.

\cite{Hatchell:2013ij} and \cite{Drabek:2012uq} highlighted 345 GHz contamination of 850\,$\micron$ due to the CO~3\hbox{--}2 
line in other Gould Belt star-forming regions. Limited $^{12}$CO and $^{13}$CO~1\hbox{--}0 data exist for the Serpens 
MWC 297 region \citep{Canto:1984dq}. A very rough estimate of the CO contamination towards the star MWC 297 can be made 
based on the published spectra. The $^{12}$CO lines are broad ($\sim 12~\hbox{ km s}^{-1}$) but do not show line wings characteristic of 
outflows.  Making the simplest assumption that the $^{12}$CO is optically thick and fills the beam in both the $J=1\hbox{--}0$ 
and $J=3\hbox{--}2$ lines, the integrated intensity of the latter will be similar to the former, $\sim 36\,\hbox{K km s}^{-1}$, 
corresponding to a CO contamination of 1.14 mJy/pixel/K km$^{-1}$ (13\,per cent of peak flux) at the position of the star MWC 297 
using the conversion in \citet{Drabek:2012uq} updated for the beam parameters in \citet{Dempsey:2013uq}. \cite{Drabek:2012uq} 
noted than regions where CO emission accounts for less that 20\,per cent of total peak emission are not consistent with outflows or 
major contamination. \cite{Manoj:2007ly} find no evidence of CO~2\hbox{--}1 and $^{13}\textrm{CO~2\hbox{--}1}$ emission within 
80\,AU of MWC 297 and conclude this depletion is caused by photoionisation due to an ultra-compact HII (UCH\textrm{II}) region 
as has been detected by \cite{Drew:1997qf} and \cite{Malbet:2007zr}. 

%these phenomenon are thought to be associated with stars sufficiently massive 
%(greater than 8$M_{\mathrm{\odot}}$) that their Kelvin-Helmholtz contraction timescale is 
%shorter than their free fall and accretion timescale \citep{Manoj:2007ly}. 
%these stars reach the main sequence and start producing ionising radiation 
%whilst still embedded within their protostellar envelope \citep{McKee:2003ys}.
%This would then lead to a compact region of highly ionised winds, as detected by 
%\cite{Malbet:2007zr; Drew:1997qf}.
 
The inferred presence of an UCH\textrm{II} region has consequences for contamination at submillimetre wavelengths 
through thermal bremsstrahlung, or free-free emission, from ionised gas with temperatures of 10,000\,K or higher. 
Free-free emission is optically thick at the longest wavelengths and has a relatively flat  power law in the optically thin 
regime at radio and far infrared wavelengths before undergoing exponential cut off at shorter wavelengths. 
\cite{Skinner:1993bh} studied free-free 3.6\,cm and 6.0\,cm radio emission from stellar winds around MWC 297 and 
found a power law of the form $S_{\nu} \propto \nu^{\alpha}$ where $\alpha$ is equal to 0.6238 in the optically thin 
regime. \cite{Sandell:2011dz} extended the study down to 3 mm and revised the spectral index to $\alpha = 1.03\pm0.02$ 
which is consistent with a collimated jet component to free-free emission. The free-free power law extends into the 
submillimetre spectrum; however, at wavelengths shorter than 2.7\,mm there is potential for a thermal dust component 
in the observed flux, so submillimetre flux is not included in the calculation of $\alpha$.

Figure~\ref{fig:contamination} displays 6\,cm radio emission from the VLA CnD configuration in conjunction with 
SCUBA-2 850\,$\micron$ data (Skinner \citeyear{Skinner:1993bh}, Sandell priv. comm). Both sets of data show 
peaks in emission which are coincident with a point source at the location of the star MWC 297 in 1\,mm and 
3\,mm data presented by \cite{Alonso-Albi:2009ve}. The peak of the SCUBA-2 850\,$\micron$ emission in 
Figure~\ref{fig:contamination} is 86\,mJy/pixel, consistent with the SCUBA 850\,$\micron$ value of 82\,mJy/pixel 
\citep{Alonso-Albi:2009ve}.
 
The VLA data also show extended emission to the north and south of MWC 297 which is consistent with polar 
winds or jets. The intensity of emission is significantly weaker than that of the UCH\textrm{II} region. Considering 
the elongated beam shape of the VLA CnD observations (21.1\arcsec $\times$ 5.2\arcsec, PA$ = -61^{\circ}.3$) 
accounts for much the E/W elongation of the emission. In addition to this, \cite{Manoj:2007ly} describe this emission 
as coming from within 80\,AU of MWC 297. This is much smaller than the JCMT beam and therefore we model the 
dominant free-free emission from MWC 297 as a point source.

\begin{figure}
\begin{center}
\includegraphics[scale=0.45]{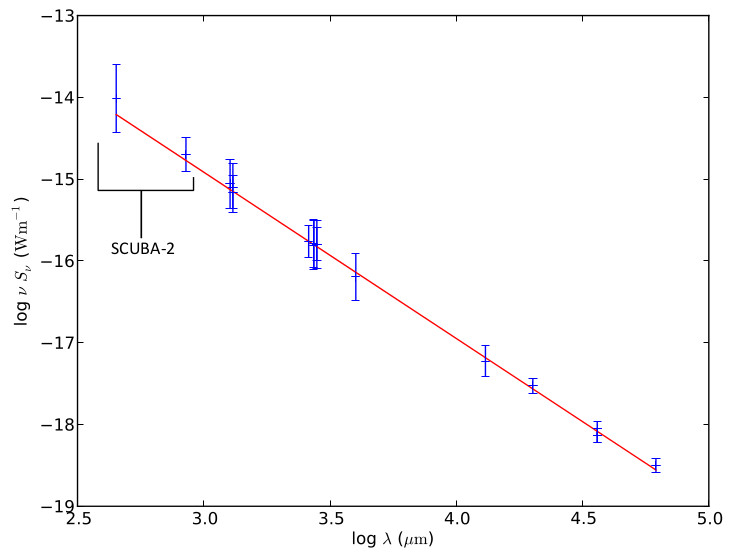}
\caption{The Spectral Energy Distribution of MWC 297 from submillimetre to radio wavelengths. SCUBA-2 fluxes (found using aperture photometry as described in Section 5.2.) are presented alongside those collated by \protect\cite{Sandell:2011dz} who fit a power law $\alpha = 1.03\pm0.02$, consistent with free-free emission from an UCH\textrm{II} region and polar jets or outflows.}
\label{fig:freefreeSED}
\end{center}
\end{figure}

By taking the revised power law least square fit to \cite{Skinner:1993bh} and \cite{Sandell:2011dz}'s results at radio and milimetre 
wavelengths and extrapolating to the submillimetre wavelengths of SCUBA-2, we are able to calculate the effect of free-free 
emission due to a point-like UCH\textrm{II} region as an integrated flux of 934$\pm$128 mJy at 450\,$\micron$ and 471$\pm$62 mJy 
at 850\,$\micron$. By convolving this point with the JCMT beam we find that free-free contamination corresponds to approximately 
73$\pm$5\,per cent and  82$\pm$4\,per cent of the 450\,$\micron$ and 850\,$\micron$ peak flux respectively in the case of MWC 297. 
Residual dust peak fluxes are $51\pm10$ mJy and $15\pm3$ mJy flux per pixel at 450\,$\micron$ and 850\,$\micron$ respectively 
and are highlighted in   Figure~\ref{fig:freefreeSED} as the flux above the free-free power law fit of $\alpha = 1.03\pm0.02$. Given our 
estimate of 13\,per cent CO contamination, dust emission could potentially account for as little as 5\,per cent of peak emission at 
850\,$\micron$.

%The original peak fluxes of MWC 297 are 188$\pm$16 mJy and 86$\pm$22 mJy at 450\,$\micron$ and 850\,$\micron$. After subtraction, the residual peak flux is below the 5$\sigma$ detection level. 

We cannot say whether any dust emission contributes at the position of MWC 297. Figure~\ref{fig:contamination} presents the 
850\,$\micron$ before and after subtraction. Figure~\ref{fig:Tcomp} presents the impact of free-free emission on temperature 
maps of the region. Even with the free-free emission subtracted, a large, extended submillimetre clump remains, though its 
peak is offset from the location of MWC 297 by 24.2\arcsec\ (approximately 6,000\,AU).

The impact of this contamination on the temperature maps is remarkable. The power law of $\alpha = 1.03\pm0.02$ that 
describes free-free emission from both an UCH\textrm{II} region and jet outflows produces greater flux at 850\,$\micron$ 
than 450\,$\micron$. Free-free dominates the flux and this results in artificially lower ratios and therefore lower temperatures. 
This is consistent with the cold spot seen in Figure \ref{fig:Tcomp}a at the location of the UCH\textrm{II} region, with a 
temperature of approximately 11\,K. We can conclude that free-free emission may contaminate submillimetre temperature 
maps where cold spots are coincident with hot OB stars.

\begin{figure*}
\begin{center}
\includegraphics[scale=0.7]{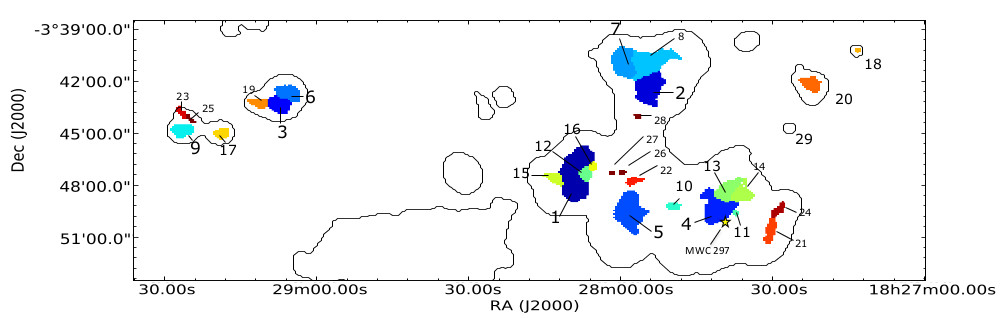}
\caption{Clumps identified in 850\,$\micron$ data with the \textsc{starlink} clump-finding algorithm \textsc{fellwalker} index numbered in order (highest to lowest following the colour scale) of integrated flux. The data reduction mask is overlaid as a black contour.}
\label{fig:clumps}
\end{center}
\end{figure*}

%%%%%%%%%%%%%%%%%%%%%%%%%%%%%%%%%%%%%%%%%%%%%%%%%%%

\section{The SCUBA-2 clump catalogue}

In this section we introduce the clump-finding algorithm \textsc{fellwalker} used to identify 
clumps in the SCUBA-2 data presented in Figure~\ref{fig:maps}. We calculate clump 
masses and compare these to their Jeans masses to determine whether or not the objects 
are unstable to gravitational collapse. 

%Identifying Structure
\subsection{Identification of structure}

%A simplified approach to a `clump' in the ISM models it as a sphere of gas with a Gaussian-like density profile. It is not feasible to constrain a real radius of the volume of this clumps and instead we turn to an artificial, \emph{effective radius}. The effective radius is smaller than reality as it measures from the centre of the clump to where the density profile drops below the noise level as opposed to where the density becomes indistinguishable from the ISM. The inevitable consequence of a smaller clump radius is a smaller integrated flux for the clump. Whilst effective radius does not provide a real measure of clamp mass, it does provide a mechanism for defining a lower limit on the mass of the clump by defining a minimum clump size. 

Clumps do not have well defined boundaries within the ISM. We use the signal to noise 
ratio to define a boundary at an \emph{effective radius}. The boundary is determined 
by the \textsc{starlink} \textsc{CUPID} package for the detection and analysis of objects 
\citep{Berry:2013uq}, specifically the \textsc{fellwalker} algorithm which assigns pixels to 
a given region based on positive gradient towards a common emission peak. This method 
has greater consistency over parameter space than other algorithms (Watson 
\citeyear{watson:2010pc}, Barry et al. 2014, submitted). \textsc{fellwalker} was developed by 
\cite{Berry:2007vn}, and the 2D version of the algorithm used here considers a pixel in the data 
above the noise level parameter and then compares its value to the adjacent pixels. \textsc{fellwalker} then moves on to 
the adjacent pixel which provides the greatest positive gradient. This process continues 
until the peak is reached - when this happens all the pixels in the `route' are assigned 
an index and the algorithm is repeated with a new pixel. All `routes' that reach the same 
peak are assigned the same index and form the `clump'.  Clump-finding algorithms, 
such as this, have been used by \cite{Johnstone:2000fk}, \cite{Hatchell:2005fk}, \cite{Kirk:2006vn} 
and \cite{Hatchell:2007qf} to define the extent of clumps for the purposes of measuring 
clump mass. 

%A simplified approach to a `clump' in the ISM models it as a sphere of gas with a Gaussian-like density profile. In this scenario a radius would exist where the density of the clump becomes indistinguishable from a constant density ISM and this forms the basis for the calculation of Jeans masses and lengths. Ideally this radius could be used to define an area from which the total flux could be observed and subsequently be used to calculate mass of the clump. Observations of clumps show they are rarely spherical or any regular volume. Density profile varies depending on how evolved the clumps and whether it has started collapsing to form a protostar or not and the ISM is not constant level. In addition to these problems, observational data comes with a maximum level of statistical noise in the flux, under which uncertainty is too high for reliable data. 

We tuned the \textsc{fellwalker} algorithm to produce a set of objects consistent 
with a by-eye decomposition, setting the following parameters; \emph{MinDip} 
= 1$\sigma$ (minimum flux between two peaks), \emph{MinPix} = 4 pixels (minimum number 
of pixels per valid clump), \emph{MaxJump} = 1 pixel (distance between clump 
peaks), \emph{FWHMBeam} = 0 (FWHM of instrument), \emph{MinHeight} = 3$\sigma$ 
(minimum height of clump peak to register as a valid clump) and \emph{Noise} = 3$\sigma$ 
(detection level). Throughout this process we used a constant noise level, 
$\sigma$, calculated via the method described by \cite{Salji:2013kx} and described in 
Section 3.1. \cite{watson:2010pc} discusses the \textsc{fellwalker} parameters in depth 
and concludes \emph{MinDip} and \emph{MaxPix} are the most influential in returning 
the maximum breakup of clouds into clumps, a subset of which will later be used to compile 
a list of protostellar cores. The $3\sigma$ level allows for the detection of the smallest 
clumps that may be missed at the $5\sigma$ level on account of insufficient pixels for 
detection as outlined above. This method also included a number of spurious clumps 
associated with high variance pixels at the maps edges. In order to avoid these we first 
masked the SCUBA-2 data with the data reduction mask shown in Figure \ref{fig:maps}. 

%We parameterise \textsc{fellwalker} to detect clumps consistent with 3$\sigma$ or greater 
%detections from 850\,$\micron$ data. The same boundaries are then used consistently 
%on the 450\,$\micron$ data. Spurious data around the edges of the map resulting from the 
%scanning procedure are removed. \cite{watson:2010pc} discusses the \textsc{fellwalker} 
%parameters in depth and concludes \emph{MinDip} and \emph{MaxPix} are the most 
%influential in returning the maximum number of clumps, and therefore identifying smaller scale 
%cores, and we set these equal to the $rms$ noise level and four pixels respectively. 
%We also found \emph{FWHMBeam} was important and setting this to zero included 
%all point sources regardless of size whereas setting \emph{MaxJump} to one pixel 
%isolated sources that were closer to each other. \emph{MinHeight} and \emph{Noise} 
%were both set to minimum statistical certainty of $3\sigma$,  \emph{MaxBad} was 
%set to 100 to stop pixels being eliminated if they were contact with `bad' data. 

Using these parameters 28 submillimetre clumps were detected in 850\,$\micron$ data 
and are presented in Figure \ref{fig:clumps}. Two sources (SMM 23 and 25) were immediately 
discarded as they were not consistent with a 5$\sigma$ detection. A further two clumps 
were split into two separate objects by the algorithm when there was no discernible peak 
in the submillimetre data. In these cases (SMM 7 \& 8 and SMM 13 \& 14) the objects were 
recombined into single object. We note that this is a side effect of having a low \emph{MinDip} 
parameter to maximise the detection of smaller clumps.  In total a sample of 23 clumps 
are presented in Table 3. We note that there is a known bias that underestimates the 
size of a clump as its peak flux approaches the cutoff level and therefore biases against 
the detection of cold, faint objects (examples might be SMM 26 and 27). Modelling 
clump profiles could be used to better estimate the full extent of these objects. However, 
as these present a minority of cases we take no further action on this issue 
\citep{Rosolowsky:2006bh}.

The \textsc{fellwalker} algorithm is insensitive to low mass, isolated objects where detections 
were limited to less than five pixels above the noise level. We find that one potential 
source was missed on account of it only exhibiting a single significant pixel above the 
5$\sigma$ noise level. Here object flux was measured with aperture photometry 
(see Section 5.2).

Due to the higher noise level of the 450\,$\micron$ data many objects detected at 850\,$\micron$ 
were not present at 450\,$\micron$. Therefore we apply the 850\,$\micron$ clump boundaries to the 450\,$\micron$ 
data when calculating integrated intensity at that wavelength to ensure consistent flux extraction at 
both wavelengths for each object.

%We are aiming to detect clumps above the 5$\sigma$ noise level, however to ensure the 
%detection of the smallest clumps that were often omitted by \textsc{fellwalker} on account of their diminutive size, 
%we lowered the noise level condition to 3$\sigma$ in increase the area which can be detected whilst 
%ensuring a degree of statistical significance. The consequence of this lower \emph{Noise} level is the inclusion of several spurious clumps 
%in the output which required removal by eye. Whilst this was feasible in the context Serpens 
%MWC 297, reproducing this method on larger regions will require \textsc{fellwalker} to be parameterised 
%to minimise anomalous results at the expense of free floating, point like YSOs. 

%IR + Other sources

%Mass

\subsection{Measurement of mass}

%Estimates for the final mass of stars forming from the cloud are calculated based on an estimate of the star formation efficiency (for a review, see \citeauthor{McKee:2007ly} \citeyear{McKee:2007ly}). 

SCUBA-2 observations of the Serpens MWC 297 region were used to calculate the masses of the \textsc{fellwalker} 
clumps. \cite{Hildebrand:1983fy} describes how the mass of a cloud can be calculated from the 
submillimetre emission of dust grains fitted to a black-body spectrum for a nominal temperature. 
We follow this standard method for calculating clump mass (for example \citeauthor{Johnstone:2000fk}
\citeyear{Johnstone:2000fk}, \citeauthor{Kirk:2006vn}\,\citeyear{Kirk:2006vn}, \citeauthor{Sadavoy:2010ve}
\citeyear{Sadavoy:2010ve} and \citeauthor{Enoch:2011lh}\,\citeyear{Enoch:2011lh}). We use flux at 
850\,$\micron$ ($S_{\mathrm{850}}$) per pixel, dust opacity ($\kappa_{\mathrm{850}}$), distance 
($d$) and a variable temperature ($T_{\mathrm{d}}$) per pixel, summing over all pixels, $i$, in 
the clump to calculate the total clump mass:

\begin{eqnarray}
M  & = 0.39 \sum_{i} S_{\mathrm{850,i}}\left [\exp \left ( \frac{17\mathrm{K}}{T_{\mathrm{d},i}} \right ) - 1 \right ]  &       \nonumber \\
\nonumber    & \times \left ( \frac{d}{250\mathrm{pc}} \right )^{2}\left ( \frac{\kappa_{\mathrm{850}}}{0.012\mathrm{\hbox{cm}^{2} \hbox{g}^{-1}}} \right )^{-1}.\\
\label{eqn:mass}
\end{eqnarray}

%Uncertainty in kappa
There is a high degree of uncertainty in the value of $\kappa_{\mathrm{850}}$. The popular OH5 model 
of opacities in dense ISM, with a specific gas to dust ratio of 161, gives 0.012 $\hbox{cm}^{2} \hbox{g}^{-1}$ at 
850\,$\micron$ \citep{Ossenkopf:1994vn}. Comparable studies suggest values of 0.01 \citep{Johnstone:2000fk}, 
0.019 \citep{Eiroa:2008ta} and 0.02 $\hbox{cm}^{2} \hbox{g}^{-1}$ \citep{Kirk:2006vn}. \cite{Henning:1995qf} find 
$\kappa_{\mathrm{850}}$ can vary by up to a factor of two. We assume an opacity of $\kappa_{\mathrm{850}}$ 
= 0.012 following \cite{Hatchell:2005fk}. This value is consistent with $\beta$ = 1.8 over a wavelength range of 
30\,$\micron$ to 1.3\,mm. We assume a distance $d = 250\pm50\hbox{ pc}$ following \cite{Sandell:2011dz} as 
outlined in Section 1.

We calculate dust masses using dust temperatures calculated for each pixel where possible. Not all the clumps 
shown in Figure~\ref{fig:clumps} have temperature data due to the noise constraints of the temperature mapping 
process and the requirement that the region is also detected at 450\,$\micron$. For those that do not, a constant clump 
temperature of 15\,K is assumed following \cite{Johnstone:2000fk} and \cite{Kirk:2006vn}. Some clumps have only 
partial temperature data. In these cases the remaining pixels are filled with a value equal to the mean of the existing 
data. In some cases (SMM 6 and 11 for example), temperature data is limited to a few pixels whereas the total clump 
area is an order of magnitude larger. As it is unlikely that such a small sample of data will accurately represent the 
whole clump, results for objects such as these should be treated with a larger degree of uncertainty. Edge effects have 
a negligible influence on clump mass as high temperatures reduce the contribution in Equation 5. Clump masses 
are listed in Table~\ref{tab:mass}. %Core masses are calculated with the same method and are shown in Table~\ref{table:cores}.  

The total mass of clumps in Serpens MWC 297 is 40$\pm$3\,$M_{\mathrm{\odot}}$. Individual clump 
masses range over 2 orders of magnitude from 0.05 to 19\,$M_{\mathrm{\odot}}$ with 29\,per cent of 
objects having a mass of 1\,$M_{\mathrm{\odot}}$ or higher. Figure~\ref{fig:clumps} shows how 
\textsc{fellwalker} divides the areas of star formation into five large-scale star-forming clouds and a small 
number of isolated objects. Of these clouds, SMM 1, 12, 15 \& 16 is the most massive at 21$\pm$2\,$M_{\mathrm{\odot}}$, 
containing 53\,per cent of all the mass detected by \textsc{fellwalker}, followed by SMM 2, 7 \& 8 at 6.6$\pm$0.3\,$M_{\mathrm{\odot}}$ 
(17\,per cent), SMM 4, 10, 11, 13, 14, 21 \& 24 at 3.3$\pm$0.1\,$M_{\mathrm{\odot}}$ (9\,per cent), 
SMM 3, 6 \& 19 at 3.1$\pm$0.1\,$M_{\mathrm{\odot}}$ (8\,per cent) and SMM 5, 22, 26 \& 27 at 
3.1$\pm$0.3\,$M_{\mathrm{\odot}}$ (8\,per cent). 

\begin{table*}
%\caption{Properties of Submillimetre clumps in MWC 297} 
\caption{Properties of submillimetre clumps in MWC 297.}
\label{tab:mass}
\begin{center}
\begin{tabular}{@{}lllccllcllc}

\hline
ID$^{a}$	&Object name$^{b}$	&	$S_{\mathrm{450}}^{c}$	&	$S_{\mathrm{850}}^{c}$	&	$M_{\mathrm{850}}^{d}$	&	$T_{\mathrm{d}}^{e}$	&	Area$^{a}$	&$M_{\mathrm{J}}^{f}$ & $M_{\mathrm{850}}/$$M_{\mathrm{J}}$ &	SGBS YSOc ID$^{g}$\\
		&				&	(Jy)					&	(Jy)					&	(M$_{\mathrm{\odot}})$	&	(K)				&	(pixels)		&(M$_{\mathrm{\odot}})$	& 	 						&				\\
\hline
\hline
SMM1	&	JCMTLSG J1828090-0349497 	&	45.7	&	11.5	&	19(2)		&	10.1(0.5)	&	358	&	2.1(0.1)	&	9.12(1.05)		&-\\
SMM2	&	JCMTLSG J1827542-0343197 	&	33.2	&	5.0	&	3.5(0.2)		&	17.9(0.9)	&	205	&	2.9(0.1)	&	1.21(0.1)		&YSOc41\\
SMM3	&	JCMTLSG J1829071-0344378 	&	11.2	&	1.9	&	1.6(0.1)		&	14.6(0.7)	&	94	&	1.58(0.08)	&	1.03(0.1)		&-\\
SMM4	&	JCMTLSG J1827405-0350257 	&	43.9	&	4.7	&	0.91(0.05)		&	46(2)	&	213	&	7.4(0.4)	&	0.12(0.01)		&-\\
SMM5	&	JCMTLSG J1827590-0350137 	&	36.8	&	5.4	&	3.3(0.3)		&	18.2(0.9)	&	265	&	3.3(0.2)	&	0.99(0.09)		&-\\
SMM6	&	JCMTLSG J1829055-0343138 	&	7.7	&	1.3	&	1.2(0.1)		&	14.2(0.7)	&	94	&	1.53(0.08)	&	0.77(0.08)		&YSOc73\\
SMM7*	&	JCMTLSG J1827586-0342557 	&	59.4	&	7.7	&	3.1(0.2)		&	25(2)	&	419	&	8.2(0.3)	&	0.37(0.03)		&-\\
%SMM8	&	JCMTLSG J1827570-0341017 	&	30.0	&	4.4	&	2.08(0.15)		&	21.7(1.1)	&	246	&	3.79(0.19)	&	0.55(0.05)		\\
SMM9	&	JCMTLSG J1829260-0345139 	&	3.1	&	0.8	&	0.67(0.06)		&	15.0(-)	&	73	&	1.43(0.07)	&	0.47(0.05)		&-\\
SMM10	&	JCMTLSG J1827501-0350437 	&	3.5	&	0.4	&	0.29(0.03)		&	15.0(-)	&	26	&	0.85(0.04)	&	0.35(0.04)		&YSOc32\\
SMM11	&	JCMTLSG J1827373-0350197 	&	1.1	&	0.2	&	0.1(0.02)		&	17.6(0.9)	&	9	&	0.59(0.03)	&	0.17(0.03)		&YSOc17\\
SMM12	&	JCMTLSG J1828074-0348437 	&	4.1	&	1.0	&	0.86(0.07)		&	15.0(-)	&	42	&	1.08(0.05)	&	0.79(0.08)		&-\\
SMM13*	&	JCMTLSG J1827393-0349257 	&	30.9	&	3.8	&	1.25(0.06)		&	28(2)	&	199	&	6.3(0.2)	&	0.20(0.01)		&-\\
%SMM14	&	JCMTLSG J1827369-0349257 	&	12.9	&	1.6	&	0.6(0.04)		&	25.4(1.3)	&	89	&	2.67(0.13)	&	0.22(0.02)\\
SMM15	&	JCMTLSG J1828126-0348197 	&	1.8	&	1.0	&	0.8(0.07)		&	15.0(-)	&	50	&	1.18(0.06)	&	0.68(0.07)		&-\\
SMM16	&	JCMTLSG J1828058-0347017 	&	2.2	&	0.4	&	0.33(0.03)		&	15.0(-)	&	18	&	0.71(0.04)	&	0.47(0.05)		&YSOc47\\
SMM17	&	JCMTLSG J1829187-0346559 	&	2.7	&	0.4	&	0.34(0.03)		&	15.0(-)	&	42	&	1.08(0.05)	&	0.32(0.03)		&-\\
SMM18	&	JCMTLSG J1827133-0341438 	&	0.2	&	0.1	&	0.05(0.02)		&	15.0(-)	&	7	&	0.44(0.02)	&	0.12(0.04)		&YSOc2\\
SMM19	&	JCMTLSG J1829107-0344378 	&	2.5	&	0.4	&	0.33(0.03)		&	15.0(-)	&	47	&	1.14(0.06)	&	0.29(0.03)		&-\\
SMM20	&	JCMTLSG J1827225-0343378 	&	1.4	&	0.7	&	0.61(0.05)		&	15.0(-)	&	73	&	1.43(0.07)	&	0.43(0.04)		&-\\
SMM21	&	JCMTLSG J1827297-0351378 	&	5.1	&	0.6	&	0.46(0.04)		&	15.0(-)	&	65	&	1.35(0.07)	&	0.34(0.04)		&-\\
SMM22	&	JCMTLSG J1827582-0348137 	&	4.7	&	0.5	&	0.1(0.01)		&	42(2)	&	31	&	2.6(0.1)	&	0.04(0.0)		&-\\
SMM24	&	JCMTLSG J1827285-0350378 	&	3.5	&	0.4	&	0.29(0.03)		&	15.0(-)	&	37	&	1.02(0.05)	&	0.29(0.03)		&-\\
SMM26	&	JCMTLSG J1827594-0348437 	&	1.5	&	0.2	&	0.15(0.02)		&	15.0(-)	&	10	&	0.53(0.03)	&	0.28(0.04)		&-\\
SMM29	&	JCMTLSG J1828022-0348377 	&	0.6	&	0.1	&	0.09(0.02)		&	15.0(-)	&	6	&	0.41(0.02)	&	0.23(0.05)		&-\\
\hline

\end{tabular}
\end{center}

\raggedright
a) Clumps as identified by the \textsc{fellwalker} algorithm. \\
b) Position of the highest value pixel in each clump (at 850\micron). \\
c) Integrated fluxes of the clumps as determined by \textsc{fellwalker}. The uncertainty at 450\,$\micron$ is 0.3\,Jy and at 850\,$\micron$ is 0.02\,Jy. There is an additional systematic error in calibration of 10.6\,per cent and 3.4\,per cent at 450\,$\micron$ and 850\,$\micron$.\\
d) As calculated with equation \ref{eqn:mass}. Errors in brackets are calculated from error in total flux, described in c., and error in mean temperature of 5\,per cent. These results do not include the systematic error in distance (20\,per cent) and opacity (100\,per cent). \\
e) Mean temperature as calculated from the temperature maps (Figure~\ref{fig:mwc}). Where no temperature data is available an arbitrary value of 15K(-) is assigned that is consistent with the literature.  \\
f) As calculated with equation \ref{eqn:sadavoy}. These results have a systematic error uncertainty due to distance of 20\,per cent.\\
g) Where a \textsc{fellwalker} source is coincident with a SGBS YSOc, that object is listed here. A complete list is presented in Table \ref{table:SGBS_YSO_small}.\\
Objects indicated with * have been merged with an adjacent object which was incorrectly identified as a separate clump by \textsc{fellwalker}. 
\end{table*}

%Jeans stability

\subsection{Clump stability}

\begin{figure}
\begin{center}
\includegraphics[scale=0.45]{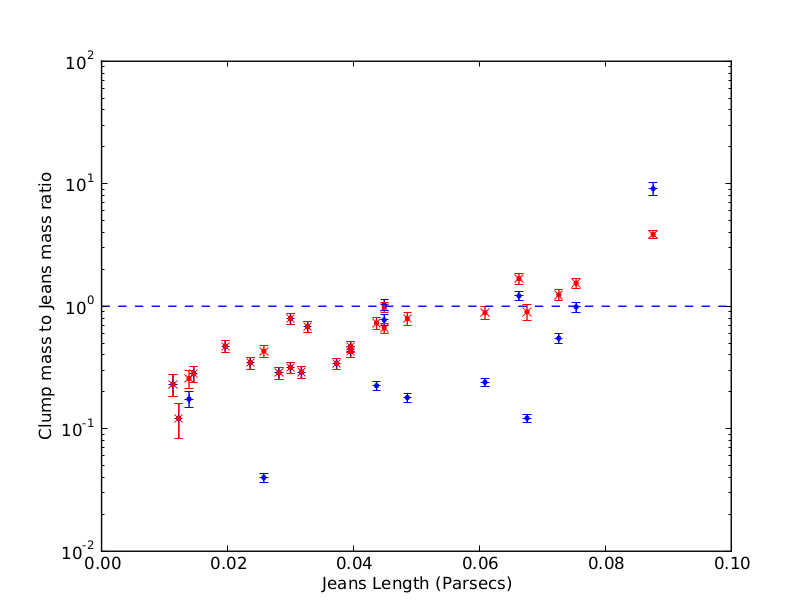}
\caption{Jeans stability plotted against Jeans length. All clumps with $M_{\mathrm{850}}/M_{\mathrm{J}} > 1$, as shown by the dashed line, are expected to be undergoing collapse. Blue circles represent calculations made with real temperature data whereas red crosses indicate those made with an assumed temperature of 15\,K. Systematic error in the measurement of distance to MWC 297 accounts for 20\,per cent uncertainty on Jeans length.}
\label{fig:jeans_plot}
\end{center}
\end{figure}

%In this section we outline our method for calculating the Jeans mass, MJ of our clumps.

The Jeans instability \citep{Jeans:1902dz} describes the balance between thermal support and 
gravitational collapse in an idealised cloud of gas. $R_{\mathrm{J}}$ defines a critical length 
scale above which the cloud may collapse on a free fall timescale and star formation can 
take place. Analogously, $M_{\mathrm{J}}$ defines an upper limit of mass. Assuming a spherical 
clump has a density such that it is Jeans unstable to perturbations at the size of the clump, 
$R_{\mathrm{J}}$, then

\begin{equation}
M_{\mathrm{J}} = 1.9\left ( \frac{\bar{T}_{\mathrm{d}}}{10\mathrm{K}} \right )\left ( \frac{R_{\mathrm{J}}}{0.07\mathrm{pc}} \right )M_{\mathrm{\odot}}. 
\label{eqn:sadavoy}
\end{equation}

We use the effective radius of the clump, as determined by clump area (in pixels) from 
\textsc{fellwalker} (Table~\ref{tab:mass}), as the length scale $R_{\mathrm{J}}$. We note that 
effective radius is a lower limit on clump size. Mean temperature, $\bar{T}$, across the clump 
is calculated directly from our temperature maps. 

Whereas mass was calculated on a pixel-by-pixel basis, this is not possible for $M_J$ 
as the characteristic length scale of the Jeans instability covers the entire object. 
Instead we use a mean temperature calculated from our maps. Temperatures 
and Jeans masses of clumps are also shown in Table~\ref{tab:mass}. The masses of 
clumps calculated with the temperature data in the previous section deviates from the 
equivalent masses calculated with a uniform mean temperature (set at 15\,K) of that clump 
by 12\,per cent on average per clump which is sufficiently similar to allow this analysis.

This method is based on the work by \cite{Sadavoy:2010ve} who performed a similar analysis 
for starless cores in the Gould belt. They used the assumption of a typical cold (10K) 
molecular cloud core size of 0.07 pc \citep{di-Francesco:2007bh}. \cite{Rosolowsky:2008dq} 
determined a range of temperatures of 9\,K to 26\,K in Perseus (a similar region to 
Serpens-Aquila) from ammonia observations. This paper goes a step further and is able to 
use mean temperatures specific to each clump. We determine a mean clump temperature 
of $20\pm10$\,K. The greater uncertainty on this value is indicative of the greater diversity of 
temperatures than assumed by \cite{Sadavoy:2010ve}. 

Under the assumption that only internal  pressure can balance self-gravity, $M_J$ sets an upper 
limit on the mass of a sphere of gas for a given radius. If the observed mass, $M_{850}$, is greater 
than the calculated $M_J$, or alternatively $M_{\mathrm{850}}/$$M_{\mathrm{J}}> 1$, that would 
suggest that the clump is unstable to gravitational collapse and hence active star formation is likely 
\citep{Mairs:2014zr}. An object that has $M_{\mathrm{850}}/$$M_{\mathrm{J}}\ll1$ is currently 
stable and will not collapse (alternatively it has already collapsed and the majority of the mass is 
now contained within the protostar). Given the uncertainties present in theory and observations, the 
stability of objects where $M_{\mathrm{850}}/$$M_{\mathrm{J}}\approx 1$ is ambiguous. Figure~
\ref{fig:jeans_plot} plots $M_{\mathrm{850}}/$$M_{\mathrm{J}}$ against the Jeans length scale for 
the clumps identified in Serpens MWC 297 and reveals that at least three out of a total 22 clumps 
detected by \textsc{fellwalker} are Jeans unstable and may contain protostars. Evidence for these are 
addressed in Section 5. For comparison, $M_{\mathrm{850}}/$$M_{\mathrm{J}}$ is plotted for the 
same list of objects, assuming a single clump temperature of 15\,K (the red crosses in 
Figure~\ref{fig:jeans_plot}). We observe that in a majority of cases using a real temperature has caused 
the ratio to decrease and we therefore conclude that previous authors who have used a constant 
temperature of 15\,K have underestimated the stability of their clumps.

\section{The SCUBA-2 Confirmed YSOc catalogue}

In this section we cross-reference our list of SCUBA-2 clumps, as identified by \textsc{fellwalker}, 
with \emph{Spitzer} YSOc catalogues and produce our own SCUBA-2 confirmed YSOc catalogue 
for the Serpens MWC 297 region. 

We calculate the relative distribution of protostars to PMS stars in the region as a measure 
of dynamical evolution of YSOcs within a star-forming cluster.  We produce Spectral Energy 
Distributions (SEDs) of the YSOcs where supplementary data exist. With the addition of new 
SCUBA-2 data at 450\,$\micron$ and 850\,$\micron$ we update the classification of the 
YSOcs in the Serpens MWC 297 region.

\subsection{IR and other YSO candidates}

We pull together existing YSOc catalogues, discuss the various methods used to compile them, 
compare the distribution of objects to the SCUBA-2 submillimetre data. From here on Class 0, I and 
Flat Spectrum (FS) YSOs are referred to as protostars and Class II, Transition Disk (TD) and III YSOs 
are referred to as Pre-Main Sequence (PMS) stars. 

Three YSOc catalogues are found for the Serpens MWC 297 region, each deploying a different method 
to identify and classify YSOcs. The earliest catalogue found is of \emph{Chandra} ACIS-I X-Ray 
observations carried out by \cite{Damiani:2006ve} over an area of 16.9\arcmin\ $\times$ 8.7\arcmin\ 
centred on the star MWC 297. YSOc identification is a byproduct of the investigation into the X-ray flaring of 
the star MWC 297 and as a consequence their sample is incomplete for the whole of the Serpens MWC 297 region 
(30\,\arcmin\ diameter). They find that the star MWC 297 only accounts for 5.5 per cent of X-ray emission 
in the region. The rest is attributed to flaring low mass PMS. As \cite{Damiani:2006ve} do not make the 
distinction between YSOs and more evolved objects in their work it is not possible to use these data for 
the purposes of classification.   

\SpitzerGB\ and \SpitzerYC\ \citep{gutermuth09} used \emph{Spitzer} observations to catalogue YSOcs 
for the Serpens MWC 297 region. The details of these surveys are noted in Section 2.2. \SpitzerGB\ used 
IRAC and MIPS bands to identify Class I and II detecting a total of 76 YSOcs within a 20\,\arcmin\ 
radius of the centre of the field (Table \ref{table:SGBS_YSO_small}), whereas \cite{gutermuth09} identified 22 
YSOcs using a colour-colour method, though the coverage of \SpitzerYC\ is limited to a 15\arcmin\ square.

Where the samples overlap we find notable differences between the catalogues. \SpitzerGB\ include five 
protostars whereas \SpitzerYC\ include four. Of these samples, only three are consistent across catalogues. 
These are YSOc2, 47 and 11 presented in Table \ref{table:SGBS_YSO_small}. Similarly \SpitzerGB\ identifies 
22 PMS-stars whereas \SpitzerYC\ identified 18. Across the sample 11 are consistent in both catalogues. 
Objects that appear in both catalogues are most likely to be real YSOs. 

%The remaining protostars in this subset do not appear to be consistent with the strong submillimetre peak typically observed for Class 0/I objects. 

Of the two \emph{Spitzer} YSOc surveys, we use \SpitzerGB\ as the primary \emph{Spitzer} catalogue 
because it covers all of the SCUBA-2 mapped area.

\begin{figure*}
\begin{center}
\includegraphics[scale=0.75]{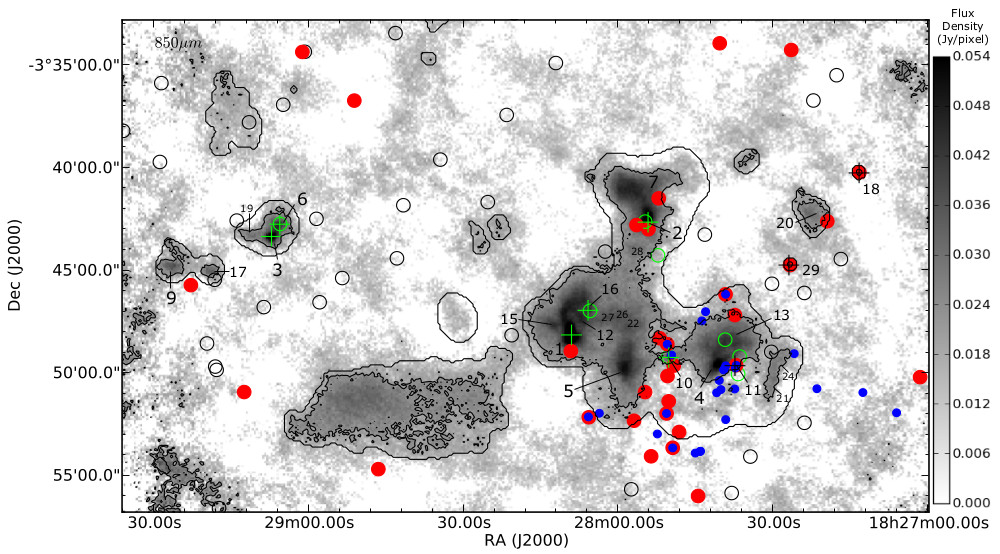}
\caption{850\,$\micron$ greyscale map of Serpens MWC 297. Outer contours mark the data reduction mask (Figure 1) and inner contours the 3$\sigma$ detection level (0.0079 Jy/pixel). 
Circular markers indicate the location of YSOcs as catalogued by \SpitzerGB and crosses indicate the location of SCUBA-2 confirmed YSOs (Table~\ref{table:cores}). 
YSOcs are coded by evolutionary classification based on their spectral indices ($\alpha_{\mathrm{IR}}$) in the \emph{Spitzer} case and by bolometric temperature, $T_{\mathrm{bol}}$, in the SCUBA-2 case (Table~\ref{table:cores}). \emph{Spitzer} YSOcs are indicated by hollow black circles (Class III), solid red circles (Class II) and green hollow circles (Class 0/I). SCUBA-2 confirmed YSOs are indicated by black crosses (Class II) and green crosses (Class 0/I). Small, solid blue circles mark the location of \protect\cite{Damiani:2006ve} X-ray sources, typically associated with Class II and III objects.}
\label{fig:YSO}
\end{center}
\end{figure*}

\begin{figure*}
\begin{center}
\includegraphics[scale=1.45]{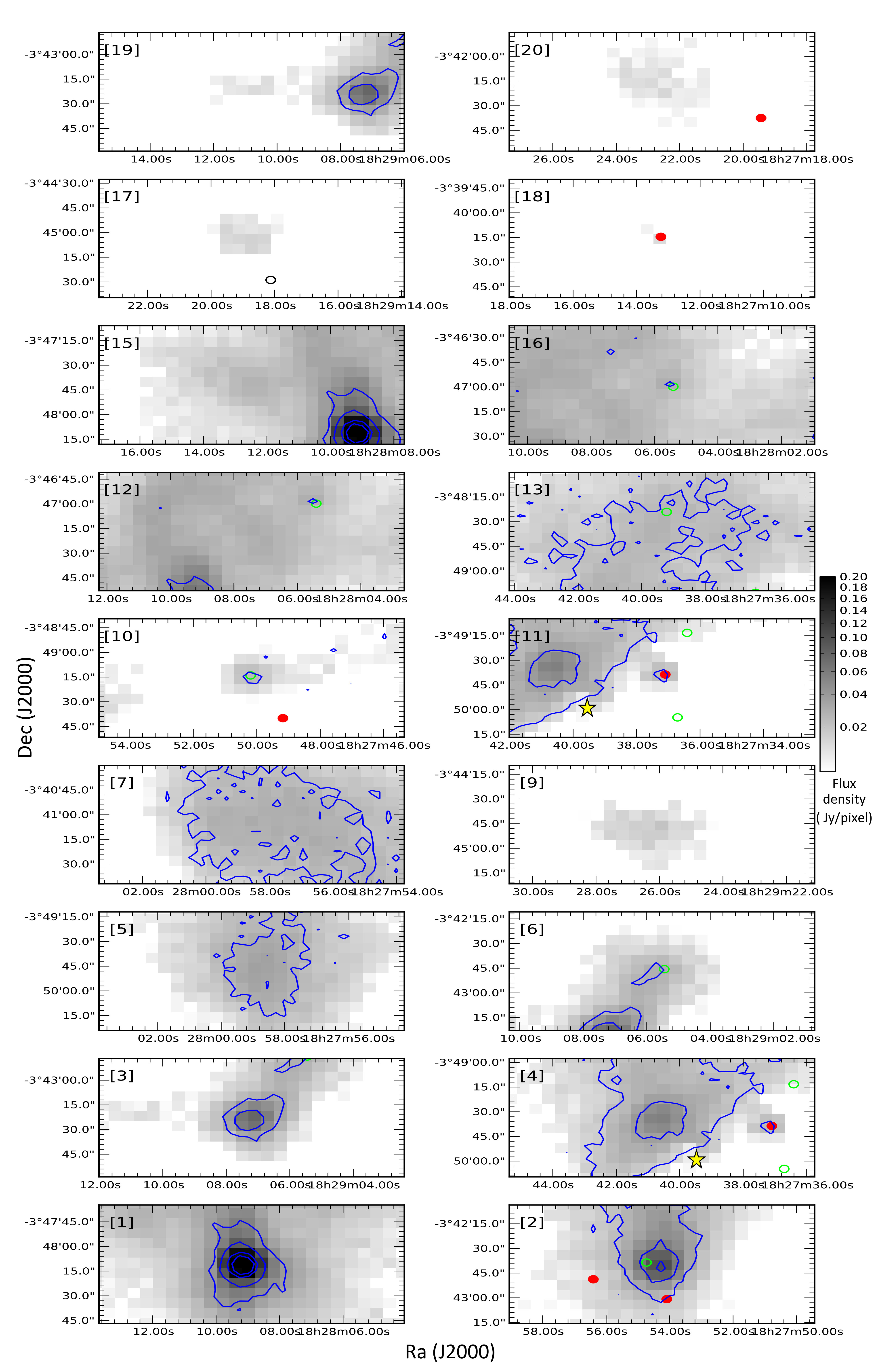}
\caption{Comparison of flux emission from \textsc{fellwalker} objects at 450\,$\micron$ (contours), 850\,$\micron$ (greyscale) and the \SpitzerGB\ YSOcs (markers). Numbers in square brackets correspond to the objects in Table~\ref{tab:mass}. Maps show contours of 450\,$\micron$ submillimetre flux at 5, 10, 20 and 30 $\sigma$ ($\sigma$ = 0.016 Jy/pixel). \emph{Spitzer} YSOcs are indicated by hollow black circles (Class III), solid red circles (Class II) and green hollow circles (Class 0/I). The star indicates the location of the star MWC 297.}
\label{fig:minimaps2}
\end{center}
\end{figure*}

All IR surveys are subject to contamination by Galactic sources (for example, field red giants) and extra-Galactic 
sources (broad line AGN). \cite{gutermuth09} calculate that this should account for less than 2\,per cent of 
sources in Serpens/Aquila. In addition to this, \cite{Connelley:2010nx} discuss how target inclination can play 
a role in classification. In Table~\ref{tab:YSO_cat} we give the total numbers of YSOcs in each catalogue by evolutionary 
class whilst in Figures~\ref{fig:YSO} and \ref{fig:minimaps2} we plot the positions and evolutionary classification of the SGBS 
YSOcs on the 850\,$\micron$ flux map. In Figure~\ref{fig:YSO} we show whether or not the Spitzer YSOcs are consistent 
with the \cite{Damiani:2006ve} X-ray sources. 

%One further catalogue was found for the region. \cite{Connelley:2010nx} uses \emph{IRTF} 2MASS NIR data to classify 
%Class I sources, by spectral index. The study lacks depth, returning a single object for this region and this being 
%MWC 297, an object which is omitted from \SpitzerGB\ due to saturation. This result should be questioned as MWC 297 
%has been observed to be a Class III or ZAMS B1.5Ve star \citep{Drew:1997qf} which are optically visible, where as Class 
%I objects are typically obscured by their natal envelopes. 

\begin{table}
\caption{YSO candidates in the MWC 297 region.}
\label{tab:YSO_cat}
\begin{center}
\begin{tabular}{c|ccccc}
	& \multicolumn{5}{ |c }{YSO Classification}	\\
	&	0/I	&	II	&	III	\\
\hline
\cite{Damiani:2006ve}			&	-	&	-	&	27	\\
\SpitzerGB$^{a}$ - \cite{gutermuth08}	&	8	&	32	&	36	\\
\SpitzerYC\ - \cite{gutermuth09}	&	4	&	16	&	2	\\
%\cite{Connelley:2010nx}			&	1	&	-	&	-	\\
\hline
Total$^{b}$					&	10	&		&	72	&		\\
\hline
\end{tabular}
\end{center}

a) Within a 20\arcmin\ radius area centred at RA(J2000) = $18^{h}$ $28^{m}$ $13^{s}.8$, Dec. (J2000) = $-03^{\circ}$ $44'$ 1.7\arcsec. \\
b) The totals account for sources which feature in multiple catalogues.

\end{table}%

\cite{Kaas:2004fx}, \cite{Winston:2007if} and \cite{harvey07} discuss how evolutionary 
class (determined by IR spectral index) and spatial distribution in a star-forming region 
are correlated, finding that Class 0/I and FS sources are concentrated towards 
the central filaments of Serpens Main region whereas Class II, TD and III sources are much more widely 
distributed. We incorporate SCUBA-2 data into this method, allowing for direct comparison of 
evolutionary class spatial distribution with H$_{\mathrm{2}}$ column density. Our method takes 
the ratio of the number of protostars to PMS stars. Ratios are calculated for the region within the 
data reduction mask (a large scale region defined as where \emph{Herschel} 500\,$\micron$ 
emission is greater than 2\,Jy/beam; see Figure~\ref{fig:maps}), and the emission `cloud' defined 
as above the 3$\sigma$ detection in SCUBA-2 850\,$\micron$, consistent with the levels set for 
\textsc{fellwalker} clump analysis in Section 4.1. In addition the ratio was calculated for the 
space outside of the data reduction mask up to the boundaries of the SCUBA-2 data in 
Figure~\ref{fig:YSO} as a control region. Table~\ref{tab:dist} shows the results for these 
corresponding areas for the YSOcs catalogues listed in Table~\ref{tab:YSO_cat} and plotted 
in Figure~\ref{fig:YSO}. 

\begin{table}
\caption{Ratios of protostars (Class 0, I, FS) to PMS stars (Class II, TD, III) in the \SpitzerGB\ and SCUBA-2 catalogue }
\label{tab:dist}
\begin{center}
\begin{tabular}{c|cc|c}
	&	Protostars	&	PMS-stars	&	Ratio	\\
\hline
Control region		&	0	&	49	&	0.0		\\
 \emph{Herschel} 2Jy beam$^{-1}$ mask	&	10	&	23	&	0.43		\\
 \emph{SCUBA-2} 3$\sigma$ mask	&	8	&	10	&	0.80		\\
\hline
\end{tabular}
\end{center}
\end{table}%

Preliminary work by \cite{Kaas:2004fx} suggested that Class I to Class II ratios were 10 
times greater within cloud regions of Serpens Main than outside them. \cite{harvey07} conducted 
a similar analysis and found ratios of 0.37 for the whole region and 1.4 and 3.0 for the cloud 
regions. Whereas our ratios are not as large (0.8), they do follow the same trend of greater 
numbers of protostars in regions of higher column density, supporting the conclusion that
protostars form in regions of high column density and then migrate away from these regions 
as they evolve into PMS-stars. 

\begin{figure*}
\begin{center}
\includegraphics[scale=1.45]{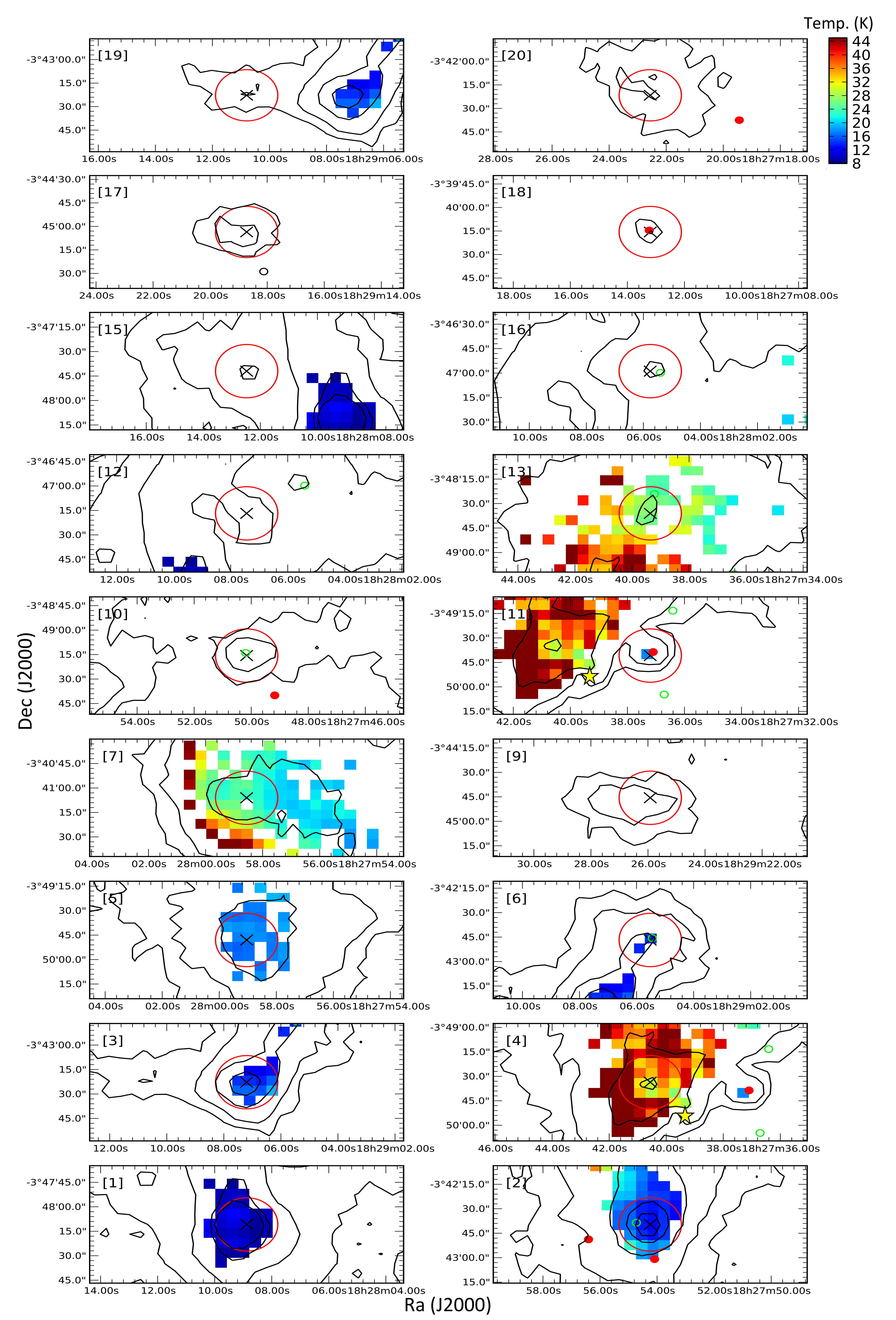}
\caption{Submillimetre clumps in Serpens MWC 297 as identified by the \textsc{fellwalker} clump-finding algorithm. Numbers in square brackets correspond to the objects in Table~\ref{tab:mass}. Maps show contours of 850\,$\micron$ submillimetre flux at 5, 10, 20 and 30 $\sigma$ ($\sigma$ = 0.0022 Jy/pixel) up to the position of peak flux (black cross). The aperture from which SED flux density was calculated is plotted as the scale size of a protostellar core (0.05\,pc). Temperature is shown where it is statistically significant and is used to calculate the masses shown in Table~\ref{table:cores}. \emph{Spitzer} YSOcs are indicated by hollow black circles (Class III), solid red circles (Class II) and green hollow circles (Class 0/I). The star indicates the location of the star MWC 297.}
\label{fig:minimaps1}
\end{center}
\end{figure*}

%SCUBA-2 catalogue

\subsection{SCUBA-2 YSO candidates}

\begin{table*}
\caption{Properties of YSO candidates in MWC 297.}
\label{table:cores} 
\begin{center}
\begin{tabular}{@{}lllcccccccc}
%{tabular}{@{}lllccllcllc}
%{@{}l{1cm}llcccccccc}

\hline
ID$^{a}$		&	$S_{\mathrm{450}}^{b}$	&	$S_{\mathrm{850}}^{b}$ 	&	$M_{\mathrm{850}}^{c}$	&	$T_{\mathrm{d}}^{d}$ &	$T_{\mathrm{bol}}^{e}$ & $\alpha_{\mathrm{IR}}^{f}$ & $L_{\mathrm{bol}}^{e}$	& $L_{\mathrm{smm}}$/$L_{\mathrm{bol}}^{e}$	&	SGBS class$^{g}$		&	Class \\
			&	(Jy)	&	(Jy)	&		(M$_{\mathrm{\odot}})$ 	&	(K) 	&	(K)	&	& (L$_{\mathrm{\odot}})$	&	per cent	&		&	 \\
\hline
\hline
S2-YSOc1		&	14.4	&	3.09	&	5.1(0.5)	&	10.3(0.5)  		&	30(3) 	&	1.65(	0.08)		&	1.1(0.1)		&	5.0(0.5)	&	`Red'			&	0\\
S2-YSOc2		&	11.1	&	1.56	&	1.3(0.1)	&	15.6(0.8) 		&	290(30)	&	0.56(0.05)  	&	2.1(0.2)		&	1.9(0.2)	&	`YSOc red'		&	I\\
S2-YSOc3		&	7.0	&	1.09	&	0.95(0.08)	&	14.8(0.7)		&	8(1)	 	&	1.4(0.7)		&	0.8(0.1)		&	3.0(0.3)	&	`Flat'	 			&	0\\ 
S2-YSOc6	 	&	4.1	&	0.68	&	0.62(0.06)	&	14.2(0.7)  		&	100(10)	&	0.30(0.05)		&	0.28(0.03)		&	5.2(0.5)	&	`YSOc'			&	0/I\\
S2-YSOc10		&	4.5	&	0.37	&	0.31(0.03)	&	15.0(-)		&	190(20)	&	0.17(0.05)		&	0.82(0.08)		&	1.8(0.2)	&	`YSOc star+dust'	&	I\\
S2-YSOc11		&	3.9	&	0.35	&	0.22(0.02)	&	17.6(0.9) 		&	780(60)	&	-0.43(0.06)	& 	3.3(0.2)		&	0.3(0.0)	&	`YSOc'			&	II\\
S2-YSOc16		&	3.7	&	0.73	&	0.60(0.05)	&	15.0(-) 		&	120(10)	&	0.9(0.3)		&	0.73(0.07)		&	1.9(0.2)	&	`star F5V'			&	I\\
S2-YSOc18		&	0.1	&	0.11	&	0.09(0.02)	&	15.0(-) 		&	820(50)	&	-0.17(0.05)	&	1.32(0.08)		&	0.1(0.0)	&	`YSOc star+dust'	&	II\\
S2-YSOc29		&	4.0	&	0.36	&	0.30(0.03)	&	15.0(-) 	 	&	860(50)	&	-0.49(0.05)	&	0.34(0.02)		&	0.1(0.0)	&	`YSOc star+dust'	&	II\\
\hline
MWC 	&	-		&	1.05	&		-	&	-	 		&	660(6)	 &	-			&	422(4) 		&	0.1(0.0)	&	`2mass'			&	III\\
  297		&\\

\end{tabular}
\end{center}

\raggedright
a) SCUBA-2 YSOcs (S2-YSOc) as identified by cross-referencing the SCUBA-2 clumps in Table~\ref{tab:mass} (Figure~\ref{fig:minimaps1}) with IR sources (Table \ref{table:SGBS_YSO_small}). \\
b) Integrated fluxes of the YSOcs determined by fixed 40\arcsec\ diameter aperture photometry. The uncertainty at 450\,$\micron$ is 0.0165\,Jy/pixel and at 850\,$\micron$ is 0.0022\,Jy/pixel. There is an additional systematic error in calibration of 10.6\,per cent and 3.4\,per cent at 450\,$\micron$ and 850\,$\micron$.\\
c) Mass as calculated with equation \ref{eqn:mass}. Errors in brackets are calculated from error in total flux, described in b., and error in mean temperature of 5\,per cent. These results do not include the systematic error in distance (20\,per cent) and opacity (factor of two). \\
d) Mean temperature as calculated from the temperature maps (Figure~\ref{fig:mwc}). Where no temperature data is available an arbitrary value of 15K(-) is assigned that is consistent with the literature.  \\
e) YSOcs are classified using the $T_{\mathrm{bol}}$, $L_{\mathrm{bol}}$ and  $L_{\mathrm{smm}}$/$L_{\mathrm{bol}}$ methods which are described in Section 5.4.\\
f) Values for spectral index are taken from the \SpitzerGB\ catalogue.\\
g) SGBS notation is described in \cite{evans09}. \\
%* Apertures for SMM source 4 and 12 were contaminated by neighbouring sources, the resulting fluxes have been scaled back to compensate for this. Original values (corresponding order) are as follows; 3.9Jy and 0.36 Jy for 450\,$\micron$ and 850\,$\micron$ in object 12 and 8.80 Jy and 0.90 Jy at 450\,$\micron$ and 850\,$\micron$ in object 4.

\end{table*}

In this section we determine which members of the SCUBA-2 clump catalogue (Table~\ref{tab:mass}) 
are starless and which host YSOs, as \textsc{fellwalker} is parameterised to identify both. The 
\textsc{fellwalker} algorithm is ideal for identifying larger scale, often irregular and extended 
clumps, but not effective for extracting the flux of individual YSOs, which are smaller. We 
extract a revised catalogue of YSOcs (Table~\ref{table:cores}) based on the position of the 
clumps listed in Table~\ref{tab:mass} and calculate the flux emission using aperture photometry 
with a fixed 40\arcsec diameter aperture.

Six clumps are found to contain SGBS YSOcs (Table \ref{table:SGBS_YSO_small}) by cross-referencing 
the SCUBA-2 clumps in Table~\ref{tab:mass} (Figure~\ref{fig:minimaps1}) with IR sources 
(Table \ref{table:SGBS_YSO_small}). Two further clumps (SMM 1 and 3) are found with 
little or no IR emission but are centrally condensed and have $M_{\mathrm{850}}/$$M_{\mathrm{J}}> 1$ 
signifying they are gravitationally unstable and may be early protostellar (Class 0) YSOcs.  

The following YSOcs-hosting clumps detected (SMM 1, 2, 3, 6, 10, 11, 16 and 18) are listed in 
Table~\ref{table:cores} as SCUBA-2 YSO candidates (S2-YSOc). The remaining clumps 
listed in Table~\ref{tab:mass} do not contain YSOcs and are considered starless. 
SMM 4 and 7 are notable as they have relatively high masses (greater than 1 $M_{\mathrm{\odot}}$) 
but are not forming stars. SMM 5 has $M_{\mathrm{850}}/$$M_{\mathrm{J}}= 1$ but there is no 
evidence for a 24\,$\micron$ source there. It could be argued that this is a prestellar object on 
the cusp of becoming protostellar.

In addition to all those submillimetre objects identified by \textsc{fellwalker}, we also include one 
additional YSOc, S2-YSOc 29, as listed in Table~\ref{table:cores} and YSOc11 in Table \ref{table:SGBS_YSO_small}. 
This object fulfils the criterion of coincidence with a strong IR source in the \emph{Spitzer} 24\,$\micron$ 
MIPS data and a corresponding Class III identification in the SGBS YSOcs catalogue. S2-YSOc 29 
registers a 5$\sigma$ detection with one pixel and resembles S2-YSOc 10 and 18 which are also believed 
to be an isolated, Class III PMS-stars with the remnants of an envelope/cold accretion disc contributing 
to their observed submillimetre flux. 

Apertures were placed over the peak positions of the \textsc{fellwalker} clumps (Table~\ref{tab:mass}) in 
addition to the \emph{Spitzer} YSOcs positions and the integrated SCUBA-2 flux calculated with the intention 
to measure the flux from any dense, protostellar core associated with the SCUBA-2 clump peak and/or Spitzer YSOc. 
We follow \cite{di-Francesco:2007bh}, \cite{Sadavoy:2010ve} and \cite{Rygl:2013ve}'s definition of a core as a 
gravitationally bound, dense object, of diameter less than 0.05 pc and set apertures at this size (40\arcsec at 250\,pc). 
Some larger scale emission is likely to be observed. However, through careful selection of aperture size we can 
assume that emission from the core dominates at this length scale. 

%Both \cite{di-Francesco:2007bh} and \cite{Sadavoy:2010ve} define their apertures on the physical properties of the source, a protostellar core of diameter 0.07 pc

Figures \ref{fig:YSO} and \ref{fig:minimaps2} show the locations of the SCUBA-2 YSOcs as well as those 
catalogued in the \SpitzerGB\ catalog. Figure~\ref{fig:minimaps1} shows the relationship between the 
submillimetre peaks and the \emph{Spitzer} YSOc position, with the SCUBA-2 fluxes for \emph{Spitzer} 
YSOcs presented in Table~\ref{tbl:spitzeradd}. The mass of the SCUBA-2 YSOcs are calculated with 
Equation~\ref{eqn:mass}, using a constant, mean temperature derived from our maps, and the results presented in 
Table~\ref{table:cores}. 
 
A small number of \emph{Spitzer} YSOcs inside the \emph{Herschel} 500\,$\micron$ data reduction mask are consistent 
with SCUBA-2 YSOcs with identical peak positions, for example in S2-YSOc 18 (Figure~\ref{fig:minimaps1}). In some 
cases, positions appear offset, for example S2-YSOc 2. This anomaly can be explained 
by virtue of the deeply embedded nature of the source and that \emph{Spitzer} might be observing IR 
emission from an outflow cavity rather than the YSO itself. 

YSOcs classified as 0/I by \emph{Spitzer} should also have evidence of a SCUBA-2 peak at the 
same position. Those \emph{Spitzer} detected protostars (YSOc16 and 38; Table~\ref{table:SGBS_YSO_small}) 
that lie outside of the 5$\sigma$ detection limit at 850\,$\micron$ and have no obvious peak in emission 
are unlikely to be YSOcs and discarded as incorrectly classified objects. 

A minority of cases detect greater than 5$\sigma$ flux but have no significant peak in emission, for 
example YSOc15 and 21. Examining these specific cases, both are classified as protostars and are 
deeply embedded within S2-YSOc13. Figure~\ref{fig:mwc}c shows how this region is near the centre of the 
reflection nebulae SH2-62 and therefore we interpret YSOc15 and 21 as IR emission from dust heated 
by the star MWC 297 and not real YSOcs. Many of the remaining \emph{Spitzer} YSOcs detect low level, 
extended SCUBA-2 flux with no significant peak. No significant flux is detected for objects outside the mask. 

%Those protostars within the 5$\sigma$ contour in Figure \ref{fig:YSO}, but with no obvious peak in emission, YSOc15 and 21, should be considered with suspicion.

\begin{table}
\caption{SCUBA-2 $40''$ aperture fluxes for the \emph{Spitzer} YSOc listed in Table~\ref{table:SGBS_YSO_small}.  A full version of this catalogue is available online.}  
\label{tbl:spitzeradd} 
\begin{center}
\begin{tabular}{@{}lccc@{}}

\hline
ID &{$S_{\mathrm{450}}$ }&{$S_{\mathrm{850}}$ }	&	S2-YSOc ID\\
&{ Jy }&{Jy }	&	\\
\hline
% output from getfluxes_MWC297.py
% YSOc number | S450 / Jy | S850 / Jy % footnote a: Extended low level emission in aperture.  No significant peak at YSOc position \($<3\sigma$\).\% footnote b: Outside data reduction mask.  No significant flux detected in initial data reduction stage \($<5\sigma$\).\
YSOc1	&$<0.72^{b}$	&$<0.065^{b}$&-\\
YSOc2	&$<0.72$	&$0.115\pm0.013$&S2-YSOc18\\ 					
YSOc3	&$<0.72^{b}$	&$<0.065^{b}$&-\\
YSOc4	&$<0.72^{b}$	&$<0.065^{b}$&-\\
YSOc5	&$<0.72^{b}$	&$<0.065^{b}$&-\\
YSOc6	&$<0.72^{b}$	&$<0.065^{b}$&-\\
YSOc7	&$<0.72^{b}$	&$<0.065^{b}$&-\\
YSOc8	&$<0.72^{b}$	&$<0.065^{b}$&-\\
YSOc9	&$<0.72^{b}$	&$<0.065^{b}$&-\\
YSOc10	&$<0.72^{b}$	&$<0.065^{b}$&-\\
YSOc11	&$<0.72$	&$0.154\pm0.013$&S2-YSOc29\\
YSOc12	&$<0.72^{b}$	&$<0.065^{b}$&-\\
YSOc13	&$1.73\pm0.14$$^{a}$	&$0.184\pm0.013$$^{a}$&-\\
YSOc14	&$<0.72^{b}$	&$<0.065^{b}$&-\\
YSOc15	&$3.30\pm0.14$$^{a}$	&$0.408\pm0.013$$^{a}$&-\\ %no peak
YSOc16	&$2.08\pm0.14$$^{a}$	&$0.071\pm0.013$$^{a}$&-\\
YSOc17	&$3.11\pm0.14$	&$0.362\pm0.013$&S2-YSOc11\\					
YSOc18	&$2.10\pm0.14$$^{a}$	&$0.263\pm0.013$$^{a}$&-\\	%no peak
YSOc19	&$<0.72^{b}$	&$<0.065^{b}$&-\\
YSOc20	&$1.11\pm0.14$$^{a}$	&$0.112\pm0.013$$^{a}$&-\\

\ldots & \ldots & \ldots & \ldots\\
\hline
\end{tabular}
\end{center}

(a) Extended low level emission in aperture.  No significant peak at YSOc position  ($>3\sigma$).   \\
(b) Outside data reduction mask.  No significant flux detected in initial data reduction stage ($<5\sigma$). \\
\end{table}

\subsection{Spectral Energy Distributions}

SEDs are powerful tools for determining the properties of a star and we use these as an aid to 
classification through measurement of the spectral index  across their IR wavebands, bolometric 
temperature and luminosity ratio \citep{evans09}.

SEDs are constructed from archival Two Micron All Sky Survey (2MASS) fluxes, \emph{Spitzer} 
fluxes, and from SCUBA-2 fluxes.  For the SCUBA-2 fluxes we conducted aperture photometry 
(as described in Section 5.2) at both 450\,$\micron$ and 850\,$\micron$ centred on the \textsc{fellwalker} 
clump peaks from Table~\ref{tab:mass}. None of our sources overlapped sufficiently to make blended emission 
a problem. 

%\cite{Berrilli:1992cr}, \cite{Hillenbrand:1992kl} and \cite{di-Francesco:1994dq} use an 
%approximately 15\arcsec\ aperture to measure flux which is representative an upper 
%limit based of the technology of the time. More modern instruments have much 
%greater mapping extent and instead 

%In two cases (4 and 12 in figures \ref{fig:minimaps1} and  
%\ref{fig:minimaps2}), the \textsc{fellwalker} peaks of a core was sufficiently close to another 
%object that they encroached on each others aperture. Our solution is to halve the size 
%of the aperture, removing the adjacent object from view. We then measure the new flux 
%and take the ratio of this and the full aperture. Those objects which are overlapping with 
%additional structure are singled out by having substantially greater ratios than the bulk of the cores. 
%these are removed from the sample and the mean ratio calculated. Assuming identical core profile 
%across our sample, we then apply this mean in reverse to cores 4 and 12, scaling flux back up to a full 
%sized aperture but with a profile that represents an isolated core, thus removing the contamination.

%Both objects were relatively weak in 450\,$\micron$ and have approximately 24per cent and 26per cent contamination 
%from neighbouring objects, 4 and 12 respectively. Similarly we get approximately 5per cent and 8per cent at 
%450\,$\micron$ and 850\,$\micron$ contamination in object 12.  [REWORD]. 

Our primary sources are IRAC and MIPS data from the SGBS. Six out of nine 
objects are identified in the \SpitzerGB\ YSOc catalogue. We access the full \SpitzerGB\ source 
catalogue, which includes sources not classified as YSOcs, and find fluxes of each of the remaining 
three objects. S2-YSOc 1 and 3 are low luminosity objects that cannot be reliably classified as a 
YSOc by \emph{Spitzer} and are therefore labelled `Red' and `Flat' following a description of 
their SEDs. Both objects have IRAC and MIPS  fluxes that are many orders of magnitude less than 
their peers. S2-YSOc 16 has been classed as a F5V star. Following the work of \cite{Alonso-Albi:2009ve} 
we bring together fluxes and present the SEDs in Figure~\ref{fig:SED} with specific cases of 
individual YSOs discussed in depth the following sections. 

Many of the following methods directly use the SEDs constructed in this section to classify YSOs by 
examining how the flux of the object varies with wavelength. 

\subsection{YSO classification}

\begin{figure*}
\begin{center}
\includegraphics[scale=1.2]{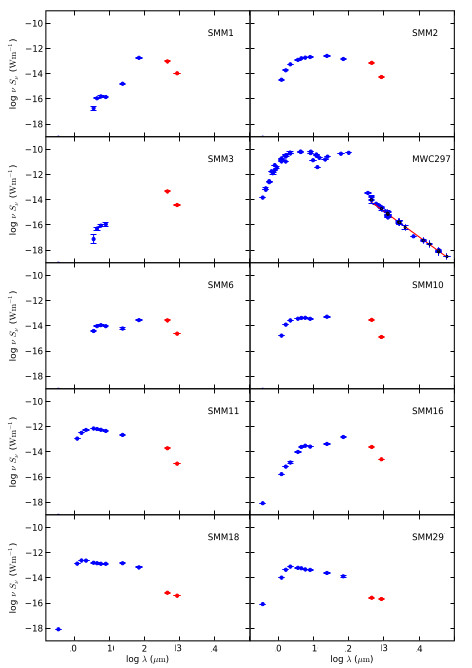}
\caption{Spectral Energy Distributions for YSOcs associated with \textsc{fellwalker} clumps (Table~\ref{table:cores}). Blue points represent archive data sourced from \emph{Spitzer} and 2MASS. Red points show new data provided by SCUBA-2 at 450\,$\micron$ and 850\,$\micron$ (note that the star MWC 297 was not identified by \textsc{fellwalker} after free-free contamination was accounted for). The straight line in MWC 297 describes free-free emission from an UCH\textrm{II} region and polar jet/wind with a spectral index $\alpha$=1.03.}
\label{fig:SED}
\end{center}
\end{figure*}

Spectral index, $\alpha_{\mathrm{IR}}$, is a direct measurement of the gradient of the SED slope 
over an range of IR wavelengths (typically 2 to 24\,$\micron$) and is expressed as 

\begin{equation}
\alpha_{\mathrm{IR}} = \frac{d \log(\lambda S_{\lambda})}{d \log(\lambda )}. \label{eqn:alpha}
\end{equation}

\cite{gutermuth08} calculated $\alpha_{\mathrm{IR}}$ from the fluxes in the \SpitzerGB\ catalogue and we 
display these results in Table \ref{table:cores} and Table~\ref{table:SGBS_YSO_small} for SGBS. 
As a classification tool for YSOs, $\alpha_{\mathrm{IR}}$ was developed by \cite{Lada:1984fk} and 
\cite{Greene:1994kl} and is summarised by \cite{evans09} who specify the boundaries between 
Class 0/I, FS, II and III as $\alpha_{\mathrm{IR}}$ = 0.3, -0.3 and -1.6.

%\begin{itemize}
%\item Class 0/I : $0.3 \leq \alpha _{IRAC}$
%\item Class Flat Spectrum : $-0.3 \leq \alpha _{IRAC} \leq 0.3$
%\item Class II : $-1.6 \leq \alpha _{IRAC} \leq -0.3$
%%\item Class III : $\alpha _{IRAC} \leq -1.6$
%\end{itemize}
$\alpha_{\mathrm{IR}}$ is one the most commonly used methods for the classification of protostars and consequently 
is one of the most criticised. Uncertainties on $\alpha_{\mathrm{IR}}$ typically vary between 10 and 20 per cent. 
However, measurements have been shown to be highly susceptible to disk geometry and source inclination 
\citep{Robitaille:2007zr} whilst extinction is known to cause $\alpha_{\mathrm{IR}}$ to appear larger. Furthermore, 
the development of $\alpha_{\mathrm{IR}}$ predates the identification of the Class 0 protostar \citep{Chandler:1990qf, 
Eiroa:1994ve, Andre:2000zr} and therefore does not distinguish between Class 0 and Class I when $\alpha_{\mathrm{IR}}$ is 
measurable (absence of  $\alpha_{\mathrm{IR}}$ has been taken in this work to define a Class 0). Via the classification 
scheme outlined above, our sample contains four Class 0/I, two FS and three Class II sources. Saturation of 
\emph{Spitzer} bands prevent measurement of $\alpha_{\mathrm{IR}}$ for MWC 297.

We calculate bolometric temperature, $T_{\mathrm{bol}}$, and luminosity, $L_{\mathrm{bol}}$, as alternative methods 
of classification of YSOs. We follow the numerical integration method of \cite{Myers:1993vn} and \cite{Enoch:2009xq} 
who calculated the discrete integral of the SED of an object for a given number of recorded fluxes. By adding SCUBA-2 
data to that from the \SpitzerGB\ source catalogue, we extend the SEDs (Figure \ref{fig:SED}) for our YSOcs into the 
submillimetre spectrum and allow for a more complete integral from which we calculate $T_{\mathrm{bol}}$, the 
temperature of a black body with the same mean frequency of the observed SED, via 

\begin{equation}
T_{\mathrm{bol}} = 1.25\times10^{-11}\bar{\nu }\,(\mathrm{K}\,\mathrm{Hz}^{-1}), \label{equ:tbol1}
\end{equation}
where $\bar{\nu}$ is the mean frequency of the whole spectrum,
\begin{equation}
\bar{\nu }=\frac{\int \nu S_{\nu} d\nu }{\int \nu d\nu }.\label{equ:tbol2}
\end{equation}
 
Classification separating boundaries for $T_{\mathrm{bol}}$ Class 0, I, II and III are 70, 350, 650 and 2800\,K \citep{Chen:1995ys}.
%\begin{itemize}
%\item Class 0 : $T_{\mathrm{bol}}< 70K$
%\item Class I : $70K \leq T_{\mathrm{bol}}\leq  350K$
%\item Class II : $350K \leq T_{\mathrm{bol}}\leq  650K$
%\item Class III : $650K \leq T_{\mathrm{bol}}\leq  2800K$
%\end{itemize}

$T_{\mathrm{bol}}$ measurements for our sources are listed in Table~\ref{table:cores}. As this method uses more available data 
it could be considered a more reliable method of classification than $\alpha_{\mathrm{IR}}$ which only covers IRAC and MIPS 
bands 2\,$\micron$ to 24\,$\micron$. Furthermore $T_{\mathrm{bol}}$ provides a quantifiable method for separating Class I and Class 0.
Similarly we calculate the ratio of submillimetre luminosity ($L_{\mathrm{smm}}$), defined as $\geq$ 350\,$\micron$ by \cite{Bontemps:1996fu}, 
to $L_{\mathrm{bol}}$ in the method described by \cite{Myers:1998ys} and \cite{Rygl:2013ve}, to classify YSOs: 

\begin{equation}
L_{\mathrm{bol}} = 4\pi d^{2}\int_{0}^{\infty}S_{\nu}  d\nu \label{eqn:Lbol},
\end{equation}
and likewise for the submillimetre luminosity,
\begin{equation}
L_{\mathrm{smm}} = 4\pi d^{2}\int_{0}^{350\micron}S_{\nu} d\nu \label{eqn:Lsmm}.
\end{equation}

This method was developed by \cite{Andre:1993ff} who originally set the Class 0/I boundary 
at 0.5 per cent (subsequently used by \cite{ Visser:2002ly} and \cite{Young:2003fk}). 
\cite{Maury:2011ys} and \cite{Rygl:2013ve} revise this upwards to 3 per cent and most recently 
\cite{Sadavoy:2014nx} has used 1 per cent outlining the lack of consensus on this issue. We follow 
the work of \cite{Rygl:2013ve} and classify objects with $L_{\mathrm{smm}}$/$L_{\mathrm{bol}}$ 
$\geq 3$ per cent as Class 0 protostars.
%\begin{itemize}
%\item Class 0 : $L_{\mathrm{smm}}$/$L_{\mathrm{bol}}$ \geq 3\,per cent$
%\item Class I or higher : $L_{\mathrm{smm}}$/$L_{\mathrm{bol}}$ \leq  3\,per cent$
%\end{itemize}
Likewise, results for $L_{\mathrm{smm}}$/$L_{\mathrm{bol}}$ are listed in Table \ref{table:cores}. 

%these methods are capable of classifying the earliest, Class 0 stage of protostellar evolution whereas 
%$\alpha_{\mathrm{IR}}$ often classifies Class 0 by lack of IR emission [CITE], a method that has become 
%increasingly less reliable as instrumentation has become more effective at detecting low luminosity sources. 
%Bolometric temperature and luminosity methods utilise the full spectrum including the submillimetre 
%wavelengths where emission is believed to be dominant in protostars.

Our sample contains two Class 0 sources, four Class I and three Class II by $T_{\mathrm{bol}}$ and 
three Class 0 to six Class I, II \& III sources by $L_{\mathrm{smm}}$/$L_{\mathrm{bol}}$ . 

Uncertainties on $L_{\mathrm{bol}}$, $L_{\mathrm{smm}}$/$L_{\mathrm{bol}}$ and $T_{\mathrm{bol}}$ 
were calculated using a Monte Carlo method. A normal distribution of fluxes, with the mean on the measured 
flux at each wavelength for each YSO with a standard deviation equal to the original error on the measurements 
was produced. From each set of fluxes our classifications were calculated and the standard deviation on 
results listed in Table~\ref{table:cores}. The size of the uncertainties are consistent with \cite{Dunham:2008fk}. 
\cite{Dunham:2008fk} and \cite{Enoch:2009xq} both study the error on $L_{\mathrm{bol}}$ and $T_{\mathrm{bol}}$ 
and conclude incompleteness of the spectrum is a major source of systematic error in results of order approximately 
31\,per cent and 21\,per cent (respectively) when compared to a complete spectrum. \cite{Enoch:2009xq} find 
that the omission of the 70\,$\micron$ flux is particularly critical when interpreting classification, leading to an 
overestimate of $L_{\mathrm{bol}}$ by 28\,per cent and underestimate of $T_{\mathrm{bol}}$ by 18\,per cent.

\begin{figure*}
\begin{center}
\includegraphics[scale=0.35]{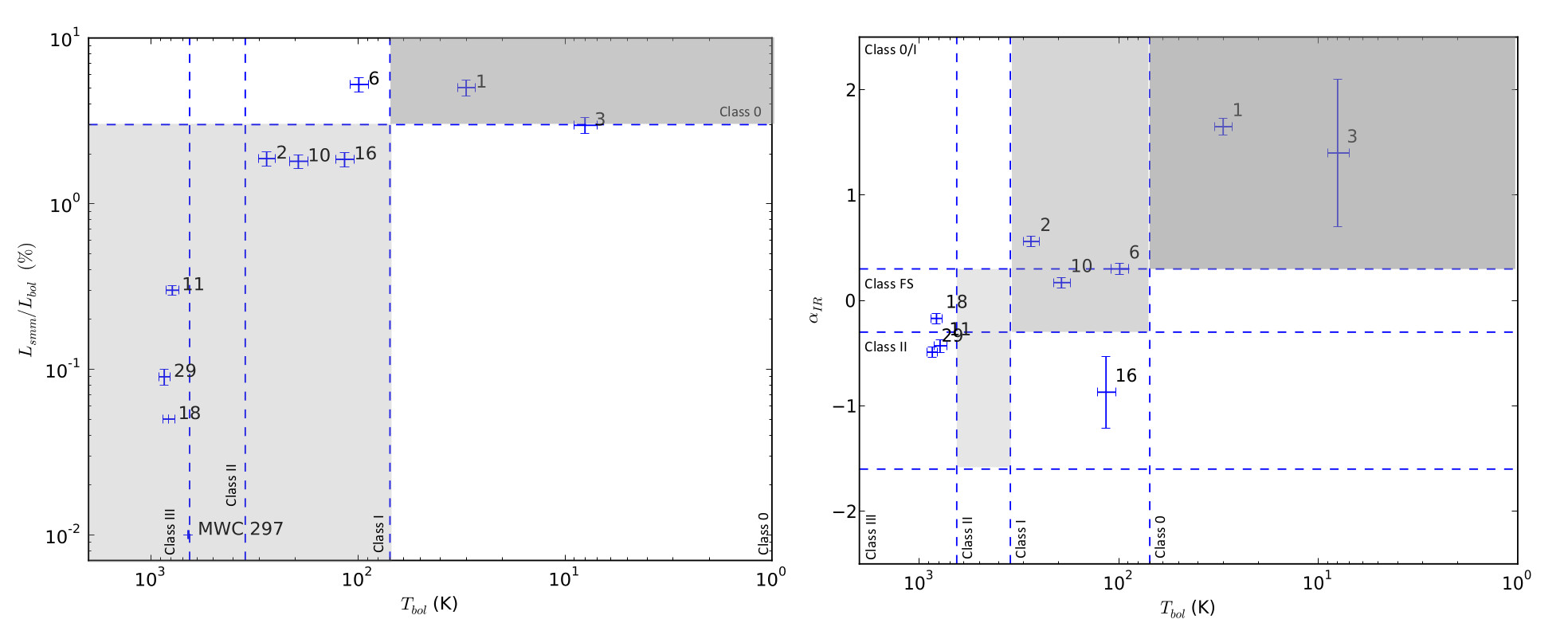}
\caption{ Bolometric temperature plotted against $L_{\mathrm{smm}}$/$L_{\mathrm{bol}}$ (\emph{left}) and $\alpha_{\mathrm{IR}}$ (\emph{right}) for the 10 YSOcs listed in Table~\ref{table:cores}. 
Dashed lines indicate the boundaries of classification of objects (greyed boxes indicating regions of class space where methods agree).}
\label{fig:class}
\end{center}
\end{figure*}

Figure~\ref{fig:class} shows a direct comparison between the $\alpha_{\mathrm{IR}}$,  
$L_{\mathrm{smm}}$/$L_{\mathrm{bol}}$ and $T_{\mathrm{bol}}$ methods of classifying 
YSOs. As outlined above, each specialises in classification at different stages of evolution 
with $T_{\mathrm{bol}}$ arguably being the most effective for classifying protostars. 
\cite{Young:2005ly} studied the merits of $T_{\mathrm{bol}}$ and $L_{\mathrm{smm}}$/$L_{\mathrm{bol}}$ 
and concluded that the latter is the more robust method for classifying Class 0 objects when 
compared to models of core collapse. However, it is also more sensitive to incompleteness 
of the submillimetre spectrum. With only two fluxes at wavelengths greater than 350\,$\micron$ 
for the majority of the YSOs in MWC 297, we must consider the results from 
$L_{\mathrm{smm}}$/$L_{\mathrm{bol}}$ to be incomplete and therefore less reliable 
than $T_{\mathrm{bol}}$. 

Out of the three objects classified as Class 0 by both $L_{\mathrm{smm}}$/$L_{\mathrm{bol}}$ 
and $T_{\mathrm{bol}}$ methods, only S2-YSOc 1 is consistent in both regimes. This object has a 
significantly positive value of $\alpha_{\mathrm{IR}}$ and so we classify this object as Class 0. 
The other two objects, S2-YSOc 3 and S2-YSOc 6, are forming in close proximity to each other but 
relatively isolated from the rest of the cloud. With a minimum separation of approximately 
10,000\,AU it seems likely that these objects formed together and therefore they are likely to 
be a similar class. S2-YSOc 3 has no noticeable IR flux at 24\,$\micron$. However, the S2-YSOc 3 SED 
(Figure \ref{fig:SED}) shows \emph{Spitzer} data consistent with emission from a heated region and so we conclude that the 
emission at 24\,$\micron$ is sufficiently weak that it does not surpass the noise level and therefore 
does not appear in Figure \ref{fig:mwc}c. Such low luminosity emission would be typical of Class 0 
and therefore we label it as such. S2-YSOc 6 has a weak, if non-negligible, detection at 24\,$\micron$ data. 
Therefore, we label it as Class 0/I. S2-YSOc 2 and 10 consistently fall into the Class I bracket by all three methods.

%SMM 2 and 10 consistently fall into the Class I bracket by all three methods. This is an interesting result as SMM2 is a 
%relatively massive, extended object whereas SMM10 is a low mass isolated core as shown in Figure~\ref{fig:minimaps1}. 
%Our physical understanding of the Class I objects raises questions about the validity of the classification of SMM10. 
%It seems this is a likely candidate where inclination is influencing the range of flux detected. 

S2-YSOc 11, 18 and 29 all represent highly evolved and largely isolated cores that are consistently 
classified as Class II/III objects and have 24\,$\micron$ detections in Figure~\ref{fig:mwc}c. 
Finally we discuss S2-YSOc16, an object labelled Class I by $T_{\mathrm{bol}}$ and by $\alpha_{\mathrm{IR}}$ 
and with a strong peak in the 24\,$\micron$ data. Figure~\ref{fig:minimaps1} shows how 
this object appears deep within an extended dust cloud. This scenario fits the definition of a Class 
I and the low mass of the object (0.60 $M_\odot$) when compared to the mass available in the 
neighbouring clumps (approximately 21 $M_\odot$) suggests that this object is early in its accretion 
life cycle.

\section{Discussion}

%[Temperature maps can be made from subregions of MWC 297 because of the quality of SCUBA-2 data]

In this paper we use SCUBA-2 450\,$\micron$ and 850\,$\micron$ data and \emph{Spitzer} 
data to investigate star formation in Serpens MWC 297 region. Taking the ratio of SCUBA-2 fluxes, we produce 
temperature maps of subregions of Serpens MWC 297 and calculate the properties of YSOs and clumps in 
the region. 

Our work builds on analytical techniques developed for SCUBA data \citep{Johnstone:2000fk, Kirk:2006vn, Sadavoy:2010ve} 
to analyse SCUBA-2 data at the same wavelengths. SCUBA-2 represents a significant improvement over 
its predecessor as it has an array of 10,000 pixels, as opposed to 128. Practically this gives the instrument a 
much wider field of view and allows larger regions to be observed quicker and to greater depth. Restricted 
to SCUBA, larger regions of star formation, for example Orion \citep{Nutter:2007ys} and Perseus 
\citep{Hatchell:2007qf}, were prioritised over the low mass Serpens MWC 297 region.

The JCMT GBS extends the coverage of the local star-forming regions over those mapped by SCUBA. SCUBA-2 
also offers much greater quality and quantity of 450\,$\micron$ data, as a result of improved array technology and 
reduction techniques pioneered by \cite{Holland:2006uq, Holland:2013fk}, \cite{Dempsey:2013uq} and 
\cite{Chapin:2013vn}. \cite{Mitchell:2001ve} is able to construct partial temperature maps from SCUBA 450\,$\micron$ 
and 850\,$\micron$ data but is limited to general statements about the region as a result of high noise estimates at 
450\,$\micron$. \cite{Reid:2005ly} go further in their use of 450\,$\micron$ data to analyse clump temperature but 
only obtain results for 54\,per cent of the clumps they detect in 850\,$\micron$. Calculated temperatures become 
increasingly unreliable at higher values to the extent they can only define a lower limit of 30\,K for temperatures 
above this value. 

%This survey is deeper than most SCUBA studies, this is demonstrated through an order of magnitude improvement in completeness with one out every three clumps we detect being the below the completeness limit of 0.4 $M_\odot$ specified by \cite{Johnstone:2000fk}.

The lower noise levels and wider coverage at 450\,$\micron$ from SCUBA-2 offer improved quality and 
quantity to the extent that temperature maps can be constructed for many features in star-forming regions.

\subsection{The state of star formation in Serpens MWC 297} 

%[MWC 297 shows evidence for active star formation within the last 3Myr]

Star formation is active and ongoing over a wide range of physical stages, from prestellar objects to Class III 
PMS-stars. We have detected 22 clumps in SCUBA-2 850\,$\micron$ data using the clump-finding algorithm 
\textsc{fellwalker} (Table~\ref{tab:mass}), from which we classify eight as YSOcs through consistency with 24\,$\micron$ 
data and the \SpitzerGB\ YSOc catalogue. We include an additional \emph{Spitzer}-detected YSOc (YSOc11) 
which was missed by \textsc{fellwalker} to provide us with a sample size of nine (Table~\ref{table:cores}), 
in addition to the 10 $M_\odot$ ZAMS star MWC 297. Seven (YSOc2, 11, 17, 32, 41, 47, 73) of these are found 
in the \SpitzerGB\ YSOc catalogues and two in the general \SpitzerGB\ source catalogue. Three Class 0, three 
Class I and three Class II sources are classified with SCUBA-2 data. 

72 Class II/III and 10 Class 0/I sources are listed in the \SpitzerGB\ catalogue for the region. We do not expect 
to detect a high proportion of the Class II objects or any Class III objects with SCUBA-2. Figure \ref{fig:YSO} 
shows how few of these objects lie within the 3$\sigma$ detection level. We do expect to detect all Class 0 
and most Class I objects with SCUBA-2 and therefore four (YSOc 15, 16, 21, 38) out of 10 Class 0/I sources 
listed in the \SpitzerGB\ catalogue that are not associated with SCUBA-2 peaks should be considered with 
scepticism. The remaining 16 objects identified by \textsc{fellwalker} are considered to be prestellar objects 
and diffuse clouds. From the SCUBA-2 catalogue every stage in star formation is represented up to stars on 
the main sequence. Given the assumed lifetime of each class, star formation has been active in this region 
for at least 3\,Myr.

%[Star formation is ongoing but will be limited in the future]

Star formation is observed at various stages in five large-scale clouds in the region which are composed 
of a number of fragmented clumps (Figure~\ref{fig:clumps}), the most evolved of which contain star forming 
cores. S2-YSOc 1 represents the most massive core we detect at 5.1$\pm$0.5\,$M_{\mathrm{\odot}}$ and is the 
most prominent object in a larger cloud of mass 21$\pm$2\,$M_{\mathrm{\odot}}$ - (see Figure~\ref{fig:minimaps1}). 
S2-YSOc 1 is the coolest YSO we have observed with mean temperature of 10.3$\pm$0.5\,K and there is no 
evidence of heating in this region. If all the mass detected in S2-YSOc 1 accretes onto the core, allowing for a 
star-formation efficiency of 30 per cent \citep{evans09}, this object may go on to form an intermediate 
mass star similar to MWC 297. 

A second cloud appears somewhat less fragmented with only two objects as opposed to four but also 
less massive with a peak core mass of 1.3$\pm$0.3\,$M_{\mathrm{\odot}}$ and total cloud mass of 
3.5\,$M_{\mathrm{\odot}}$ (Figure~\ref{fig:minimaps1} - S2-YSOc 2). Likewise a 30 per cent star-formation 
efficiency would limit the final mass to around 1 $M_{\mathrm{\odot}}$. S2-YSOc 3 and 6 (Figures~\ref{fig:minimaps1}) 
form a potentially loosely bound proto-binary composed of a Class 0 and Class I object with separation of 
10,000\,AU and masses 0.95$\pm$0.08 and 0.62$\pm$0.06\,$M_{\mathrm{\odot}}$. 
  
In addition to these deeply embedded, less evolved objects, a number of more evolved, isolated objects 
were observed. S2-YSOc 10, 18 and 29 are detached from the larger clouds and are much less luminous than the 
younger objects (Figure~\ref{fig:minimaps1}). At these stages, PMS-stars are dominated by disks rather than 
envelopes and we calculate masses of 0.31$\pm$0.03\,$M_{\mathrm{\odot}}$, 0.09$\pm$0.02\,$M_{\mathrm{\odot}}$ 
and 0.30$\pm$0.03\,$M_{\mathrm{\odot}}$ for these objects. The protostar to PMS ratios suggest that these objects 
may have been formed in a dense region and later ejected or that the associated molecular cloud was larger in the 
past. Typical core migration speeds of 1\,pc per Myr are consistent with the size of the observed region 
(30\arcmin\ diameter) and birth of these objects in one of the large clouds, most likely that associated with the star 
MWC 297 as it is the most evolved. S2-YSOc 11 and 16 are likely transition cores between Class I and II 
stages (Figure~\ref{fig:minimaps1}). 

%[Remaining dense cores have low mass]

The remaining objects are not considered to be star-forming. The most massive of these are SMM 5 and 
7 at 3.5$\pm$0.3 and 3.1$\pm$0.2\,$M_{\mathrm{\odot}}$ (see Figure~\ref{fig:minimaps1}). 
We calculate free fall timescales of 2.1 and 1.8\,Myrs for these objects. These are significantly larger 
than the typical protostellar timescale of 0.5\,Myr are therefore unlikely to form stars without accreting 
mass or cooling further. The mean temperature of starless clumps is over twice that of star-forming cores 
(32$\pm$4\,K to 15$\pm$2\,K). Our observed core temperature is consistent with the assumption made in 
Section 4.2 and used by \cite{Johnstone:2000fk} and \cite{Kirk:2006vn}. The remaining objects all have 
masses less than 1\,$M_{\mathrm{\odot}}$ and are too diffuse to produce reliable temperature data. If these 
objects go onto to form stars, they are unlikely to form anything more massive than a brown dwarf. 

%[summary discussion]

A global analysis of the region reveals that, of a total cloud mass of 40\,$M_{\mathrm{\odot}}$, only 
12.5\,$M_{\mathrm{\odot}}$ is not currently associated with ongoing star formation. Assuming a mean YSO 
mass of 0.5\,$M_{\mathrm{\odot}}$ based of IMF observations \citep{Chabrier:2005nx, evans09}, and given 
a mass of MWC 297 of 10\,$M_{\odot}$ \citep{Drew:1997qf}, we conclude that the total stellar 
(Class II or higher) mass of the region is 46\,$M_{\odot}$. To date, approximately 85 per cent of the 
original cloud mass has gone into forming stars. From this we conclude that once this current generation 
of stars are formed, there is unlikely to be any further massive star formation without further mass accreting 
from the diffuse ISM and as a result we envisage a large distribution of low mass objects with the massive 
MWC 297 system dominating the region. 

\subsection{What does SCUBA-2 tell us about the star MWC 297?}

 %[What do we already know about MWC 297]
 
The B1.5Ve star MWC 297 is a well known object. We comment on its relevant features and refer the reader to 
\cite{Sandell:2011dz} for a comprehensive review the star's properties. 
 
%[MWC 297 is located in Physically associated with the star-forming clouds]
 
MWC 297 is considered to be physically associated with the YSOcs within a 1\arcmin\ radius identified in \SpitzerGB\ and the additional 
YSO catalogues identified in Table~\ref{tab:YSO_cat} and displayed in Figure~\ref{fig:YSO}. MWC 297, objects 2MASS J18273854-0350108 
(undetected in SCUBA-2) and 2MASS J18273670-0350047 (detected as S2-YSOc 11 in SCUBA-2) were found to have a mean group 
velocity of 0.01\arcsec per year \citep{Roeser:2008zr, Zacharias:2012uq, Zacharias:2013kx} providing evidence they were formed from 
same cloud. Further evidence in 24\,$\micron$ data shown in Figure~\ref{fig:mwc}c shows how emission from warm dust heated by MWC 297, 
associated with SH2-62, is consistent with the location of dust clouds in the SCUBA-2 data. The angular distance between MWC 297 and the 
nearest clump (SMM 4) detected in SCUBA-2 amounts to a minimum physical separation of 5,000\,AU, approximately half the size of our definition 
of a core (0.05 pc, \citeauthor{Rygl:2013ve} \citeyear{Rygl:2013ve}). 
 
%[SCUBA-2 detections are consistent with free-free and residual]

We determine that free-free emission from an UCH\textrm{II} region and polar jets/winds associated with MWC 297 contaminates the 450\,$\micron$ 
and 850\,$\micron$ data \citep{Skinner:1993bh}. The nature of the free-free emission from the outflow has been debated by various authors. 
\cite{Malbet:2007zr} and \cite{Manoj:2007ly} argue for ionised stellar winds that dominate at higher latitudes, whereas \cite{Skinner:1993bh} 
and \cite{Sandell:2011dz} provide evidence for an additional source of free-free emission in the form of highly collimated polar jets. Jets are 
typically associated with less evolved objects where luminosity is dominated by accretion processes whereas MWC 297 is considered 
to be a Class III / ZAMS star where the majority of the disk has fallen onto the star or been dissipated by winds. X-ray flares are thought to be a 
signature of episodic accretion and \cite{Damiani:2006ve} detect a number of X-rays flares from the Serpens MWC 297 region but find that 
only 5.5\,per cent of total flaring is directly associated with MWC 297, suggesting that accretion onto it is minimal. The majority of X-ray emission is 
associated with additional YSOs and the companion of MWC 297, \emph{OSCA}, an A2V star identified by \cite{Habart:2003fk} and 
\cite{Vink:2005uq} at a separation of 850\,AU. 

Figure~\ref{fig:freefreeSED} and Figure~\ref{fig:contamination} show that free-free emission due to an UCH\textrm{II} region and polar winds/jets 
is responsible for the majority of flux from the star MWC 297. Original peak fluxes of 188$\pm$16\,mJy and 86$\pm$22\,mJy are reduced to 51$\pm$11\,mJy 
and 15$\pm$4\,mJy at 450\,$\micron$ and 850\,$\micron$ respectively. The 5$\sigma$ level of 82\,mJy and 11\,mJy means that flux is too uncertain to 
be detected at 450$\micron$ and therefore it is not possible to calculate reliable temperatures of the residual circumstellar envelope/disk around 
the star. The assumption of point-like free-free emission may add further uncertainty to the residual flux.

%[SCUBA-2 fluxes are not consistent with previous claims of an accretion disk - now explained by free-free]

Previous observations have interpreted a submillimetre source consistent with the location of MWC 297 as an accretion disk or circumstellar envelope 
\citep{di-Francesco:1994dq, Drew:1997qf,di-Francesco:1998fk}. We believe that these observations can now be explained as free-free emission. 
\cite{Manoj:2007ly} constrain the disk radius with radio observations to 80\,AU and calculate a disk mass of M = 0.07\,$M_{\mathrm{\odot}}$. These results are 
supported by \cite{Alonso-Albi:2009ve} who conclude that this `exceptionally low' disk mass is partly due to photoionisation by an UCH\textrm{II} region. 
Further work by \cite{Alonso-Albi:2009ve} argues for the presence of a cold circumstellar envelope. Free-free does not account for emission at 70\,$\micron$ 
and 100\,$\micron$ as shown in the SED for MWC 297 (Figure~\ref{fig:SED}) due to the exponential cutoff of the free-free power law as emission becomes 
optically thick at shorter wavelengths. 

Our results do not rule out the presence of a disk or residual envelope following subtraction of the free-free emission, but they do confirm that any residual 
disk is low mass, though with a high degree of uncertainty as the submillimetre flux observed at the position of MWC 297 likely 
contains a component from the clump SMM 4 whichs overlaps this location. Temperature information about MWC 297 is also limited by the diminished 
size of the residual emission. We note that throughout this paper we have assumed a constant value of $\beta$ = 1.8. We have argued this a fair 
assumption for the ISM and extended envelope but this does not hold for the local environment of the protostar where the value of $\beta$ is known to 
be lower, leading to higher dust temperatures (see Figure~\ref{fig:SR_temp}).

Based on these observations we suggest the following arrangement whereby we are observing both the B star, MWC 297, and the companion A star, 
\emph{OSCA}. MWC 297 has evolved further to the extent that it is producing the UCH\textrm{II} region observed. We find it unlikely that such a system 
could still be accreting matter on a large scale, or that the magnetic fields required to produce collimated jets could survive the UCH\textrm{II} region, 
and therefore we associate the jet emission observed by \cite{Skinner:1993bh} to \emph{OSCA}, an object that may be less evolved and more likely to 
still be in the accreting phase. Further evidence for active accretion onto \emph{OSCA} has been provided by \cite{Damiani:2006ve} who found substantial 
X-ray flaring from the object. A more massive disk structure would likely exist around the lower mass, and therefore less evolved, \emph{OSCA} than 
MWC 297 and therefore this is likely the source of any residual SCUBA-2 flux and \emph{Spitzer} MIPS flux observed in the combined SED.  The separation 
at 850\,AU is too small to resolve the two objects with the JCMT beam. 

\subsection{Is there evidence for radiative feedback in Serpens MWC 297?}

%[SMM4 is influenced] \cite{Damiani:2006ve}

The star MWC 297 is directly associated with the star-forming clumps identified in the SCUBA-2 data and the B star is 
directly heating those objects, none more so than SMM 4 where our result suggests that MWC 297 is directly 
influencing its evolution. A mean temperature of 46$\pm$2\,K for SMM 4 was calculated, almost a factor of three 
times higher than the typical clump temperature of 15\,K. The standard deviation of pixels of this clump is high at 11\,K. 
The clump is warmest around the exterior with temperatures peaking above 55\,K (potentially contributed to by edge effects) 
but it appears to have a cooler centre of 29\,K (Figure~\ref{fig:minimaps1}). This is warmer than the mean temperature (18\,K) 
of all the other clumps (discounting SMM 22 on account of its small size) detected by \textsc{fellwalker}. Heating of this 
object is not internal and the ISRF is not sufficient to produce such high  temperatures. Only MWC 297 can provide 
sufficient external heating. 

SMM 4 has a dust mass of 0.91$\pm$0.05 $M_{\mathrm{\odot}}$ but is the fourth most luminous clump in the region with a well 
defined, centrally condensed core. Raised temperatures mean that the object is gravitationally stable with a $M_{\mathrm{850}}/$$M_{\mathrm{J}}$
ratio of 0.12. From these results we conclude that, in the past, SMM 4 may have begun collapse on a similar timescale 
to MWC 297; however upon the B star producing sufficient radiation, MWC 297 has directly heated the larger part of the 
neighbouring clump to the extent that gravitational collapse is no longer possible, in effect suppressing, or even halting, 
the star formation process. Whether or not the low mass of SMM 4 or the power of MWC 297 is the limiting factor in this 
process remains unknown.

%[Other cores are cold]

The majority of the other clumps detected show little or no external heating and no objects show evidence of 
internal heating. Table~\ref{tab:mass} outlines a range of mean clump dust temperatures, between 10 and 46\,K, 
across the region. This is wider than the range of 12 to 20\,K assumed by \cite{Motte:1998ys} for Ophiuchus. 
Examining the mean temperatures of the Class 0 objects we find values of 12.6$\pm$0.9\,K, below the assumed 
15\,K used by \cite{Johnstone:2000fk} and \cite{Kirk:2006vn} but within the range of \cite{Motte:1998ys}. Of the six 
Class 0/I objects, two (S2-YSOc 2 and 10) lie within the nebulosity whereas the remainder lie in regions with little significant 
emission from large scale heated dust as shown in Figure~\ref{fig:minimaps1}. None of the YSOcs show significant heating. 
However, use of a constant $\beta$ may not hold towards the centre of a protostar and our use of $\beta$ = 1.8 specifically 
for large structures maybe be systematically underestimating temperatures in these regions.

Starless object SMM 7 shows heating (Figure~\ref{fig:minimaps1}) along its eastern edge which is not consistent 
with the 24\,$\micron$ emission. We suspect we are observing the `edge effect' artefact produced in the map 
making process and this consequently increases temperature to 25$\pm$2\,K. Conversely SMM 5 shows evidence 
of heated gas along its western edge in 24\,$\micron$ emission (Figure~\ref{fig:mwc}c) but is relatively cool and 
homogenous in Figure~\ref{fig:minimaps1} with a mean temperature of 18.2$\pm$0.9\,K. Prohibitively high noise 
in the 450\,$\micron$ data prevent wider examination of this feature.

%Such low values infer grain growth and the earliest stages of a protoplanetary disk. Planet construction is not typically associated with O or B stars due to heating and ionisation of the disk [CITE]. \cite{Malbet:2007zr} has found evidence of puffing up the inner parts of the disk due to heating and it is difficult to reconcile these conditions with those that favour grain growth, though the two disk model of \cite{Alonso-Albi:2009ve} offers room for planetary formation. 
%At this angle, MWC 297 would resemble a UX Orionis object \citep{Pontoppidan:2007fr}, where by the optically thick disk and circumstellar envelope cast a shadow over the light of the central star.

%%%%%%%%%%%%%%%%%%%%%%%%%%%%%%%%%%%%%%%%%%%%%%%%%%%

\section{Conclusion}

We observed Serpens MWC 297 region with SCUBA-2 at 450\,$\micron$ and 850\,$\micron$ as part of the JCMT 
Gould Belt Survey of nearby star-forming regions. The observations covers a 30\arcmin\ diameter circular region centred 
on RA(J2000) = $18^{h}$ $28^{m}$ $13{\farcs}8$, Dec. (J2000) = $-03^{\circ}$ $44'$ 1.7\arcsec\ including the 
B1.5Ve Herbig Be star MWC 297 and a collection of local dense clouds. We use the clump-finding algorithm 
\textsc{fellwalker} to identity submillimetre clumps in the data and compare our catalogue to YSOc catalogues produced 
by the \emph{Spitzer} Gould Belt Survey (\SpitzerGB), and to \emph{Spitzer} 24\,$\micron$ data of the region. 
The latter shows heating of surrounding clouds associated with the star MWC 297 and the optical nebula SH2-62,  providing 
evidence that the two are physically located in space.  

We account for sources of submillimetre contamination, finding an insignificant CO contamination estimated at 
13\,per cent but a significant amount from free-free emission as the result of an ultra-compact HII region and polar winds/jets 
associated with the star MWC 297. We use the ratio of 450\,$\micron$ and 850\,$\micron$ to build maps of dust temperature 
for Serpens MWC 297 with the aim of investigating evidence for radiative heating in the region. To do this we employed 
a method whereby each dataset is convolved with both the primary and secondary beam components of the JCMT 
beam at the other wavelength to achieve like resolution of 19.9\arcsec\ before calculating the flux ratio and consequently temperature. 
\\
\\
Our key results are:
\begin{enumerate}
\item Our temperature method uses both the primary and secondary 
components of the JCMT beam as this better reflects the shape of the 
real beam. The two component model decreases temperatures between 
5 and 9 per cent in the warmest and coolest regions respectively. 
\item We detect 22 clumps. By cross referencing this list with \emph{Spitzer} 
YSOcs and a comparison of mass to Jeans mass as a test of 
gravitational stability, we identify nine YSOcs
\item We calculated masses based on calculated temperatures (as 
opposed to an assumed value) across the whole region. Clump masses 
range between 0.02-19\,M$\odot$ and core masses range between 
0.09-5.1\,M$\odot$. Starless clumps are consistently warmer than star 
forming cores with mean temperatures of 32$\pm$4\,K compared to 
15$\pm$2\,K. 
\item We classify the YSOs using $T_{\mathrm{bol}}$ and 
$L_{\mathrm{smm}}$/$L_{\mathrm{bol}}$ as two Class 0, one 
Class 0/I, three Class I and three Class II sources. 30\,per cent of 
Class 0/I objects and 8\,per cent of Class II objects catalogued in 
\SpitzerGB\ were also detected by SCUBA-2. No Class III objects 
were detected by SCUBA-2. SCUBA-2 detected two potential Class 0 
and one Class I/II YSOcs that were not included in the \SpitzerGB\ 
YSOc catalogue.   
\item We modelled free-free emission from MWC 297 as a point-source 
with a spectral index of $\alpha = 1.03\pm0.02$. This contamination 
accounted for 73\,per cent and 83\,per cent of peak flux at 450\,$\micron$ 
and 850\,$\micron$ respectively. Residual peak fluxes were 51$\pm$10\,mJy 
and 15$\pm$3\,mJy respectively. The residual submillimetre emission for 
MWC 297 was insufficiently bright to be distinguishable from a larger 
clump (SMM 4) projected behind it on the sky.
\item We conclude that radiative heating from one generation of stars 
is directly influencing the formation of another, but we note that the effect 
is not large across the region. Our findings suggest that clump SMM 4 
had begun collapsing before radiative heating from MWC 297 raised 
the temperatures to 46$\pm$2\,K, to the extent that gravitational 
collapse is now suppressed or even halted. 
\end{enumerate}

Serpens MWC 297 region represents a low mass star-forming region with a limited number of YSOcs. We believe that in the future, 
this region will become dominated by the HII region associated with the star MWC 297. The expansion and shock front of this region 
will likely play an important role in the subsequent evolution of the cores and clumps we have detected. Further work will look 
at expanding these methods to produce temperature maps for larger regions within Serpens-Aquila, with a particular eye to 
possible free-free contamination where OB stars are observed to be forming.
\\

\emph{The James Clerk Maxwell Telescope has historically been operated by the Joint Astronomy Centre on behalf of the Science and 
Technology Facilities Council of the United Kingdom, the National Research Council of Canada and the Netherlands Organisation 
for Scientific Research. Additional funds for the construction of SCUBA-2 were provided by the Canada Foundation for Innovation. 
The identification number for the  program under which the SCUBA-2 data used in this paper is MJLSG33. This work was supported 
by a STFC studentship (Rumble) and the Exeter STFC consolidated grant (Hatchell). We would like to thank G$\ddot{o}$ran Sandell for the 
contribution of VLA data and the referee for their helpful feedback throughout the publishing process.}

%\clearpage
%%%%%%%%%%%%%%%%%%%%%%%%%%%%%%%%%%%%%%%%%%%%%%%%%%%

%\begin{thebibliography}{99}
\bibliographystyle{mn2e}
\bibliography{bibliography}

\begin{thebibliography}{}

\bibitem[\protect\citeauthoryear{{Alonso-Albi}, {Fuente}, {Bachiller}, {Neri},
  {Planesas}, {Testi}, {Bern{\'e}} \& {Joblin}}{{Alonso-Albi}
  et~al.}{2009}]{Alonso-Albi:2009ve}
{Alonso-Albi} T.,  {Fuente} A.,  {Bachiller} R.,  {Neri} R.,  {Planesas} P.,
  {Testi} L.,  {Bern{\'e}} O.,    {Joblin} C.,  2009, \aap, 497, 117

\bibitem[\protect\citeauthoryear{{Andr{\'e}}, {Men'shchikov}, {Bontemps},
  {K{\"o}nyves}, {Motte}, {Schneider}, {Didelon}, {Minier}, {Saraceno} \&
  {Ward-Thompson}}{{Andr{\'e}} et~al.}{2010}]{Andre:2010kx}
{Andr{\'e}} P.,  {Men'shchikov} A.,  {Bontemps} S.,  {K{\"o}nyves} V.,  {Motte}
  F.,  {Schneider} N.,  {Didelon} P.,  {Minier} V.,  {Saraceno} P.,
  {Ward-Thompson} D.,  2010, \aap, 518, L102

\bibitem[\protect\citeauthoryear{{Andr{\'e}} \& {Motte}}{{Andr{\'e}} \&
  {Motte}}{2000}]{Andre:2000zr}
{Andr{\'e}} P.,  {Motte} F.,  2000, in {Favata} F.,  {Kaas} A.,   {Wilson} A.,
  eds, Star Formation from the Small to the Large Scale Vol.~445 of ESA Special
  Publication, {FIRST and the Earliest Stages of Star Formation}.
p.~219

\bibitem[\protect\citeauthoryear{{Andr{\'e}}, {Ward-Thompson} \&
  {Barsony}}{{Andr{\'e}} et~al.}{1993}]{Andre:1993ff}
{Andr{\'e}} P.,  {Ward-Thompson} D.,    {Barsony} M.,  1993, \apj, 406, 122

\bibitem[\protect\citeauthoryear{{Arce} \& {Goodman}}{{Arce} \&
  {Goodman}}{1999}]{Arce:1999bh}
{Arce} H.~G.,  {Goodman} A.~A.,  1999, \apj, 517, 264

\bibitem[\protect\citeauthoryear{{Bate}}{{Bate}}{2009}]{Bate:2009uq}
{Bate} M.~R.,  2009, \mnras, 392, 1363

\bibitem[\protect\citeauthoryear{{Bergner}, {Kozlov}, {Krivtsov},
  {Miroshnichenko}, {Yudin}, {Yutanov}, {Dzhakusheva}, {Kuratov} \&
  {Mukanov}}{{Bergner} et~al.}{1988}]{Bergner:1988bh}
{Bergner} Y.~K.,  {Kozlov} V.~P.,  {Krivtsov} A.~A.,  {Miroshnichenko} A.~S.,
  {Yudin} R.~V.,  {Yutanov} N.~Y.,  {Dzhakusheva} K.~G.,  {Kuratov} K.~S.,
  {Mukanov} B.~D.,  1988, Astrofizika, 28, 529

\bibitem[\protect\citeauthoryear{{Berrilli}, {Corciulo}, {Ingrosso},
  {Lorenzetti}, {Nisini} \& {Strafella}}{{Berrilli}
  et~al.}{1992}]{Berrilli:1992cr}
{Berrilli} F.,  {Corciulo} G.,  {Ingrosso} G.,  {Lorenzetti} D.,  {Nisini} B.,
    {Strafella} F.,  1992, \apj, 398, 254

\bibitem[\protect\citeauthoryear{{Berry}, {Reinhold}, {Jenness} \&
  {Economou}}{{Berry} et~al.}{2007}]{Berry:2007vn}
{Berry} D.~S.,  {Reinhold} K.,  {Jenness} T.,    {Economou} F.,  2007, in
  {Shaw} R.~A.,  {Hill} F.,   {Bell} D.~J.,  eds, Astronomical Data Analysis
  Software and Systems XVI Vol.~376 of Astronomical Society of the Pacific
  Conference Series, {CUPID: A Clump Identification and Analysis Package}.
p.~425

\bibitem[\protect\citeauthoryear{{Berry}, {Reinhold}, {Jenness} \&
  {Economou}}{{Berry} et~al.}{2013}]{Berry:2013uq}
{Berry} D.~S.,  {Reinhold} K.,  {Jenness} T.,    {Economou} F., , 2013, {CUPID:
  Clump Identification and Analysis Package}

\bibitem[\protect\citeauthoryear{{Bonnor}}{{Bonnor}}{1956}]{Bonnor:1956vn}
{Bonnor} W.~B.,  1956, \mnras, 116, 351

\bibitem[\protect\citeauthoryear{{Bontemps}, {Andr{\'e}}, {K{\"o}nyves},
  {Men'shchikov}, {Schneider}, {Maury}, {Peretto}, {Arzoumanian}, {Attard},
  {Motte} \& {Minier}}{{Bontemps} et~al.}{2010}]{Bontemps:2010fk}
{Bontemps} S.,  {Andr{\'e}} P.,  {K{\"o}nyves} V.,  {Men'shchikov} A.,
  {Schneider} N.,  {Maury} A.,  {Peretto} N.,  {Arzoumanian} D.,  {Attard} M.,
  {Motte} F.,    {Minier} V.,  2010, \aap, 518, L85

\bibitem[\protect\citeauthoryear{{Bontemps}, {Andre}, {Terebey} \&
  {Cabrit}}{{Bontemps} et~al.}{1996}]{Bontemps:1996fu}
{Bontemps} S.,  {Andre} P.,  {Terebey} S.,    {Cabrit} S.,  1996, \aap, 311,
  858

\bibitem[\protect\citeauthoryear{{Calvet} \& {Gullbring}}{{Calvet} \&
  {Gullbring}}{1998}]{calvet98}
{Calvet} N.,  {Gullbring} E.,  1998, \apj, 509, 802

\bibitem[\protect\citeauthoryear{{Canto}, {Rodriguez}, {Calvet} \&
  {Levreault}}{{Canto} et~al.}{1984}]{Canto:1984dq}
{Canto} J.,  {Rodriguez} L.~F.,  {Calvet} N.,    {Levreault} R.~M.,  1984,
  \apj, 282, 631

\bibitem[\protect\citeauthoryear{{Chabrier}}{{Chabrier}}{2005}]{Chabrier:2005nx}
{Chabrier} G.,  2005, in {Corbelli} E.,  {Palla} F.,   {Zinnecker} H.,  eds,
  The Initial Mass Function 50 Years Later Vol.~327 of Astrophysics and Space
  Science Library, {The Initial Mass Function: from Salpeter 1955 to 2005}.
p.~41

\bibitem[\protect\citeauthoryear{{Chandler}, {Gear}, {Sandell}, {Hayashi},
  {Duncan}, {Griffin} \& {Hazella}}{{Chandler} et~al.}{1990}]{Chandler:1990qf}
{Chandler} C.~J.,  {Gear} W.~K.,  {Sandell} G.,  {Hayashi} S.,  {Duncan} W.~D.,
   {Griffin} M.~J.,    {Hazella} S.,  1990, \mnras, 243, 330

\bibitem[\protect\citeauthoryear{{Chapin}, {Berry}, {Gibb}, {Jenness}, {Scott},
  {Tilanus}, {Economou} \& {Holland}}{{Chapin} et~al.}{2013}]{Chapin:2013vn}
{Chapin} E.~L.,  {Berry} D.~S.,  {Gibb} A.~G.,  {Jenness} T.,  {Scott} D.,
  {Tilanus} R.~P.~J.,  {Economou} F.,    {Holland} W.~S.,  2013, \mnras, 430,
  2545

\bibitem[\protect\citeauthoryear{{Chen}, {Myers}, {Ladd} \& {Wood}}{{Chen}
  et~al.}{1995}]{Chen:1995ys}
{Chen} H.,  {Myers} P.~C.,  {Ladd} E.~F.,    {Wood} D.~O.~S.,  1995, \apj, 445,
  377

\bibitem[\protect\citeauthoryear{{Connelley} \& {Greene}}{{Connelley} \&
  {Greene}}{2010}]{Connelley:2010nx}
{Connelley} M.~S.,  {Greene} T.~P.,  2010, \aj, 140, 1214

\bibitem[\protect\citeauthoryear{{Curtis}, {Richer}, {Swift} \&
  {Williams}}{{Curtis} et~al.}{2010}]{Curtis:2010zr}
{Curtis} E.~I.,  {Richer} J.~S.,  {Swift} J.~J.,    {Williams} J.~P.,  2010,
  \mnras, 408, 1516

\bibitem[\protect\citeauthoryear{{Damiani}, {Micela} \& {Sciortino}}{{Damiani}
  et~al.}{2006}]{Damiani:2006ve}
{Damiani} F.,  {Micela} G.,    {Sciortino} S.,  2006, \aap, 447, 1041

\bibitem[\protect\citeauthoryear{{Deharveng}, {Zavagno}, {Anderson}, {Motte},
  {Abergel}, {Andr{\'e}}, {Bontemps}, {Leleu}, {Roussel} \&
  {Russeil}}{{Deharveng} et~al.}{2012}]{Deharveng:2012fk}
{Deharveng} L.,  {Zavagno} A.,  {Anderson} L.~D.,  {Motte} F.,  {Abergel} A.,
  {Andr{\'e}} P.,  {Bontemps} S.,  {Leleu} G.,  {Roussel} H.,    {Russeil} D.,
  2012, \aap, 546, A74

\bibitem[\protect\citeauthoryear{{Dempsey}, {Friberg}, {Jenness}, {Tilanus},
  {Thomas}, {Holland}, {Bintley}, {Berry}, {Chapin}, {Chrysostomou}, {Davis},
  {Gibb}, {Parsons} \& {Robson}}{{Dempsey} et~al.}{2013}]{Dempsey:2013uq}
{Dempsey} J.~T.,  {Friberg} P.,  {Jenness} T.,  {Tilanus} R.~P.~J.,  {Thomas}
  H.~S.,  {Holland} W.~S.,  {Bintley} D.,  {Berry} D.~S.,  {Chapin} E.~L.,
  {Chrysostomou} A.,  {Davis} G.~R.,  {Gibb} A.~G.,  {Parsons} H.,    {Robson}
  E.~I.,  2013, \mnras, 430, 2534

\bibitem[\protect\citeauthoryear{{Di Francesco}, {Evans} II, {Caselli},
  {Myers}, {Shirley}, {Aikawa} \& {Tafalla}}{{Di Francesco}
  et~al.}{2007}]{di-Francesco:2007bh}
{Di Francesco} J.,  {Evans} II N.~J.,  {Caselli} P.,  {Myers} P.~C.,  {Shirley}
  Y.,  {Aikawa} Y.,    {Tafalla} M.,  2007, Protostars and Planets V, pp 17--32

\bibitem[\protect\citeauthoryear{{Di Francesco}, {Evans} II, {Harvey}, {Mundy}
  \& {Butner}}{{Di Francesco} et~al.}{1994}]{di-Francesco:1994dq}
{Di Francesco} J.,  {Evans} II N.~J.,  {Harvey} P.~M.,  {Mundy} L.~G.,
  {Butner} H.~M.,  1994, \apj, 432, 710

\bibitem[\protect\citeauthoryear{{Di Francesco}, {Evans} II, {Harvey}, {Mundy}
  \& {Butner}}{{Di Francesco} et~al.}{1998}]{di-Francesco:1998fk}
{Di Francesco} J.,  {Evans} II N.~J.,  {Harvey} P.~M.,  {Mundy} L.~G.,
  {Butner} H.~M.,  1998, \apj, 509, 324

\bibitem[\protect\citeauthoryear{{Drabek}, {Hatchell}, {Friberg}, {Richer},
  {Graves}, {Buckle}, {Nutter}, {Johnstone} \& {Di Francesco}}{{Drabek}
  et~al.}{2012}]{Drabek:2012uq}
{Drabek} E.,  {Hatchell} J.,  {Friberg} P.,  {Richer} J.,  {Graves} S.,
  {Buckle} J.~V.,  {Nutter} D.,  {Johnstone} D.,    {Di Francesco} J.,  2012,
  \mnras, 426, 23

\bibitem[\protect\citeauthoryear{{Drew}, {Busfield}, {Hoare}, {Murdoch},
  {Nixon} \& {Oudmaijer}}{{Drew} et~al.}{1997}]{Drew:1997qf}
{Drew} J.~E.,  {Busfield} G.,  {Hoare} M.~G.,  {Murdoch} K.~A.,  {Nixon} C.~A.,
     {Oudmaijer} R.~D.,  1997, \mnras, 286, 538

\bibitem[\protect\citeauthoryear{{Dunham}, {Crapsi}, {Evans} II, {Bourke},
  {Huard}, {Myers} \& {Kauffmann}}{{Dunham} et~al.}{2008}]{Dunham:2008fk}
{Dunham} M.~M.,  {Crapsi} A.,  {Evans} II N.~J.,  {Bourke} T.~L.,  {Huard}
  T.~L.,  {Myers} P.~C.,    {Kauffmann} J.,  2008, \apjs, 179, 249

\bibitem[\protect\citeauthoryear{{Dzib}, {Loinard}, {Mioduszewski}, {Boden},
  {Rodr{\'{\i}}guez} \& {Torres}}{{Dzib} et~al.}{2010}]{Dzib:2010dq}
{Dzib} S.,  {Loinard} L.,  {Mioduszewski} A.~J.,  {Boden} A.~F.,
  {Rodr{\'{\i}}guez} L.~F.,    {Torres} R.~M.,  2010, \apj, 718, 610

\bibitem[\protect\citeauthoryear{{Dzib}, {Loinard}, {Mioduszewski}, {Boden},
  {Rodr{\'{\i}}guez} \& {Torres}}{{Dzib} et~al.}{2011}]{Dzib:2011cr}
{Dzib} S.,  {Loinard} L.,  {Mioduszewski} A.~J.,  {Boden} A.~F.,
  {Rodr{\'{\i}}guez} L.~F.,    {Torres} R.~M.,  2011, in Revista Mexicana de
  Astronomia y Astrofisica Conference Series Vol.~40 of Revista Mexicana de
  Astronomia y Astrofisica Conference Series, {VLBA astrometry of the AeBe star
  EC 95 in Serpens}.
pp 231--232

\bibitem[\protect\citeauthoryear{{Ebert}}{{Ebert}}{1955}]{Ebert:1955vn}
{Ebert} R.,  1955, \zap, 37, 217

\bibitem[\protect\citeauthoryear{{Eiroa}, {Casali}, {Miranda} \&
  {Ortiz}}{{Eiroa} et~al.}{1994}]{Eiroa:1994ve}
{Eiroa} C.,  {Casali} M.~M.,  {Miranda} L.~F.,    {Ortiz} E.,  1994, \aap, 290,
  599

\bibitem[\protect\citeauthoryear{{Eiroa}, {Djupvik} \& {Casali}}{{Eiroa}
  et~al.}{2008}]{Eiroa:2008ta}
{Eiroa} C.,  {Djupvik} A.~A.,    {Casali} M.~M.,  2008, in {Reipurth} B.,  ed.,
  , Handbook of Star Forming Regions, Volume II.
Monograph, p.~693

\bibitem[\protect\citeauthoryear{{Enoch}, {Corder}, {Duch{\^e}ne}, {Bock},
  {Bolatto}, {Culverhouse}, {Kwon}, {Lamb}, {Leitch}, {Marrone}, {Muchovej},
  {P{\'e}rez}, {Scott}, {Teuben}, {Wright} \& {Zauderer}}{{Enoch}
  et~al.}{2011}]{Enoch:2011lh}
{Enoch} M.~L.,  {Corder} S.,  {Duch{\^e}ne} G.,  {Bock} D.~C.,  {Bolatto}
  A.~D.,  {Culverhouse} T.~L.,  {Kwon} W.,  {Lamb} J.~W.,  {Leitch} E.~M.,
  {Marrone} D.~P.,  {Muchovej} S.~J.,  {P{\'e}rez} L.~M.,  {Scott} S.~L.,
  {Teuben} P.~J.,  {Wright} M.~C.~H.,    {Zauderer} B.~A.,  2011, \apjs, 195,
  21

\bibitem[\protect\citeauthoryear{{Enoch}, {Evans} II, {Sargent} \&
  {Glenn}}{{Enoch} et~al.}{2009}]{Enoch:2009xq}
{Enoch} M.~L.,  {Evans} II N.~J.,  {Sargent} A.~I.,    {Glenn} J.,  2009, \apj,
  692, 973

\bibitem[\protect\citeauthoryear{{Evans}, {Allen}, {Blake}, {Boogert},
  {Bourke}, {Harvey}, {Kessler}, {Koerner}, {Lee}, {Mundy}, {Myers}, {Padgett},
  {Pontoppidan}, {Sargent}, {Stapelfeldt}, {van Dishoeck}, {Young} \&
  {Young}}{{Evans} et~al.}{2003}]{c2d}
{Evans} N.~J.~{\sc II}.,  {Allen} L.~E.,  {Blake} G.~A.,  {Boogert} A.~C.~A.,
  {Bourke} T.,  {Harvey} P.~M.,  {Kessler} J.~E.,  {Koerner} D.~W.,  {Lee}
  C.~W.,  {Mundy} L.~G.,  {Myers} P.~C.,  {Padgett} D.~L.,  {Pontoppidan} K.,
  {Sargent} A.~I.,  {Stapelfeldt} K.~R.,  {van Dishoeck} E.~F.,  {Young} C.~H.,
     {Young} K.~E.,  2003, \pasp, 115, 965

\bibitem[\protect\citeauthoryear{{Evans}, {Dunham}, {J{\o}rgensen}, {Enoch},
  {Mer{\'{\i}}n}, {van Dishoeck}, {Alcal{\'a}}, {Myers} \&
  {Stapelfeldt}}{{Evans} et~al.}{2009}]{evans09}
{Evans} N.~J.~{\sc II}.,  {Dunham} M.~M.,  {J{\o}rgensen} J.~K.,  {Enoch}
  M.~L.,  {Mer{\'{\i}}n} B.,  {van Dishoeck} E.~F.,  {Alcal{\'a}} J.~M.,
  {Myers} P.~C.,    {Stapelfeldt} K.~R.,  2009, \apjs, 181, 321

\bibitem[\protect\citeauthoryear{{Evans}, {Harvey}, {Dunham}, {Huard}, {Mundy},
  {Lai}, {Chapman}, {Brooke}, {Enoch} \& {Stapelfeldt}}{{Evans}
  et~al.}{2007}]{c2ddel}
{Evans} N.~J.~{\sc II}.,  {Harvey} P.~M.,  {Dunham} M.,  {Huard} T.,  {Mundy}
  L.,  {Lai} S.-p.,  {Chapman} N.,  {Brooke} T.,  {Enoch} M.,    {Stapelfeldt}
  K.,  2007, Technical report, Final Delivery of Data from the c2d Legacy
  Project: IRAC and MIPS.
Spitzer Science Center, Pasadena, CA

\bibitem[\protect\citeauthoryear{{Fazio}, {Hora}, {Allen}, {Ashby}, {Barmby},
  {Deutsch}, {Huang} \& {Kleiner}}{{Fazio} et~al.}{2004}]{fazio04}
{Fazio} G.~G.,  {Hora} J.~L.,  {Allen} L.~E.,  {Ashby} M.~L.~N.,  {Barmby} P.,
  {Deutsch} L.~K.,  {Huang} J.-S.,    {Kleiner} S.,  2004, \apjs, 154, 10

\bibitem[\protect\citeauthoryear{{Font}, {Mitchell} \& {Sandell}}{{Font}
  et~al.}{2001}]{Font:2001cr}
{Font} A.~S.,  {Mitchell} G.~F.,    {Sandell} G.,  2001, \apj, 555, 950

\bibitem[\protect\citeauthoryear{{Greene}, {Wilking}, {Andre}, {Young} \&
  {Lada}}{{Greene} et~al.}{1994}]{Greene:1994kl}
{Greene} T.~P.,  {Wilking} B.~A.,  {Andre} P.,  {Young} E.~T.,    {Lada} C.~J.,
   1994, \apj, 434, 614

\bibitem[\protect\citeauthoryear{{Gutermuth}, {Bourke}, {Allen}, {Myers},
  {Megeath}, {Matthews}, {J{\o}rgensen}, {Di Francesco}, {Ward-Thompson},
  {Huard}, {Brooke}, {Dunham}, {Cieza}, {Harvey} \& {Chapman}}{{Gutermuth}
  et~al.}{2008}]{gutermuth08}
{Gutermuth} R.~A.,  {Bourke} T.~L.,  {Allen} L.~E.,  {Myers} P.~C.,  {Megeath}
  S.~T.,  {Matthews} B.~C.,  {J{\o}rgensen} J.~K.,  {Di Francesco} J.,
  {Ward-Thompson} D.,  {Huard} T.~L.,  {Brooke} T.~Y.,  {Dunham} M.~M.,
  {Cieza} L.~A.,  {Harvey} P.~M.,    {Chapman} N.~L.,  2008, \apjl, 673, L151

\bibitem[\protect\citeauthoryear{{Gutermuth}, {Megeath}, {Myers}, {Allen},
  {Pipher} \& {Fazio}}{{Gutermuth} et~al.}{2009}]{gutermuth09}
{Gutermuth} R.~A.,  {Megeath} S.~T.,  {Myers} P.~C.,  {Allen} L.~E.,  {Pipher}
  J.~L.,    {Fazio} G.~G.,  2009, \apjs, 184, 18

\bibitem[\protect\citeauthoryear{{Habart}, {Testi}, {Natta} \&
  {Vanzi}}{{Habart} et~al.}{2003}]{Habart:2003fk}
{Habart} E.,  {Testi} L.,  {Natta} A.,    {Vanzi} L.,  2003, \aap, 400, 575

\bibitem[\protect\citeauthoryear{{Harvey}, {Mer{\'{\i}}n}, {Huard}, {Rebull},
  {Chapman}, {Evans} II \& {Myers}}{{Harvey} et~al.}{2007}]{harvey07}
{Harvey} P.,  {Mer{\'{\i}}n} B.,  {Huard} T.~L.,  {Rebull} L.~M.,  {Chapman}
  N.,  {Evans} II N.~J.,    {Myers} P.~C.,  2007, \apj, 663, 1149

\bibitem[\protect\citeauthoryear{{Harvey}, {Huard}, {J{\o}rgensen},
  {Gutermuth}, {Mamajek}, {Bourke}, {Mer{\'{\i}}n}, {Cieza}, {Brooke},
  {Chapman}, {Alcal{\'a}}, {Allen}, {Evans} II, {Di Francesco} \&
  {Kirk}}{{Harvey} et~al.}{2008}]{harvey08}
{Harvey} P.~M.,  {Huard} T.~L.,  {J{\o}rgensen} J.~K.,  {Gutermuth} R.~A.,
  {Mamajek} E.~E.,  {Bourke} T.~L.,  {Mer{\'{\i}}n} B.,  {Cieza} L.,  {Brooke}
  T.,  {Chapman} N.,  {Alcal{\'a}} J.~M.,  {Allen} L.~E.,  {Evans} II N.~J.,
  {Di Francesco} J.,    {Kirk} J.~M.,  2008, \apj, 680, 495

\bibitem[\protect\citeauthoryear{{Hatchell}, {Fuller}, {Richer}, {Harries} \&
  {Ladd}}{{Hatchell} et~al.}{2007}]{Hatchell:2007qf}
{Hatchell} J.,  {Fuller} G.~A.,  {Richer} J.~S.,  {Harries} T.~J.,    {Ladd}
  E.~F.,  2007, \aap, 468, 1009

\bibitem[\protect\citeauthoryear{{Hatchell}, {Richer}, {Fuller}, {Qualtrough},
  {Ladd} \& {Chandler}}{{Hatchell} et~al.}{2005}]{Hatchell:2005fk}
{Hatchell} J.,  {Richer} J.~S.,  {Fuller} G.~A.,  {Qualtrough} C.~J.,  {Ladd}
  E.~F.,    {Chandler} C.~J.,  2005, \aap, 440, 151

\bibitem[\protect\citeauthoryear{{Hatchell}, {Terebey}, {Huard}, {Mamajek},
  {Allen}, {Bourke}, {Dunham}, {Gutermuth}, {Harvey}, {J{\o}rgensen},
  {Mer{\'{\i}}n}, {Noriega-Crespo} \& {Peterson}}{{Hatchell}
  et~al.}{2012}]{hatchell12}
{Hatchell} J.,  {Terebey} S.,  {Huard} T.,  {Mamajek} E.~E.,  {Allen} L.,
  {Bourke} T.~L.,  {Dunham} M.~M.,  {Gutermuth} R.,  {Harvey} P.~M.,
  {J{\o}rgensen} J.~K.,  {Mer{\'{\i}}n} B.,  {Noriega-Crespo} A.,    {Peterson}
  D.~E.,  2012, \apj, 754, 104

\bibitem[\protect\citeauthoryear{{Hatchell}, {Wilson}, {Drabek}, {Curtis},
  {Richer}, {Nutter}, {Di Francesco}, {Ward-Thompson} \& {JCMT GBS
  Consortium}}{{Hatchell} et~al.}{2013}]{Hatchell:2013ij}
{Hatchell} J.,  {Wilson} T.,  {Drabek} E.,  {Curtis} E.,  {Richer} J.,
  {Nutter} D.,  {Di Francesco} J.,  {Ward-Thompson} D.,    {JCMT GBS
  Consortium} 2013, \mnras, 429, L10

\bibitem[\protect\citeauthoryear{{Hennebelle} \& {Chabrier}}{{Hennebelle} \&
  {Chabrier}}{2011}]{Hennebelle:2011ly}
{Hennebelle} P.,  {Chabrier} G.,  2011, \apjl, 743, L29

\bibitem[\protect\citeauthoryear{{Henning} \& {Sablotny}}{{Henning} \&
  {Sablotny}}{1995}]{Henning:1995qf}
{Henning} T.,  {Sablotny} R.~M.,  1995, Advances in Space Research, 16, 17

\bibitem[\protect\citeauthoryear{{Herbig}}{{Herbig}}{1960}]{Herbig:1960eu}
{Herbig} G.~H.,  1960, \apjs, 4, 337

\bibitem[\protect\citeauthoryear{{Hildebrand}}{{Hildebrand}}{1983}]{Hildebrand:1983fy}
{Hildebrand} R.~H.,  1983, QJRAS, 24, 267

\bibitem[\protect\citeauthoryear{{Hillenbrand}, {Strom}, {Vrba} \&
  {Keene}}{{Hillenbrand} et~al.}{1992}]{Hillenbrand:1992kl}
{Hillenbrand} L.~A.,  {Strom} S.~E.,  {Vrba} F.~J.,    {Keene} J.,  1992, \apj,
  397, 613

\bibitem[\protect\citeauthoryear{{Holland}, {MacIntosh}, {Fairley}, {Kelly},
  {Montgomery}, {Gostick}, {Atad-Ettedgui}, {Ellis}, {Robson}, {Hollister} \&
  {Woodcraft}}{{Holland} et~al.}{2006}]{Holland:2006uq}
{Holland} W.,  {MacIntosh} M.,  {Fairley} A.,  {Kelly} D.,  {Montgomery} D.,
  {Gostick} D.,  {Atad-Ettedgui} E.,  {Ellis} M.,  {Robson} I.,  {Hollister}
  M.,    {Woodcraft} A.,  2006, in Society of Photo-Optical Instrumentation
  Engineers (SPIE) Conference Series Vol.~6275 of Society of Photo-Optical
  Instrumentation Engineers (SPIE) Conference Series, {SCUBA-2: a 10,000-pixel
  submillimeter camera for the James Clerk Maxwell Telescope}

\bibitem[\protect\citeauthoryear{{Holland}, {Bintley}, {Chapin},
  {Chrysostomou}, {Davis}, {Dempsey}, {Duncan}, {Fich}, {Friberg}, {Halpern},
  {Irwin}, {Jenness}, {Kelly}, {MacIntosh} \& {Robson}}{{Holland}
  et~al.}{2013}]{Holland:2013fk}
{Holland} W.~S.,  {Bintley} D.,  {Chapin} E.~L.,  {Chrysostomou} A.,  {Davis}
  G.~R.,  {Dempsey} J.~T.,  {Duncan} W.~D.,  {Fich} M.,  {Friberg} P.,
  {Halpern} M.,  {Irwin} K.~D.,  {Jenness} T.,  {Kelly} B.~D.,  {MacIntosh}
  M.~J.,    {Robson} E.~I.,  2013, \mnras, 430, 2513

\bibitem[\protect\citeauthoryear{{Huttemeister}, {Wilson}, {Henkel} \&
  {Mauersberger}}{{Huttemeister} et~al.}{1993}]{Huttemeister:1993ve}
{Huttemeister} S.,  {Wilson} T.~L.,  {Henkel} C.,    {Mauersberger} R.,  1993,
  \aap, 276, 445

\bibitem[\protect\citeauthoryear{{Jeans}}{{Jeans}}{1902}]{Jeans:1902dz}
{Jeans} J.~H.,  1902, Royal Society of London Philosophical Transactions Series
  A, 199, 1

\bibitem[\protect\citeauthoryear{{Jenness}, {Chapin}, {Berry}, {Gibb},
  {Tilanus}, {Balfour}, {Tilanus} \& {Currie}}{{Jenness}
  et~al.}{2013}]{Jenness:2013fk}
{Jenness} T.,  {Chapin} E.~L.,  {Berry} D.~S.,  {Gibb} A.~G.,  {Tilanus}
  R.~P.~J.,  {Balfour} J.,  {Tilanus} V.,    {Currie} M.~J., , 2013, {SMURF:
  SubMillimeter User Reduction Facility}

\bibitem[\protect\citeauthoryear{{Johnstone}, {Wilson}, {Moriarty-Schieven},
  {Joncas}, {Smith}, {Gregersen} \& {Fich}}{{Johnstone}
  et~al.}{2000}]{Johnstone:2000fk}
{Johnstone} D.,  {Wilson} C.~D.,  {Moriarty-Schieven} G.,  {Joncas} G.,
  {Smith} G.,  {Gregersen} E.,    {Fich} M.,  2000, \apj, 545, 327

\bibitem[\protect\citeauthoryear{{Juvela}, {Ristorcelli}, {Pelkonen},
  {Marshall}, {Montier}, {Bernard}, {Paladini} \& {Lunttila}}{{Juvela}
  et~al.}{2011}]{Juvela:2011ys}
{Juvela} M.,  {Ristorcelli} I.,  {Pelkonen} V.-M.,  {Marshall} D.~J.,
  {Montier} L.~A.,  {Bernard} J.-P.,  {Paladini} R.,    {Lunttila} T.,  2011,
  \aap, 527, A111

\bibitem[\protect\citeauthoryear{{Kaas}, {Olofsson}, {bontemps}, {Andr{\'e}},
  {Nordh}, {Huldtgren}, {Prusti}, {Persi}, {Delgado}, {Motte} \&
  {Abergel}}{{Kaas} et~al.}{2004}]{Kaas:2004fx}
{Kaas} A.~A.,  {Olofsson} G.,  {bontemps} S.,  {Andr{\'e}} P.,  {Nordh} L.,
  {Huldtgren} M.,  {Prusti} T.,  {Persi} P.,  {Delgado} A.~J.,  {Motte} F.,
  {Abergel} A.,  2004, \aap, 421, 623

\bibitem[\protect\citeauthoryear{{Kirk}, {Johnstone} \& {Di Francesco}}{{Kirk}
  et~al.}{2006}]{Kirk:2006vn}
{Kirk} H.,  {Johnstone} D.,    {Di Francesco} J.,  2006, \apj, 646, 1009

\bibitem[\protect\citeauthoryear{{Kirk}, {Ward-Thompson}, {Di Francesco},
  {Bourke}, {Evans}, {Mer{\'{\i}}n}, {Allen}, {Cieza}, {Dunham}, {Harvey},
  {Huard}, {J{\o}rgensen}, {Miller}, {Noriega-Crespo}, {Peterson}, {Ray} \&
  {Rebull}}{{Kirk} et~al.}{2009}]{kirk09}
{Kirk} J.~M.,  {Ward-Thompson} D.,  {Di Francesco} J.,  {Bourke} T.~L.,
  {Evans} N.~J.,  {Mer{\'{\i}}n} B.,  {Allen} L.~E.,  {Cieza} L.~A.,  {Dunham}
  M.~M.,  {Harvey} P.,  {Huard} T.,  {J{\o}rgensen} J.~K.,  {Miller} J.~F.,
  {Noriega-Crespo} A.,  {Peterson} D.,  {Ray} T.~P.,    {Rebull} L.~M.,  2009,
  \apjs, 185, 198

\bibitem[\protect\citeauthoryear{{Koenig}, {Allen}, {Gutermuth}, {Hora},
  {Brunt} \& {Muzerolle}}{{Koenig} et~al.}{2008}]{Koenig:2008jo}
{Koenig} X.~P.,  {Allen} L.~E.,  {Gutermuth} R.~A.,  {Hora} J.~L.,  {Brunt}
  C.~M.,    {Muzerolle} J.,  2008, \apj, 688, 1142

\bibitem[\protect\citeauthoryear{{K{\"o}nyves}, {Andr{\'e}}, {Men'shchikov},
  {Schneider}, {Arzoumanian}, {Bontemps}, {Attard}, {Motte} \&
  {Didelon}}{{K{\"o}nyves} et~al.}{2010}]{Konyves:2010oq}
{K{\"o}nyves} V.,  {Andr{\'e}} P.,  {Men'shchikov} A.,  {Schneider} N.,
  {Arzoumanian} D.,  {Bontemps} S.,  {Attard} M.,  {Motte} F.,    {Didelon} P.,
   2010, \aap, 518, L106

\bibitem[\protect\citeauthoryear{{Kraemer}, {Jackson}, {Kassis}, {Deutsch},
  {Hora}, {Simon}, {Hoffmann}, {Fazio}, {Dayal}, {Bania}, {Clemens} \&
  {Heyer}}{{Kraemer} et~al.}{2003}]{Kraemer:2003uf}
{Kraemer} K.~E.,  {Jackson} J.~M.,  {Kassis} M.,  {Deutsch} L.~K.,  {Hora}
  J.~L.,  {Simon} R.,  {Hoffmann} W.~F.,  {Fazio} G.~G.,  {Dayal} A.,  {Bania}
  T.~M.,  {Clemens} D.~P.,    {Heyer} M.~H.,  2003, \apj, 588, 918

\bibitem[\protect\citeauthoryear{{Lada} \& {Wilking}}{{Lada} \&
  {Wilking}}{1984}]{Lada:1984fk}
{Lada} C.~J.,  {Wilking} B.~A.,  1984, \apj, 287, 610

\bibitem[\protect\citeauthoryear{{Ladd}, {Myers} \& {Goodman}}{{Ladd}
  et~al.}{1994}]{Ladd:1994ly}
{Ladd} E.~F.,  {Myers} P.~C.,    {Goodman} A.~A.,  1994, \apj, 433, 117

\bibitem[\protect\citeauthoryear{{Mairs}, {Johnstone}, {Offner} \&
  {Schnee}}{{Mairs} et~al.}{2014}]{Mairs:2014zr}
{Mairs} S.,  {Johnstone} D.,  {Offner} S.~S.~R.,    {Schnee} S.,  2014, \apj,
  783, 60

\bibitem[\protect\citeauthoryear{{Malbet}, {Benisty}, {de Wit}, {Kraus},
  {Meilland}, {Millour}, {Tatulli}, {Berger}, {Chesneau}, {Hofmann}, {Isella},
  {Natta}, {Petrov}, {Preibisch}, {Stee}, {Testi}, {Weigelt}, {Antonelli} \&
  {Beckmann}}{{Malbet} et~al.}{2007}]{Malbet:2007zr}
{Malbet} F.,  {Benisty} M.,  {de Wit} W.-J.,  {Kraus} S.,  {Meilland} A.,
  {Millour} F.,  {Tatulli} E.,  {Berger} J.-P.,  {Chesneau} O.,  {Hofmann}
  K.-H.,  {Isella} A.,  {Natta} A.,  {Petrov} R.~G.,  {Preibisch} T.,  {Stee}
  P.,  {Testi} L.,  {Weigelt} G.,  {Antonelli} P.,    {Beckmann} U.,  2007,
  \aap, 464, 43

\bibitem[\protect\citeauthoryear{{Mannings}}{{Mannings}}{1994}]{Mannings:1994kx}
{Mannings} V.,  1994, \mnras, 271, 587

\bibitem[\protect\citeauthoryear{{Manoj}, {Ho}, {Ohashi}, {Zhang}, {Hasegawa},
  {Chen}, {Bhatt} \& {Ashok}}{{Manoj} et~al.}{2007}]{Manoj:2007ly}
{Manoj} P.,  {Ho} P.~T.~P.,  {Ohashi} N.,  {Zhang} Q.,  {Hasegawa} T.,  {Chen}
  H.-R.,  {Bhatt} H.~C.,    {Ashok} N.~M.,  2007, \apjl, 667, L187

\bibitem[\protect\citeauthoryear{{Mathis}, {Mezger} \& {Panagia}}{{Mathis}
  et~al.}{1983}]{Mathis:1983dq}
{Mathis} J.~S.,  {Mezger} P.~G.,    {Panagia} N.,  1983, \aap, 128, 212

\bibitem[\protect\citeauthoryear{{Maury}, {Andr{\'e}}, {Men'shchikov},
  {K{\"o}nyves} \& {Bontemps}}{{Maury} et~al.}{2011}]{Maury:2011ys}
{Maury} A.~J.,  {Andr{\'e}} P.,  {Men'shchikov} A.,  {K{\"o}nyves} V.,
  {Bontemps} S.,  2011, \aap, 535, A77

\bibitem[\protect\citeauthoryear{{Men'shchikov}, {Andr{\'e}}, {Didelon},
  {K{\"o}nyves}, {Schneider}, {Motte}, {Bontemps}, {Arzoumanian}, {Attard},
  {Abergel}, {Baluteau}, {Bernard} \& {Cambr{\'e}sy}}{{Men'shchikov}
  et~al.}{2010}]{Menshchikov:2010kl}
{Men'shchikov} A.,  {Andr{\'e}} P.,  {Didelon} P.,  {K{\"o}nyves} V.,
  {Schneider} N.,  {Motte} F.,  {Bontemps} S.,  {Arzoumanian} D.,  {Attard} M.,
   {Abergel} A.,  {Baluteau} J.-P.,  {Bernard} J.-P.,    {Cambr{\'e}sy} L.,
  2010, \aap, 518, L103

\bibitem[\protect\citeauthoryear{{Mitchell}, {Johnstone}, {Moriarty-Schieven},
  {Fich} \& {Tothill}}{{Mitchell} et~al.}{2001}]{Mitchell:2001ve}
{Mitchell} G.~F.,  {Johnstone} D.,  {Moriarty-Schieven} G.,  {Fich} M.,
  {Tothill} N.~F.~H.,  2001, \apj, 556, 215

\bibitem[\protect\citeauthoryear{{Motte}, {Andre} \& {Neri}}{{Motte}
  et~al.}{1998}]{Motte:1998ys}
{Motte} F.,  {Andre} P.,    {Neri} R.,  1998, \aap, 336, 150

\bibitem[\protect\citeauthoryear{{Myers}, {Adams}, {Chen} \& {Schaff}}{{Myers}
  et~al.}{1998}]{Myers:1998ys}
{Myers} P.~C.,  {Adams} F.~C.,  {Chen} H.,    {Schaff} E.,  1998, \apj, 492,
  703

\bibitem[\protect\citeauthoryear{{Myers} \& {Ladd}}{{Myers} \&
  {Ladd}}{1993}]{Myers:1993vn}
{Myers} P.~C.,  {Ladd} E.~F.,  1993, \apjl, 413, L47

\bibitem[\protect\citeauthoryear{{Nutter} \& {Ward-Thompson}}{{Nutter} \&
  {Ward-Thompson}}{2007}]{Nutter:2007ys}
{Nutter} D.,  {Ward-Thompson} D.,  2007, \mnras, 374, 1413

\bibitem[\protect\citeauthoryear{{Offner}, {Klein}, {McKee} \&
  {Krumholz}}{{Offner} et~al.}{2009}]{Offner:2009pt}
{Offner} S.~S.~R.,  {Klein} R.~I.,  {McKee} C.~F.,    {Krumholz} M.~R.,  2009,
  \apj, 703, 131

\bibitem[\protect\citeauthoryear{{Ossenkopf} \& {Henning}}{{Ossenkopf} \&
  {Henning}}{1994}]{Ossenkopf:1994vn}
{Ossenkopf} V.,  {Henning} T.,  1994, \aap, 291, 943

\bibitem[\protect\citeauthoryear{{Peterson}, {Caratti o Garatti}, {Bourke},
  {Forbrich}, {Gutermuth}, {J{\o}rgensen}, {Allen}, {Patten}, {Dunham},
  {Harvey}, {Mer{\'{\i}}n}, {Chapman}, {Cieza}, {Huard}, {Knez}, {Prager} \&
  {Evans}}{{Peterson} et~al.}{2011}]{peterson11}
{Peterson} D.~E.,  {Caratti o Garatti} A.,  {Bourke} T.~L.,  {Forbrich} J.,
  {Gutermuth} R.~A.,  {J{\o}rgensen} J.~K.,  {Allen} L.~E.,  {Patten} B.~M.,
  {Dunham} M.~M.,  {Harvey} P.~M.,  {Mer{\'{\i}}n} B.,  {Chapman} N.~L.,
  {Cieza} L.~A.,  {Huard} T.~L.,  {Knez} C.,  {Prager} B.,    {Evans} N.~J.,
  2011, \apjs, 194, 43

\bibitem[\protect\citeauthoryear{{Pilbratt}, {Riedinger}, {Passvogel}, {Crone},
  {Doyle}, {Gageur}, {Heras}, {Jewell}, {Metcalfe}, {Ott} \&
  {Schmidt}}{{Pilbratt} et~al.}{2010}]{Pilbratt:2010fk}
{Pilbratt} G.~L.,  {Riedinger} J.~R.,  {Passvogel} T.,  {Crone} G.,  {Doyle}
  D.,  {Gageur} U.,  {Heras} A.~M.,  {Jewell} C.,  {Metcalfe} L.,  {Ott} S.,
  {Schmidt} M.,  2010, \aap, 518, L1

\bibitem[\protect\citeauthoryear{{Reid} \& {Wilson}}{{Reid} \&
  {Wilson}}{2005}]{Reid:2005ly}
{Reid} M.~A.,  {Wilson} C.~D.,  2005, \apj, 625, 891

\bibitem[\protect\citeauthoryear{{Rieke}, {Young}, {Engelbracht}, {Kelly},
  {Low}, {Haller} \& {Beeman}}{{Rieke} et~al.}{2004}]{rieke04}
{Rieke} G.~H.,  {Young} E.~T.,  {Engelbracht} C.~W.,  {Kelly} D.~M.,  {Low}
  F.~J.,  {Haller} E.~E.,    {Beeman} J.~W.,  2004, \apjs, 154, 25

\bibitem[\protect\citeauthoryear{{Robitaille}, {Whitney}, {Indebetouw} \&
  {Wood}}{{Robitaille} et~al.}{2007}]{Robitaille:2007zr}
{Robitaille} T.~P.,  {Whitney} B.~A.,  {Indebetouw} R.,    {Wood} K.,  2007,
  \apjs, 169, 328

\bibitem[\protect\citeauthoryear{{Roeser}, {Schilbach}, {Schwan}, {Kharchenko},
  {Piskunov} \& {Scholz}}{{Roeser} et~al.}{2008}]{Roeser:2008zr}
{Roeser} S.,  {Schilbach} E.,  {Schwan} H.,  {Kharchenko} N.~V.,  {Piskunov}
  A.~E.,    {Scholz} R.-D.,  2008, VizieR Online Data Catalog, 1312, 0

\bibitem[\protect\citeauthoryear{{Rosolowsky} \& {Leroy}}{{Rosolowsky} \&
  {Leroy}}{2006}]{Rosolowsky:2006bh}
{Rosolowsky} E.,  {Leroy} A.,  2006, \pasp, 118, 590

\bibitem[\protect\citeauthoryear{{Rosolowsky}, {Pineda}, {Foster}, {Borkin},
  {Kauffmann}, {Caselli}, {Myers} \& {Goodman}}{{Rosolowsky}
  et~al.}{2008}]{Rosolowsky:2008dq}
{Rosolowsky} E.~W.,  {Pineda} J.~E.,  {Foster} J.~B.,  {Borkin} M.~A.,
  {Kauffmann} J.,  {Caselli} P.,  {Myers} P.~C.,    {Goodman} A.~A.,  2008,
  \apjs, 175, 509

\bibitem[\protect\citeauthoryear{{Rygl}, {Benedettini}, {Schisano}, {Elia},
  {Molinari}, {Pezzuto}, {Andr{\'e}}, {Bernard}, {White}, {Polychroni},
  {Bontemps}, {Cox} \& {Di Francesco}}{{Rygl} et~al.}{2013}]{Rygl:2013ve}
{Rygl} K.~L.~J.,  {Benedettini} M.,  {Schisano} E.,  {Elia} D.,  {Molinari} S.,
   {Pezzuto} S.,  {Andr{\'e}} P.,  {Bernard} J.~P.,  {White} G.~J.,
  {Polychroni} D.,  {Bontemps} S.,  {Cox} N.~L.~J.,    {Di Francesco} J.,
  2013, \aap, 549, L1

\bibitem[\protect\citeauthoryear{{Sadavoy}, {Di Francesco}, {Andr{\'e}},
  {Pezzuto}, {Bernard}, {Maury}, {Men'shchikov} \& {Motte}}{{Sadavoy}
  et~al.}{2014}]{Sadavoy:2014nx}
{Sadavoy} S.~I.,  {Di Francesco} J.,  {Andr{\'e}} P.,  {Pezzuto} S.,  {Bernard}
  J.-P.,  {Maury} A.,  {Men'shchikov} A.,    {Motte} F.,  2014, \apjl, 787, L18

\bibitem[\protect\citeauthoryear{{Sadavoy}, {Di Francesco}, {Bontemps},
  {Megeath}, {Rebull}, {Allgaier}, {Carey}, {Gutermuth}, {Hora}, {Huard},
  {McCabe}, {Muzerolle}, {Noriega-Crespo}, {Padgett} \& {Terebey}}{{Sadavoy}
  et~al.}{2010}]{Sadavoy:2010ve}
{Sadavoy} S.~I.,  {Di Francesco} J.,  {Bontemps} S.,  {Megeath} S.~T.,
  {Rebull} L.~M.,  {Allgaier} E.,  {Carey} S.,  {Gutermuth} R.,  {Hora} J.,
  {Huard} T.,  {McCabe} C.-E.,  {Muzerolle} J.,  {Noriega-Crespo} A.,
  {Padgett} D.,    {Terebey} S.,  2010, \apj, 710, 1247

\bibitem[\protect\citeauthoryear{{Sadavoy}, {Di Francesco}, {Johnstone},
  {Currie}, {Drabek}, {Hatchell}, {Nutter}, {Andr{\'e}}, {Arzoumanian} \&
  {Benedettini}}{{Sadavoy} et~al.}{2013}]{Sadavoy:2013qf}
{Sadavoy} S.~I.,  {Di Francesco} J.,  {Johnstone} D.,  {Currie} M.~J.,
  {Drabek} E.,  {Hatchell} J.,  {Nutter} D.,  {Andr{\'e}} P.,  {Arzoumanian}
  D.,    {Benedettini} M.,  2013, \apj, 767, 126

\bibitem[\protect\citeauthoryear{{Salji}}{{Salji}}{2014}]{Salji:2013kx}
{Salji} C.,  2014, in Protostars and Planets VI Posters {Filament
  identification and characterisation in Gould Belt Clouds}.
p.~19

\bibitem[\protect\citeauthoryear{{Sandell}, {Weintraub} \&
  {Hamidouche}}{{Sandell} et~al.}{2011}]{Sandell:2011dz}
{Sandell} G.,  {Weintraub} D.~A.,    {Hamidouche} M.,  2011, \apj, 727, 26

\bibitem[\protect\citeauthoryear{{Schnee}, {Mason}, {Di Francesco}, {Friesen},
  {Li}, {Sadavoy} \& {Stanke}}{{Schnee} et~al.}{2014}]{Schnee:2014uq}
{Schnee} S.,  {Mason} B.,  {Di Francesco} J.,  {Friesen} R.,  {Li} D.,
  {Sadavoy} S.,    {Stanke} T.,  2014, ArXiv e-prints:

\bibitem[\protect\citeauthoryear{{Schnee}, {Ridge}, {Goodman} \& {Li}}{{Schnee}
  et~al.}{2005}]{Schnee:2005zr}
{Schnee} S.~L.,  {Ridge} N.~A.,  {Goodman} A.~A.,    {Li} J.~G.,  2005, \apj,
  634, 442

\bibitem[\protect\citeauthoryear{{Sharpless}}{{Sharpless}}{1959}]{Sharpless:1959hc}
{Sharpless} S.,  1959, \apjs, 4, 257

\bibitem[\protect\citeauthoryear{{Shirley}, {Evans} II \& {Rawlings}}{{Shirley}
  et~al.}{2002}]{Shirley:2002vn}
{Shirley} Y.~L.,  {Evans} II N.~J.,    {Rawlings} J.~M.~C.,  2002, \apj, 575,
  337

\bibitem[\protect\citeauthoryear{{Shirley}, {Evans} II, {Rawlings} \&
  {Gregersen}}{{Shirley} et~al.}{2000}]{Shirley:2000uq}
{Shirley} Y.~L.,  {Evans} II N.~J.,  {Rawlings} J.~M.~C.,    {Gregersen} E.~M.,
   2000, \apjs, 131, 249

\bibitem[\protect\citeauthoryear{{Skinner}, {Brown} \& {Stewart}}{{Skinner}
  et~al.}{1993}]{Skinner:1993bh}
{Skinner} S.~L.,  {Brown} A.,    {Stewart} R.~T.,  1993, \apjs, 87, 217

\bibitem[\protect\citeauthoryear{{Spezzi}, {Vernazza}, {Mer{\'{\i}}n}, {Allen},
  {Evans} II, {J{\o}rgensen}, {Bourke}, {Cieza}, {Dunham}, {Harvey}, {Huard},
  {Peterson}, {Tothill} \& {The Gould's Belt Team}}{{Spezzi}
  et~al.}{2011}]{spezzi11}
{Spezzi} L.,  {Vernazza} P.,  {Mer{\'{\i}}n} B.,  {Allen} L.~E.,  {Evans} II
  N.~J.,  {J{\o}rgensen} J.~K.,  {Bourke} T.~L.,  {Cieza} L.~A.,  {Dunham}
  M.~M.,  {Harvey} P.~M.,  {Huard} T.~L.,  {Peterson} D.,  {Tothill} N.~F.~H.,
    {The Gould's Belt Team} 2011, \apj, 730, 65

\bibitem[\protect\citeauthoryear{{Strai{\v z}ys}, {{\v C}ernis} \& {Barta{\v
  s}i{\=u}t{\.e}}}{{Strai{\v z}ys} et~al.}{2003}]{Straizys:2003nx}
{Strai{\v z}ys} V.,  {{\v C}ernis} K.,    {Barta{\v s}i{\=u}t{\.e}} S.,  2003,
  \aap, 405, 585

\bibitem[\protect\citeauthoryear{{Stutz}, {Launhardt}, {Linz}, {Krause},
  {Henning}, {Kainulainen}, {Nielbock}, {Steinacker} \& {Andr{\'e}}}{{Stutz}
  et~al.}{2010}]{Stutz:2010hq}
{Stutz} A.,  {Launhardt} R.,  {Linz} H.,  {Krause} O.,  {Henning} T.,
  {Kainulainen} J.,  {Nielbock} M.,  {Steinacker} J.,    {Andr{\'e}} P.,  2010,
  \aap, 518, L87

\bibitem[\protect\citeauthoryear{{Ubach}, {Maddison}, {Wright}, {Wilner},
  {Lommen} \& {Koribalski}}{{Ubach} et~al.}{2012}]{Ubach:2012fk}
{Ubach} C.,  {Maddison} S.~T.,  {Wright} C.~M.,  {Wilner} D.~J.,  {Lommen}
  D.~J.~P.,    {Koribalski} B.,  2012, \mnras, 425, 3137

\bibitem[\protect\citeauthoryear{{Vink}, {O'Neill}, {Els} \& {Drew}}{{Vink}
  et~al.}{2005}]{Vink:2005uq}
{Vink} J.~S.,  {O'Neill} P.~M.,  {Els} S.~G.,    {Drew} J.~E.,  2005, \aap,
  438, L21

\bibitem[\protect\citeauthoryear{{Visser}, {Richer} \& {Chandler}}{{Visser}
  et~al.}{2002}]{Visser:2002ly}
{Visser} A.~E.,  {Richer} J.~S.,    {Chandler} C.~J.,  2002, \aj, 124, 2756

\bibitem[\protect\citeauthoryear{{Ward-Thompson}, {Di Francesco}, {Hatchell},
  {Hogerheijde}, {Nutter}, {Bastien}, {Basu}, {Bonnell}, {Bowey}, {Brunt} \&
  {Buckle}}{{Ward-Thompson} et~al.}{2007}]{WardThompson:2007ve}
{Ward-Thompson} D.,  {Di Francesco} J.,  {Hatchell} J.,  {Hogerheijde} M.~R.,
  {Nutter} D.,  {Bastien} P.,  {Basu} S.,  {Bonnell} I.,  {Bowey} J.,  {Brunt}
  C.,    {Buckle} J.,  2007, \pasp, 119, 855

\bibitem[\protect\citeauthoryear{Watson}{Watson}{2010}]{watson:2010pc}
Watson M.,  2010, MSc thesis, University of Hertfordshire.

\bibitem[\protect\citeauthoryear{{Winston}, {Megeath}, {Wolk}, {Muzerolle},
  {Gutermuth}, {Hora}, {Allen}, {Spitzbart}, {Myers} \& {Fazio}}{{Winston}
  et~al.}{2007}]{Winston:2007if}
{Winston} E.,  {Megeath} S.~T.,  {Wolk} S.~J.,  {Muzerolle} J.,  {Gutermuth}
  R.,  {Hora} J.~L.,  {Allen} L.~E.,  {Spitzbart} B.,  {Myers} P.,    {Fazio}
  G.~G.,  2007, \apj, 669, 493

\bibitem[\protect\citeauthoryear{{Wood}, {Myers} \& {Daugherty}}{{Wood}
  et~al.}{1994}]{Wood:1994qf}
{Wood} D.~O.~S.,  {Myers} P.~C.,    {Daugherty} D.~A.,  1994, \apjs, 95, 457

\bibitem[\protect\citeauthoryear{{Young}, {Shirley}, {Evans} II \&
  {Rawlings}}{{Young} et~al.}{2003}]{Young:2003fk}
{Young} C.~H.,  {Shirley} Y.~L.,  {Evans} II N.~J.,    {Rawlings} J.~M.~C.,
  2003, \apjs, 145, 111

\bibitem[\protect\citeauthoryear{{Young}, {Harvey}, {Brooke}, {Chapman},
  {Kauffmann}, {Bertoldi}, {Lai}, {Alcal{\'a}}, {Bourke}, {Spiesman}, {Allen},
  {Blake} \& {Evans} II}{{Young} et~al.}{2005}]{Young:2005ly}
{Young} K.~E.,  {Harvey} P.~M.,  {Brooke} T.~Y.,  {Chapman} N.,  {Kauffmann}
  J.,  {Bertoldi} F.,  {Lai} S.-P.,  {Alcal{\'a}} J.,  {Bourke} T.~L.,
  {Spiesman} W.,  {Allen} L.~E.,  {Blake} G.~A.,    {Evans} II N.~J.,  2005,
  \apj, 628, 283

\bibitem[\protect\citeauthoryear{{Zacharias}, {Finch}, {Girard}, {Henden},
  {Bartlett}, {Monet} \& {Zacharias}}{{Zacharias}
  et~al.}{2012}]{Zacharias:2012uq}
{Zacharias} N.,  {Finch} C.~T.,  {Girard} T.~M.,  {Henden} A.,  {Bartlett}
  J.~L.,  {Monet} D.~G.,    {Zacharias} M.~I.,  2012, VizieR Online Data
  Catalog, 1322, 0

\bibitem[\protect\citeauthoryear{{Zacharias}, {Finch}, {Girard}, {Henden},
  {Bartlett}, {Monet} \& {Zacharias}}{{Zacharias}
  et~al.}{2013}]{Zacharias:2013kx}
{Zacharias} N.,  {Finch} C.~T.,  {Girard} T.~M.,  {Henden} A.,  {Bartlett}
  J.~L.,  {Monet} D.~G.,    {Zacharias} M.~I.,  2013, \aj, 145, 44

\bibitem[\protect\citeauthoryear{{Ziener} \& {Eisl{\"o}ffel}}{{Ziener} \&
  {Eisl{\"o}ffel}}{1999}]{Ziener:1999kl}
{Ziener} R.,  {Eisl{\"o}ffel} J.,  1999, \aap, 347, 565

\end{thebibliography}

%\end{thebibliography}

\clearpage
%%%%%%%%%%%%%%%%%%%%%%%%%%%%%%%%%%%%%%%%%%%%%%%%%%%

\end{document}